# Roadmap on spatiotemporal light fields

**Yijie Shen[1,29,*], Qiwen Zhan[2,3,29,*], Logan G. Wright[4,5], Demetrios N. Christodoulides[6], Frank W. Wise[5], Alan E. Willner[7], Zhe Zhao[7], Kai-heng Zou[7], Chen-Ting Liao[8], Carlos Hernández-García[9], Margaret Murnane[8], Miguel A. Porras[10], Andy Chong[11], Chenhao Wan[2,12], Konstantin Y. Bliokh[13], Murat Yessenov[6], Ayman F. Abouraddy[6], Liang Jie Wong[14], Michael Go[14], Suraj Kumar[14], Cheng Guo[15], Shanhui Fan[16], Nikitas Papasimakis[1], Nikolay I. Zheludev[1,17], Lu Chen[18,19], Wenqi Zhu[18,19], Amit Agrawal[18,19], Spencer W. Jolly[20], Christophe Dorrer[21], Benjamín Alonso[9], Ignacio Lopez-Quintas[9], Miguel López-Ripa[9], Íñigo J. Sola[9], Yiqi Fang[22], Qihuang Gong[22,23], Yunquan Liu[22,23], Junyi Huang[24], Hongliang Zhang[24], Zhichao Ruan[24], Mickael Mounaix[25], Nicolas K. Fontaine[26], Joel Carpenter[25], Ahmed H. Dorrah[27], Federico Capasso[27], and Andrew Forbes[28]**

1. Optoelectronics Research Centre & Centre for Photonic Metamaterials, University of Southampton, Southampton SO17 1BJ, UK
2. School of Optical-Electrical and Computer Engineering, University of Shanghai for Science and Technology, Shanghai 200093, China
3. Zhangjiang Laboratory, 100 Haike Road, Shanghai, 201204, China
4. Physics & Informatics Laboratories, NTT Research, Inc., 940 Stewart Drive, Sunnyvale, California 94085, USA
5. School of Applied and Engineering Physics, Cornell University, Ithaca, NY 14853, USA
6. College of Optics & Photonics-CREOL, University of Central Florida, Orlando, FL 32816, USA
7. Department of Electrical Engineering, University of Southern California, Los Angeles, CA 90089, USA
8. JILA and Department of Physics, University of Colorado and NIST, Boulder, CO 80309, USA
9. Grupo de Investigación en Aplicaciones del Láser y Fotónica, Departamento de Física Aplicada, Universidad de Salamanca, 37008 Salamanca, Spain
10. Grupo de Sistemas Complejos, ETSIME, Universidad Politécnica de Madrid, Rios Rosas 21, 28003 Madrid, Spain
11. Physics Department, Pusan National Umiversity, Geumjeong-gu, Busan 46241, South Korea
12. School of Optical and Electronic Information and Wuhan National Laboratory for Optoelectronics, Huazhong University of Science and Technology, Wuhan, Hubei 430074, China
13. Theoretical Quantum Physics Laboratory, RIKEN Cluster for Pioneering Research, Wako-shi, Saitama 351-0198, Japan
14. School of Electrical and Electronic Engineering, Nanyang Technological University, 50 Nanyang Avenue, Singapore 639798, Singapore
15. Department of Applied Physics, Stanford University, Stanford, California 94305, USA
16. Ginzton Laboratory and Department of Electrical Engineering, Stanford University, Stanford, California 94305, USA
17. Centre for Disruptive Photonic Technologies, School of Physical and Mathematical Sciences and The Photonics Institute, Nanyang Technological University, Singapore 637378, Singapore
18. National Institute of Standards and Technology, Gaithersburg, MD 20899, USA
19. Maryland NanoCenter, University of Maryland, College Park, MD 20742, USA
20. Service OPERA-Photonique, Université Libre de Bruxelles (ULB), Brussels, Belgium
21. Laboratory for Laser Energetics, University of Rochester, 250 East River Road, Rochester, NY 14623-1299, USA
22. State Key Laboratory for Mesoscopic Physics and Collaborative Innovation Center of Quantum Matter, School of Physics, Peking University, Beijing 100871, China
23. Beijing Academy of Quantum Information Sciences, Beijing 100193, China
24. Interdisciplinary Center of Quantum Information, State Key Laboratory of Modern Optical Instrumentation, and Zhejiang Province Key Laboratory of Quantum Technology and Device, Department of Physics, Zhejiang University, Hangzhou 310027, China
25. School of Information Technology and Electrical Engineering, The University of Queensland, Brisbane, QLD, 4072, Australia
26. Nokia Bell Labs, 600 Mountain Ave, New Providence, NJ 07974, USA
27. Harvard John A. Paulson School of Engineering and Applied Sciences, Harvard University, Cambridge, MA 02138, USA.
28. School of Physics, University of the Witwatersrand, Private Bag 3, Wits 2050, South Africa
29. Guest editors of the Roadmap.

*Emails: y.shen@soton.ac.uk (Y.S.); qwzhan@usst.edu.cn (Q.Z.)

## Abstract

Spatiotemporal sculpturing of light pulse with ultimately sophisticated structures represents the holy grail of the human everlasting pursue of ultra-fast information transmission and processing as well as ultra-intense energy concentration and extraction. It also holds the key to unlock new extraordinary fundamental physical effects. Traditionally, spatiotemporal light pulses are always treated as spatiotemporally separable wave packet as solution of the Maxwell's equations. In the past decade, however, more generalized forms of spatiotemporally nonseparable solution started to emerge with growing importance for their striking physical effects. This roadmap intends to highlight the recent advances in the creation and control of increasingly complex spatiotemporally sculptured pulses, from spatiotemporally separable to complex nonseparable states, with diverse geometric and topological structures, presenting a bird's eye viewpoint on the zoology of spatiotemporal light fields and the outlook of future trends and open challenges.

## Contents

# 1. Introduction: Composing symphony of light

Yijie Shen, University of Southampton (UK)
Qiwen Zhan, University of Shanghai for Science and Technology (China)

A famous French impressionist musician, Claude-Achilles Debussy, has a famous saying "*Music is the arithmetic of sounds as optics is the geometry of light.*" Indeed, music and optics are inextricably connected through their physical similarity, despite of the fact that they are subjects categorized as art and science respectively. However, in Debussy's age, geometric optics was the mainstream understanding of the nature of light. Soon after the emergency of wave optics, more parallel analogues between music and optics have been unveiled, since fundamentally both the sound and light are wave phenomena. The importance of electromagnetic wave to optics is same as the acoustic wave to music. Therefore, a piece of music played by an instrument is analogue to a temporal optical pulse (Fig.1a and Fig.1b). For convenience of composition, the composer use frequency notation, the stave, to abstractly describe the temporal music (Fig.1c), which can be seem as the process of Fourier transformation (the music player translation of the score is just the inverse Fourier transformer). Similarly, in optics, to characterize a chirped pulse, we usually measure it in the spectral domain to record the pulse information (Fig.1d), rather than directly capture the very fast temporal signal. In retrospect, music evolved from ancient simple solo by a single instrument to the contemporary complex symphony by spatially distributed instruments recorded by complex full score (Fig.1e and Fig.1f). In parallel, recent advances of optics enable researchers to spatiotemporally sculpture light pulses and correspondingly characterize them with novel spectroscopic technology in the spatio-frequency domain (Fig.1g and Fig.1h), opening a whole new chapter of modern optics.

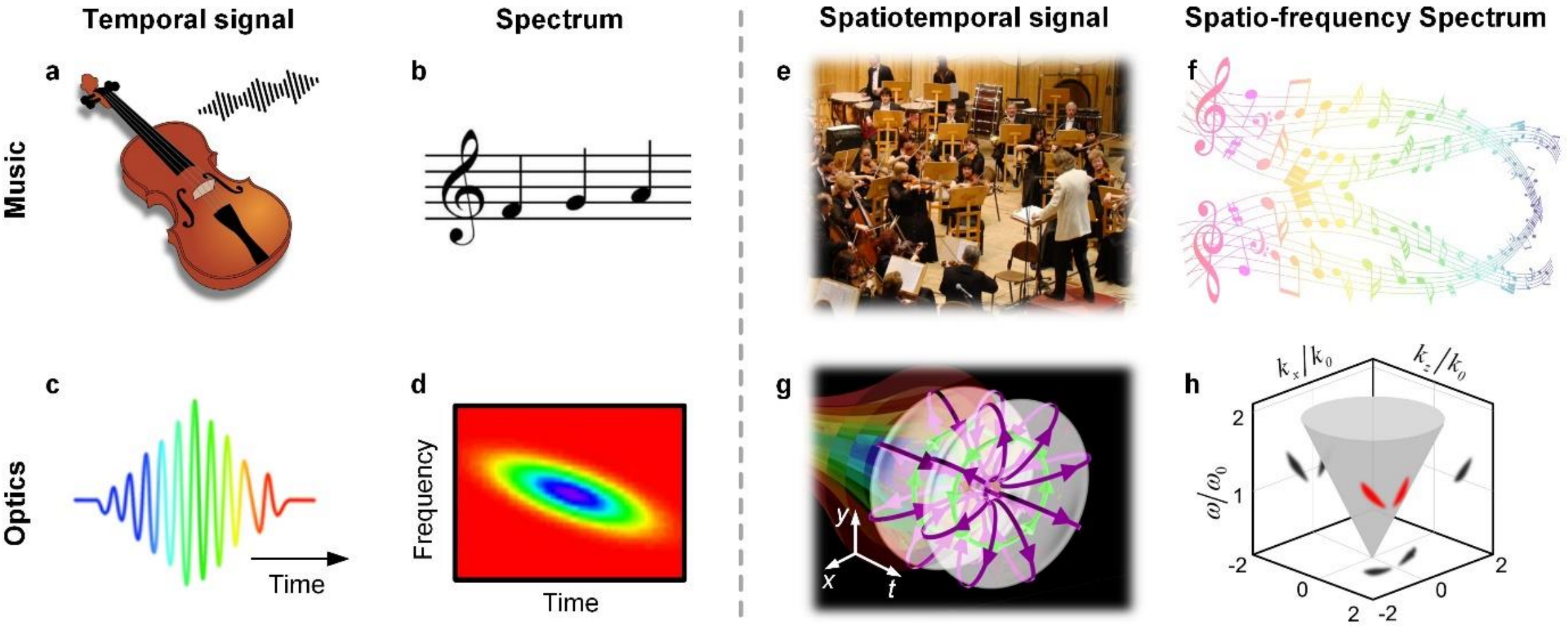

**Figure 1.** a,b, A violin can play a piece of music as a temporal signal (a), which can be characterized by the notes on a staff (b). c,d, The temporal evolution of an optical chirped pulse (c) can be characterized by its Fourier spectrum in the frequency domain (d). e,f, A band of spatially distributed music instruments is needed to compose a complex symphony (e), which can be characterized by a full score including all parts of musical composition (f). g,h, The spatiotemporal evolution of a smoke-ring-like toroidal structured pulse (g), the spatio-frequency description of which is a complex distribution located on the light cone. Images (a,e,f) come from pixabay.com.

Structuring light in all its degrees of freedom is steadily gaining traction [1], extending our familiar 2D transverse mode pattern to 3D spatial control, and to include spatiotemporal control for 4D even higher-dimensional forms [2]. Temporal shaping of light, especially in the ultrafast domain, has already benefited diverse applications, from material manufacturing to ultraprecise metrology. For example, the locking of carrier envelope phase of optical pulse can result in frequency comb as a powerful tool for ultraprecise metrology [3]. The shaping of ultrafast laser pulse into burst profiles enables the improved effect of ablation-cooled material removal for advanced laser processing [4]. On the other hand, spatial shaping of light also attracted growing attention for arbitrary tailoring of light patterns for diverse purposes, with recent advances of the exploiting of geometric transformation under general symmetry to reveal the 'hidden' degrees of freedom of light [5]. When temporal shaping meets spatial shaping with complex space-time nonseparability to control, spatiotemporally structured light comes into play, injecting new vitality to photonic science and applications [6]. For instance, the generation of toroidal space-time nonseparable pulse has promised to excite new forms of multipole moments in matter and free space [7]. The ultrafast vortex or vector pulses can be used to manufacture helical micro- or nano-structure for producing chiral functional materials [8], emulate topological textures of quasiparticles in condensed matters [9], and to excite new nonlinear effect in light-matter interactions [10]. Therefore, sculpturing light in the spatiotemporal domain with increasingly complex topological structures represent the most cutting-edge frontier for both fundamental physics and applied optics.

This roadmap highlights the rapidly growing body of works on the generation and characterization of various spatiotemporally structured light pulses as well as the associated novel physics. We will make an attempt to summarize a unified framework for their classification, generation and characterization, from prior spatiotemporally separable pulses to nontrivial spatiotemporally nonseparable pulses. We also review the widespread and potential applications of spatiotemporal light pulse along with a perspective on new opportunities for both fundamental and applied science.

## 2. Spatiotemporal mode-locking

Logan G. Wright, Cornell University and NTT Research, Inc.
Demetrios N. Christodoulides, CREOL/University of Central Florida
Frank W. Wise, Cornell University

### Status

In a mode-locked light source, fields in many modes self-synchronize to oscillate collectively, leading to a beam of pulses of light that is, by any reasonable human standards, utterly astounding. The pulses are short. In a sense, they can be as “short” as the proton is small (1 femtometer). The regular train of pulses equivalently defines a “comb” of discrete frequency “teeth” in the spectral domain. This frequency comb is equally remarkable: if a 10 THz bandwidth optical comb is compared to a 10 cm long hair comb, the optical comb’s teeth correspond to about 1 nm thickness. These unique qualities have uniquely enabled an ever-expanding list of methods and applications, for which there have been many excellent reviews and books, and at least 3 Nobel prizes so far.

Thus far, virtually all optical mode-locking has been one-dimensional: the modes that lock together all share a roughly identical transverse (x,y) shape, and as a result the physics can be accurately described by factoring out those spatial dimensions, considering nonlinear optical dynamics in just one space dimension (z) plus time (t). One of the grand successes of mode-locking physics has been to understand and classify stable mode-locked pulses within certain families of one-dimensional dissipative Kerr solitons, stationary solutions to dissipative nonlinear wave equations [1]. In the last several years, we and many other researchers have started to explore a more general scenario, so-called *spatiotemporal mode-locking (STML)* [2]–[6]. In STML, the modes that lock together can be of a much richer form, and the mode-locked pulses formed are multimode, or spatiotemporal, dissipative solitons.

There are at least three reasons to be excited about STML, which will be elaborated on in this article.

1. STML represents arguably the least understood regime of mode-locking physics. Better understanding of STML is thus especially likely to (eventually) lead to breakthrough discoveries or inventions in mode-locked light sources that translate to even more astounding, application-enabling photonic capabilities.
2. The ultrafast, high-dimensional nonlinear physics involved in multidimensional mode-locking is both an ideal laboratory testing ground, and an ideal candidate, for emerging data-driven control and nonlinear inverse design techniques.
3. Light sources based on spatiotemporal nonlinear optical wave physics are uniquely qualified to produce spatiotemporally structured light with extremely high intensity.

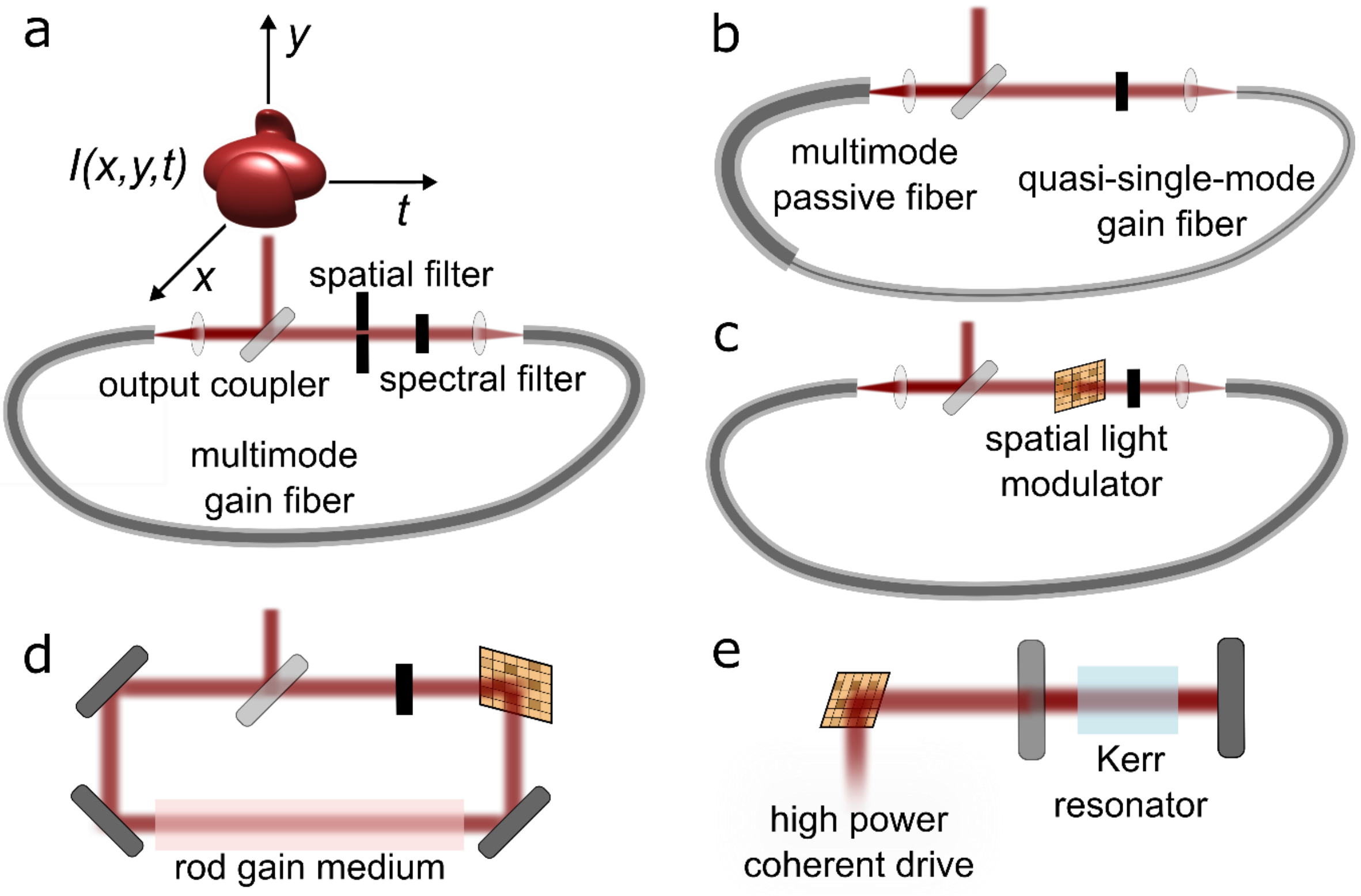

**Figure 1.** Settings for realizing spatiotemporal mode-locking. (a) A fiber ring resonator based on an active multimode fiber, typically with a graded-index profile, which provides gain. (b) A more experimentally accessible quasi-multimode ring fiber resonator. Inclusion of a highly multimode passive fiber leads to spatiotemporal dynamics, although a strong spatial filtering effect due to the single-mode active fiber constrains the richness of this behaviour. (c) Inclusion of a spatial light modulator, or other reconfigurable elements in the resonator may permit adaptive, high-dimensional control of mode-locking physics. (d) Realizing spatiotemporal mode-locking in systems that do not include fiber, such as a rod-like gain medium. (e ) Kerr resonators should also support spatiotemporal mode-locking, and richer control over the intracavity physics could be realized by spatial and/or temporal shaping of the coherent drive.

## Current and Future Challenges

A first challenge (rather specific and technical) for the science of STML is that the multimode graded-index gain fiber utilized in the first experimental demonstration [2] (Fig. 1a) is not commercially available, making direct replication of that specific laser design challenging. Some aspects of multidimensional mode-locking physics can be explored in a more accessible cavity design, based on few- or single-mode step-index gain fiber, and a large section of passive graded-index multimode fiber [2] (Fig. 1b). This cavity, made entirely of off-the-shelf components, is suitable for exploring a rather constrained subspace of multidimensional dissipative soliton physics, and for developing techniques to observe and control it. It is, however, unlikely to be the basis for major breakthroughs in mode-locked laser performance, especially if the performance targets are maximum peak intensity or pulse energy.

This challenge is related to the more exciting issue of achieving better control over STML and/or observing it in more controllable settings. In multimode fibers, manufacturing disorder and mechanical perturbations cause a complicated coupling between transverse modes. Historically, observing multidimensional solitons in nature has been extremely challenging, owing to their fragility to even minor perturbations [7]. Yet, we have found that in real disordered multimode lasers, a wide range of qualitatively diverse multimode dissipative solitons form spontaneously from noise, and then stably persist [3]. While the fact these stable spatiotemporal pulses form at all challenges many historical notions of multidimensional nonlinear waves [7], we expect that researchers will need to learn how to better control these pulses' spatiotemporal structure before they will have widespread utility. To both better understand and control multidimensional mode-locking, it will be essential to design and introduce new optical systems and mode-locking strategies in which disorder is either strongly suppressed, and/or in which the cavity's configuration can be adjusted rapidly and along many different degrees of freedom.

Finally, our theoretical understanding of the physics of mode-locking in multimode optical cavities is still at an early stage, especially in comparison to the accuracy with which we can describe one-dimensional mode-locking. While advances in control and simplification of experimental prototypes will help, the greatest opportunities presented by multidimensional mode-locking physics are implicit in their demand for new theoretical techniques. While complexity has never stopped physicists from developing predictive models, and progress on predictive models of STML is ongoing, today we have fewer excuses than ever. We are amidst an explosion of progress in data-driven and computationally intensive modelling and control techniques, such as deep learning [8], inverse design [9], and reinforcement learning. These provide researchers with an unprecedented toolbox to understand - and control - high-dimensional nonlinear processes like STML.

**Advances in Science and Technology to Meet Challenges**

To address the first and second challenge, we recommend that researchers explore new cavities and photonic systems for multimode mode-locking. Researchers may design and fabricate new multimode gain fibers suitable for STML. Fibers currently being developed for mode-division multiplexed telecommunications are suitable, for example. Multidimensional nonlinear wave physics may also advance lasers based on step-index multimode gain media, such as the multimode transient dynamics that have been predicted to dramatically increase achievable mode-locked pulse energy [3], [5]. Another especially promising route is STML in cavities or regenerative amplifiers that do not consist of fiber at all, but rather rod-like or thin-disk gain media (Fig. 1d). In fully multimode lasers with sufficiently high gain per pass (whether fiber or otherwise), including elements like spatial light modulators within the cavity should afford researchers with rich, high-dimensional control (Fig. 1c). Mode-locking mechanisms beyond nonlinear polarization rotation, such as Kerr-lens-inspired[10] or Mamyshev saturable absorbers, are likely to prove more useful and flexible in the long run. Finally, still greater control and performance may be achieved with STML not in lasers, but in coherently driven resonators, either in integrated photonics [11] or in fiber [12] or bulk resonators (Fig. 1e).

At its heart, the theoretical challenge of STML is an inverse problem: How can we cause (by fixed cavity designs or adjustable control parameters) coherent mode-locked pulses to self-organize with the qualities (shape, energy, stability) we desire? This inverse problem is challenging because its underlying physics are high-dimensional, nonlinear, stochastic, and dynamic. On the other hand, it is not so challenging that it seems impossible. In fact, we believe STML's inverse problem is a uniquely accessible, and even uniquely enticing, hard inverse problem.

Why? For one, while the STML inverse problem is significantly more challenging than any problem attempted in inverse photonic design so far [9], it is not obviously harder than recent feats of data-driven models, such as the mastery of ImageNet, text generation, and the games of Go or Starcraft II (These feats are all effectively solutions to high-dimensional, stochastic nonlinear inverse problems). Just as games of Go may be simulated quickly, experiments with reconfigurable multimode lasers can be executed rapidly, facilitating the training of data-intensive schemes like deep reinforcement learning. Second, in contrast to Starcraft and Go, multimode lasers are within or just beyond the reach of traditional physics models. For this reason, solving STML's inverse problem represents a uniquely apt opportunity to develop symbiotic methods that combine traditional physics with computation and data-intensive machine learning. Success in this pursuit will have far-reaching implications not just in new light sources with tailored spatiotemporal features, but more broadly in the control of complex nonlinear systems throughout science and engineering, such as fluid dynamics, chemical reactions, and magnetohydrodynamics.

**Concluding Remarks**

As other articles in this and other Roadmaps lay bare, perhaps the single most promising frontier for future developments in light and light-based applications is the control of light in many dimensions and degrees of freedom. This is particularly true for high-intensity light, which is used to drive and control extreme processes with multifaceted spatiotemporal scales, like the dynamics of plasmas, free electrons, and x-rays. To shape high-intensity light in space and time, it seems unlikely that it will be possible to linearly filter it from an unstructured pulse, since such multidimensional pulse carving will inevitably be lossy, nonlinear, and prone to damage. Instead, it seems more promising (or at least highly complementary) to learn how to shape light directly from the source. Addressing this challenge of tailored light from the source is the inverse problem of STML: *How can we design or control a complex, multimode nonlinear photonic system like a laser so that it self-organizes coherent fields with the spatiotemporal qualities we desire?* This STML inverse problem is enticing as a route to extremely high intensity, spatiotemporally tailored light (and all its applications), and as an accessible challenge problem for physics-informed, data-driven control of complex nonlinear physics.

**Acknowledgements**
*The authors wish to thank NTT Research for their financial and technical support. This effort was sponsored, in part, by the Department of the Navy, Office of Naval Research under ONR award number N00014-20-1-2789.*

## 3. Spatiotemporally structured light by optical-frequency-comb lines

Alan E. Willner, Kaiheng Zou, and Zhe Zhao
University of Southern California

### Status

Structured light has gained increasing interest recently due to its unique amplitude and phase spatial distributions that can be used in various applications [1]. Such optical beams could carry a pure single spatial mode or a coherent superposition of many spatial modes, and these modes can be from one of various types of modal basis sets (e.g., Laguerre Gaussian (LG), Hermite Gaussian (HG)). Moreover, each mode: (a) is orthogonal to all other modes within that set, and (b) has a complex weighted coefficient.

For a single-frequency beam at a given propagation distance: (a) a single-mode beam has a static spatial distribution of that mode [1], whereas (b) a beam composed of many weighted modes can be tailored to have nearly any static amplitude and phase spatial distribution [2]. This tailoring is accomplished by the constructive and destructive interference that occurs among the different orthogonal modes [2].

The above beams are generated from a single-frequency source. However, spatially structured beams that are generated from ***multiple*** frequencies can exhibit interesting dynamic (i.e., time-variant) behavior at a given propagation distance [3]. Such spatiotemporal (ST) beams can be achieved by various methods [3], with one example being through the synthesis of multiple coherent frequency lines from an optical comb and each line carrying a different combination of weighted orthogonal spatial modes [4]. Depending on the elapsed time, the comb frequency spacing ($\Delta f$) will induce a time-dependent phase shift $\Delta\varphi(t)=2\pi\Delta f t$ between neighboring lines. A phase shift at a given propagation distance will be frequency dependent, such that the instantaneous constructive or destructive interference between a spatial mode on one frequency with a spatial mode on a different frequency would change with time (see Fig. 1) [4]. Therefore, when combining multiple modes on each of multiple frequencies, time-variant spatial profiles can be generated that result in a spatiotemporal beam with a dynamic spatial distribution [4].

The transmission of multiple frequencies each with multiple modes can produce either dynamic spatiotemporal beams or pulses (see Fig. 2); such pulses are often referred to as wave packets [3]. Examples of such structured-light beams or pulses include those that carry orbital-angular-momentum (OAM). Light carrying OAM can be represented as a subset of LG modes, and can be characterized as having: (i) a phasefront that "twists" in the azimuthal direction as it propagates, (ii) the OAM value, $\ell$, is the number of 2π phase shifts in the azimuthal direction, and (iii) the intensity profile has a ring structure with a central null [2]. Moreover, the index $p$+1 of LG modes indicates the number of intensity rings in the radial direction.

Examples of dynamic spatiotemporal light include the following:

(i) Spatiotemporal beams carrying OAM with time-dependent beam radii (Fig. 2a) [5]: A beam carries a single OAM value, and its beam radius expands and contracts periodically at a given propagation distance. Such a beam is generated by synthesizing comb lines carrying complex-weighted LG modes with the same single azimuthal index $\ell$ but with multiple radial indices $p$. The time-varying frequency- and mode-dependent constructive and destructive interference between the $p$ modes results in the dynamic oscillating behavior.

(ii) Spatiotemporal beams carrying two forms of orbital angular momenta (Fig. 2b) [4]: A beam can carry two forms of orbital angular momenta when viewed at a given propagation distance. The beam's momenta can be tailored to include: (1) ***rotation***: the beam rotates around its own beam center, and (2) ***revolution***: the rotating beam can also revolve around some external axis point from which it is spatially offset. Note that this is similar to the way the Earth revolves around the Sun yet simultaneously rotates around its own axis. Such a beam is generated by combining multiple frequency lines, with each line carrying complex-weighted LG modes composed of a single $\ell$ value and multiple p values. By choosing proper weights, the dynamic frequency- and mode-dependent constructive and destructive interference produces an LG mode with a spatial offset that has rotation around the mode center and revolution around the central axis.

(iii) Near-diffraction-free space-time OAM wave packets with tunable group velocities (Fig. 2c) [6]: An ST pulse can be created that carries OAM can be created that has reduced diffraction and a tunable group velocity between superluminal and subluminal. Such pulses are generated by synthesizing multiple comb lines, each line carrying a single Bessel mode with an identical $\ell$ value but a different Bessel radial spatial frequency $k_r$. Reduced diffraction is achieved by the use of Bessel modes that naturally have such a characteristic. The intensity peak of such an ST pulse occurs when the different Bessel modes on different frequencies constructively interfere. Due to the dynamic phase delay between the neighboring Bessel modes, the constructive interference occurs at different times as the pulse propagates. This movement of the pulse intensity peak results in a group velocity that can be controlled by selecting the specific modes and frequencies.

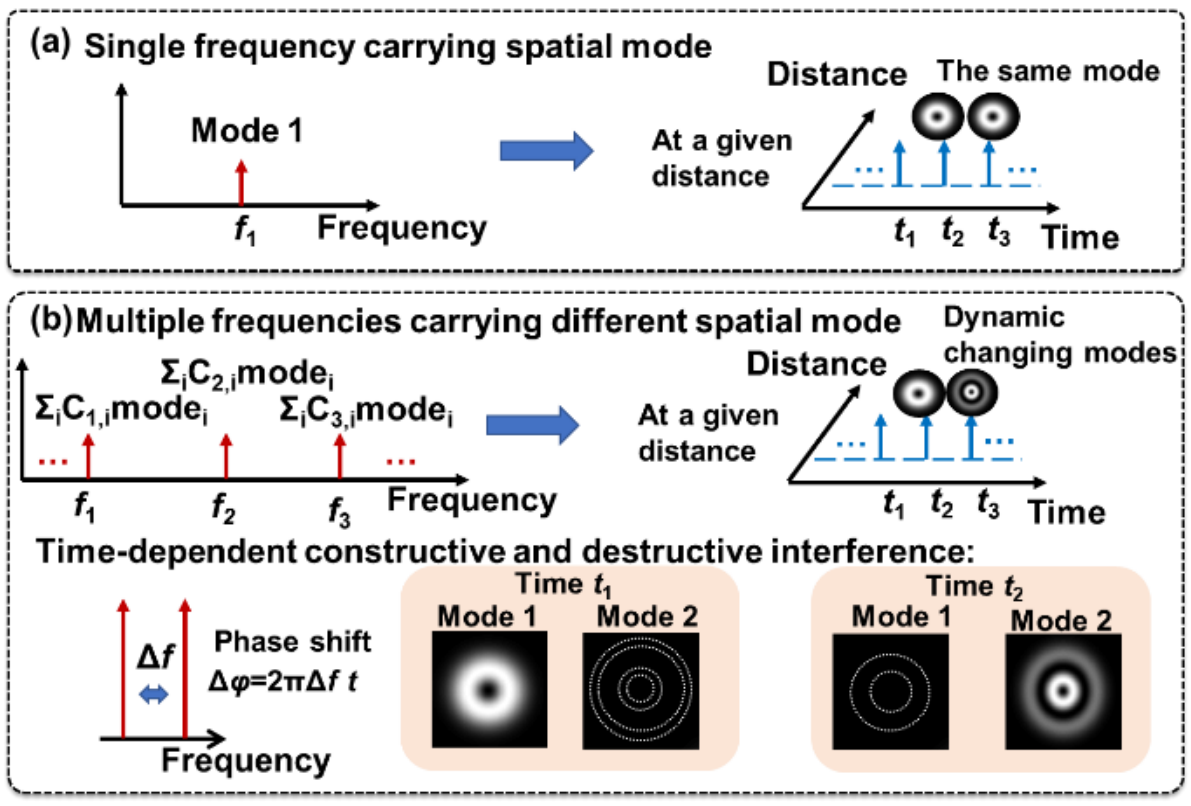

**Figure 1.** (a) Typically, the structured light is generated from a single-frequency optical source and has a static amplitude and phase spatial distribution at a given propagation distance. (b) One example of methods generating the dynamic spatiotemporal beams is synthesizing optical frequency comb lines, each carrying a superpositions of spatial modes.

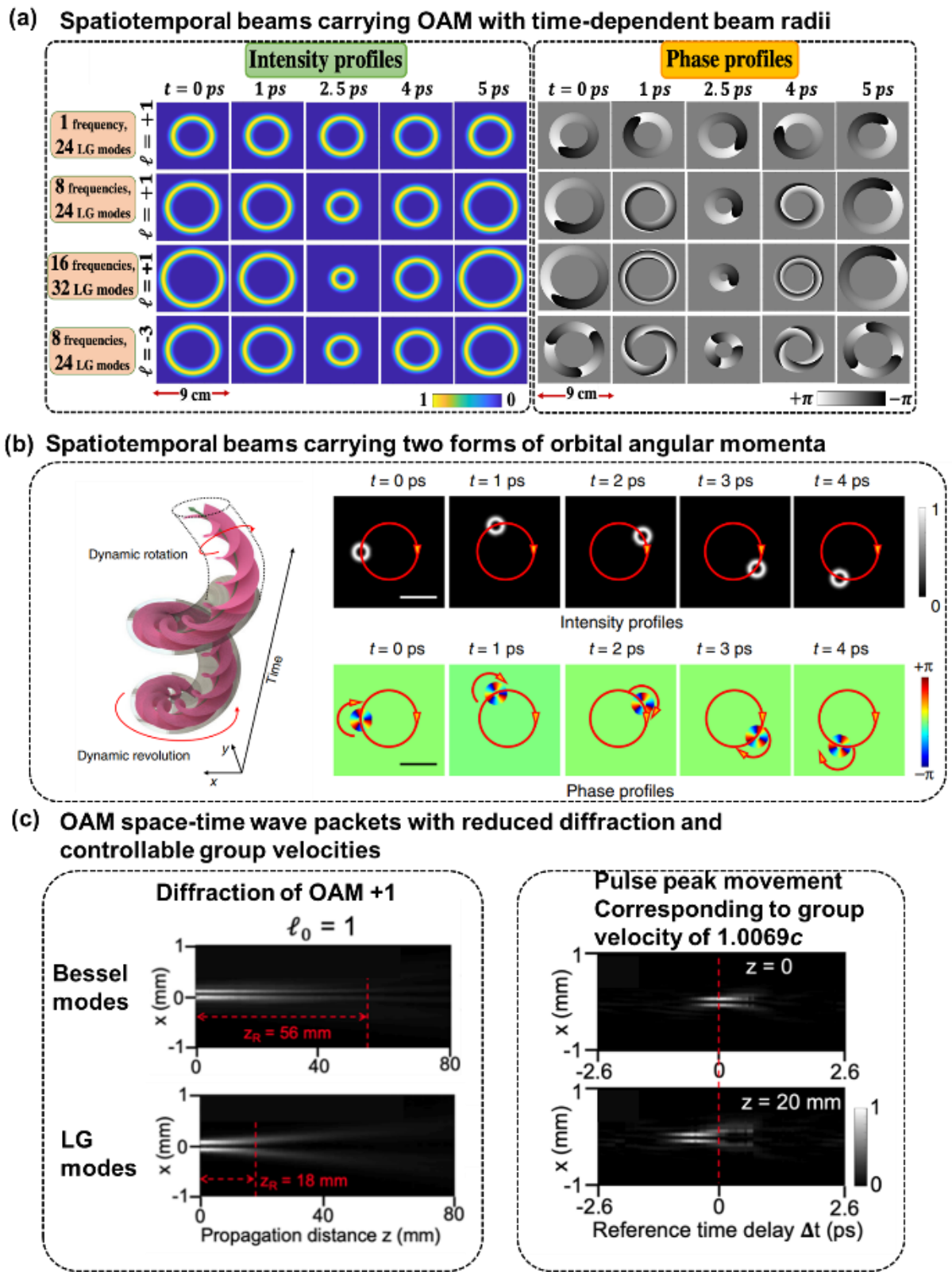

**Figure 2.** Examples of dynamic spatiotemporal structured light generated by synthesizing optical frequency comb lines, each carrying different single or superpositions of spatial modes. (a) Simulated spatiotemporal beams carrying OAM with time-dependent beam radii. (b) Simulated spatiotemporal beams carrying two forms of orbital angular momenta. (c) Experimentally measured near-diffraction-free space-time OAM wave packets with tunable group velocities

## Current and Future Challenges

There are several challenges in the generation and propagation of dynamic spatiotemporal structured light with high modal purity, including the following:

(1) ***Diffraction***: Diffraction might affect the dynamic spatial properties when viewed at different propagation distances. This is because: (i) higher-order modes diffract faster than lower-order modes, thus producing a different coherent spatial interference pattern depending on propagation distance; and (ii) the same mode on different frequencies will diffract differently, thus producing a different dynamic spatial interference pattern depending on propagation distance. This frequency-dependent modal diffraction can be significant when the frequency spacing increases and the propagation distance grows beyond the Rayleigh range [4]. In such cases, the resulting modal purity of the spatiotemporal light can decrease. To alleviate this challenge, it might be possible to use Bessel modes with limited diffraction to reduce this effect [6].

(2) ***Modal Basis Set***: Different modal basis sets can be carried on multiple frequency comb lines to achieve different dynamic properties. For example, circularly-symmetric LG and Bessel modes may be well-suited for dynamic angular and radial movement, whereas rectangularly-symmetric HG modes may be well-suited for dynamic lateral movement. Other spatial modes might also be explored for various dynamic behavior. Moreover, apart from modes with structured phase, beams with spatially tailored polarization states (e.g., vector modes) could also be further explored for achieving interesting dynamic spatiotemporal behavior.

(3) ***Comb Source***: The properties of the optical frequency comb source can affect the generated spatiotemporal beams. Specific examples include: (i) The frequency spacing can determine the speed of the dynamic change as the spatial distribution depends on the relative phase shift $\Delta\varphi(t)=2\pi\Delta ft$; (ii) The number of comb lines can affect the quality of the spatiotemporal light, with more comb lines providing higher purity and dynamic range; (iii) A highly coherent comb is desirable, since the phase noise on the comb lines could induce phase distortion on the modes, thus degrading the generated spatiotemporal light.

**Advances in Science and Technology to Meet Challenges**

There are various potential advances in generating, propagating, and using dynamic spatiotemporal light composed of multiple modes on multiple frequencies. These include:
(1) ***Spatial Modulation***: In general, spatial modulation should be performed on multiple comb-line frequencies to generate spatiotemporal light. Such spatial modulation can be achieved several ways, including the use of spatial light modulators [6], metasurfaces [7] and multi-plane light converters [8]. Some desirable features of future spatial modulation include rapidly tunable and reconfigurable operation for many modes, broad bandwidth covering a large number of comb lines, compact size, low cost, and high purity of complex weighted modes.
(2) ***Photonic Integrated Circuits***: Photonic integrated circuits (PIC) could reduce the size, weight, complexity, power consumption, and cost of spatiotemporal light generation. It might be desirable to have a single device for spatiotemporal light generation that includes multiple elements (e.g., a comb source, wavelength (de)muxes, and spatial amplitude and phase modulators of the individual frequencies. This vision could encompass technologies such as: (i) integrated Kerr frequency combs based on optical microresonators [9], (ii) integrated spatial mode converters [10], and (iii) tunable optical gratings. Integration of these various types of devices is quite challenging, but future deployment would likely be significantly accelerated with PIC implementation.
(3) ***Applications***: Single-frequency-based structured light has shown various potential applications, such as imaging, sensing, and communications [1]. At present, it is interesting but still unclear as to the potential applications of such dynamic ST light beams and pulses. Is it possible to perform fast ST-based imaging and sensing, taking advantage of the dynamic spatial distributions? Moreover, is it possible to utilize dynamic ST light in a highly reconfigurable communication system? Many more unanswered questions exist, and yet it is exciting to ponder the uses of this novel light.

**Concluding Remarks**

Unlike single-frequency-based structured light, spatiotemporal beams and pulses can exhibit unique dynamic spatial properties due to the time-dependent interference of various spatial modes that are carried on multiple comb-line frequencies. This can be considered as a “toolkit” for generating dynamic light distributions with tailorable properties. It is exciting to envision the future innovations of more sophisticated spatiotemporal light and their potential unique applications.

**Acknowledgements**

*ONR through a MURI N00014-20-1-2789; Vannevar Bush Faculty Fellowship by the Basic Research Office of the ASD/R&E and funded by ONR (N00014-16-1-2813); DURIP (FA9550-20-1-0152).*

## 4. Space-time coupling in ultrafast vortices

Miguel A. Porras, Complex Systems Group, Technical University of Madrid

**Status.** Space-time couplings (STCs) may be intrinsic to the nature of waves, and as such unavoidable. They become relevant in few-cycle, ultrafast pulses (UFPs) and ultrafast vortices (UFVs). Even if one ``prepares" a space-time (ST) separable field $E(t,x,y)=P(t)B(x,y)$ on a certain aperture, of frequency spectrum $\widehat{E}\,(\mathrm{w},x,y)=\ \hat{P}(\mathrm{w})\ B(x,y)$, each color propagates differently, transforming into a nonseparable spectrum $\hat{E}'(\mathrm{w},x,y)=\ \hat{P}(\mathrm{w})\ B'(\mathrm{w},x,y)$, and thus a nonseparable field $E'(t,x,y)$ upon propagation. The associated STC effects are known for decades, but are still often ignored. It took two decades to definitively confirm that the carrier-envelope-phase (CEP) in a focal volume differs from Gouy's phase [1] because of these STCs, to mention a crucial parameter in field-sensitive light-matter interactions. In UFPs, STCs such as sub- and superluminality, spatial chirps, pulse front curvature and broadening, all in a vacuum, are only large with UFP durations approaching their minimum attainable value, nominally $\Delta t=0$. This is why they have only been observed in single- and sub-cycle THz pulses [2].

For UFVs, an azimuthal-temporal coupling has turned out to impose a minimum duration above zero, e.g., $\Delta t\geq\sqrt{|l|}/\mathrm{w}_0$, where $\mathrm{w}_0$ is the central frequency and $l$ the topological charge (TC), for pulsed Laguerre-Gauss (LG) beams [3]. STCs also become large when the UFV duration approaches the minimum duration, as from Fig. 1 (a) to (b), with the greatly enhanced transverse chirp in (c), occurring at minimum durations of one or more cycles for large TCs. The limit $\Delta t\approx\sqrt{|l|}/\mathrm{w}_0$ has not yet been reached, but experiments approximate it from afar. Small transverse chirps have been reported in [4].

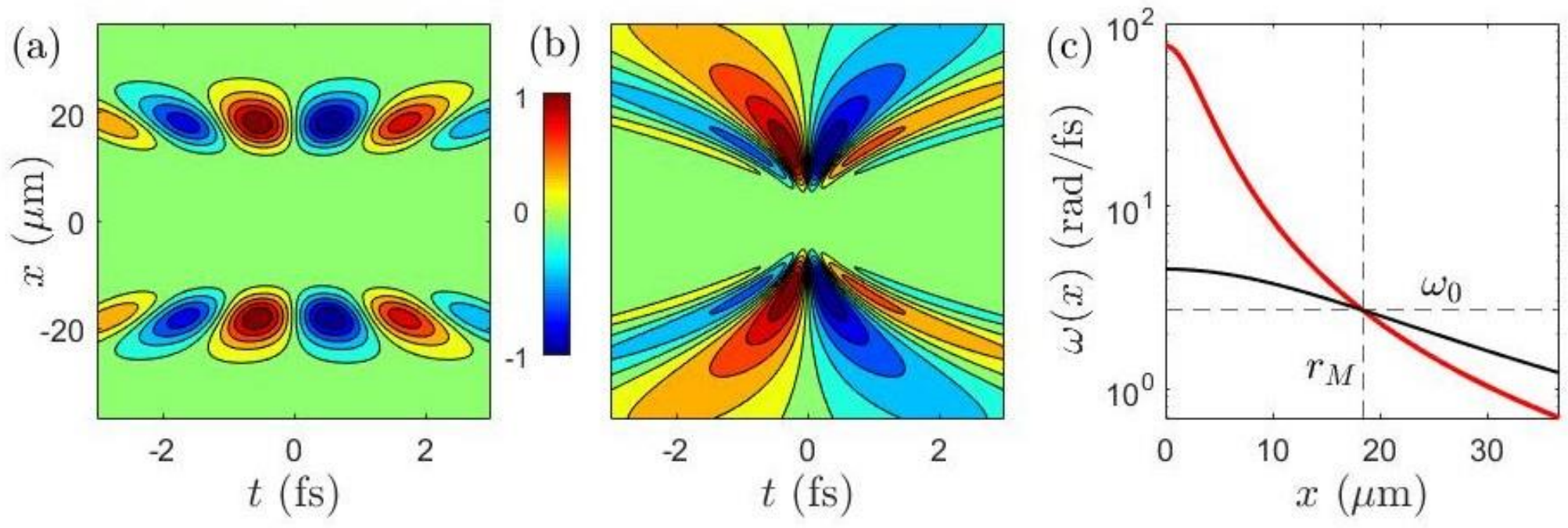

**Figure 1.** *STCs in UFVs become huge as the limit* $\Delta t\approx\sqrt{|l|}/\omega_0$ *is approached*. (a) Nearly separable real field ($y$ = 0 section) of isodiffracting ($z_R$ = constant = 0.11 mm) pulsed LG beam of central frequency $\omega_0$= 2.69 $\mathrm{fs}^{-1}$ ($\lambda$=700 nm), TC $l$ = 27, and duration (a) $\Delta t\gg\sqrt{|l|}/\omega_0$. (b) The same but for $\Delta t\approx\sqrt{|l|}/\omega_0$. (c) Mean frequency at each distance $x$ from the vortex center. The moderate transverse chirp for (a) (black) becomes huge for (b) (red) (note the logarithmic vertical scale). The lines indicate the radius $r_M$ of maximum energy density, or bright ring, and its mean frequency $\omega_0$.

It turned out that the most important of these STCs are controlled by a single parameter pertaining the femtosecond laser source, at least for high quality UFPs and UFVs [1,5]. The $g_0$-factor characterizes the w-dependence of the Rayleigh range of the w-dependent beam profile $\widehat{E}(\mathrm{w},x,y)$. Depending on the $g_0$-factor of the broadband source creating the UFV, different colors are focused differently [Fig. 2(a)], which translates into different STCs, as in the focused UFVs in Fig. 2(b) differing in duration and transverse chirp. Measurements of $g_0$ would predict these STCs, but they are scarce. The only systematic measurement yielded values between $-1$ and $-2$, and predicted indeed the observed STCs [1].

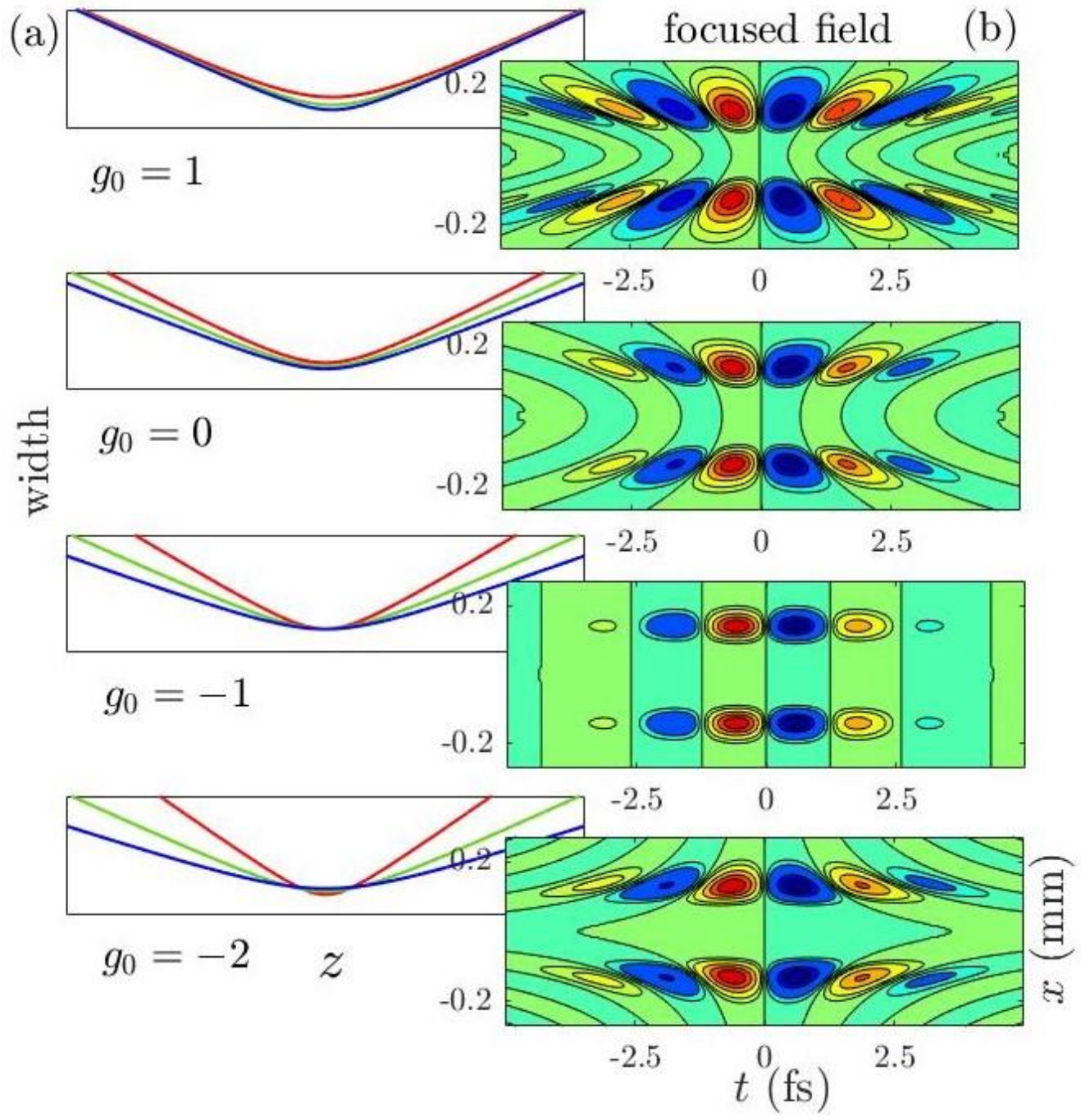

**Figure 2.** (a) *Chromatic focusing of broadband Gaussian or LG beam.* $g_0$= +1 means constant width on the focusing element (placed on the left), $g_0$= 0 constant Rayleigh range (isodiffraction), $g_0$= -1 constant divergence (constant width at the focus), and $g_0$= -2 is a generic situation close to experiments [1]. (b) ST structure of the focused LG pulses of Gaussian width 0.944 mm, $\omega_0$= 2.69 fs$^{-1}$ ($\lambda$=700 nm), TC $l$ = 15, bandwidth compatible with single-cycle duration, and with the respective values of $g_0$. The color scale range from +1 (red) to -1 (blue).

The knowledge and control of these intrinsic STCs is imperative in experiments with UFPs and UFVs. They ultimately and inevitably affect all broadband laser-matter interactions. Ref. [6], e.g., highlights the strong dependence of electron acceleration on $g_0$ . Its determination will open the route to control and optimization of these interactions.

**Current and Future Challenges.** In the quest towards ever shorter, higher TC, and higher quality pulses, these intrinsic STCs will eventually dominate. The current techniques of ST resolved characterization of ultrashort pulses such as spatially-resolved Fourier-transform spectrometry, the SEA-F-SPIDER technique, Hartmann wavefront sensors and others, are more than enough for the observation of these STCs, as done for the 5.5-fs UFV in [7], and even for EUV vortices in [8].

For UFPs without OAM the observation of intrinsic STCs requires the minimization of other STCs arising from their generation or shaping devices that may mask the former, or to enhance the intrinsic STCs by lowering their duration down to single-cycle or sub-cycle pulses. Indeed, the few unambiguous observations involved almost perfectly focused Gaussian beams [1] and single and sub-cycle THz pulses [2].

With UFVs, there is more flexibility, but also added difficulty. Durations need to be decreased and/or TCs need to be increased to get somewhat closer to the limit $\Delta t \approx \sqrt{|l|}/\mathrm{w}_0$ while maintaining good vortex quality. The shortest UFVs generated are close to 2 cycles, but with low TC [7,9]. UFVs with TCs reaching the hundred are no longer uncommon in high-harmonic generation experiments [8], or directly from laser sources, but they are quasi-monochromatic. The challenge is to do both together. As a reference, a single-cycle UFV must carry TC $l$ = 27 for the intrinsic STC effects to be huge. This is far from current technical achievements, but STCs would also be large with, say, double duration or half TC, as in Fig. 2. The small but observable STCs in [4] belong to an UFVs with even smaller TC $l = 5$ and larger duration $\Delta t = 6$ fs.

Techniques of characterization of the ST-resolved optical field are well-established, but they are still complex and time-consuming. A measurement of the $g_0$-factor of the femtosecond source with which the UFVs is generated is simpler and provides a first approximation to the expected STCs, as corroborated in [1]. Systematic measurements of the $g_0$-factor of different sources would fill the gap in the knowledge of typical values. The THz source in [2] appears to be characterized by $g_0 \approx +1$, high-power femtosecond laser sources are conjectured to have also $g_0 \approx +1$ [6], but negative values were measured after post-compression at hollow core fibers [1]. A measurement of $g_0$ only requires a spectrally resolved measurement of the Rayleigh range, or equivalently, a spectrally resolved measurement of the beam width [1], and would allow us to discern, for example, which of the focused UFVs in Fig. 2(b) would interact with matter in an experiment.

**Advances in Science and Technology to Meet Challenges.** Demands of high temporal resolution and control of OAM content in applications such as superdense optical communications, ultrafast spectroscopy, or ultrafast quantum interference experiments, will push research towards a new generation of UFV generators reaching the fundamental limits. The optimization of these applications will press towards a full characterization of the intrinsic STCs in UFVs.

The generation of sub-2-cycle UFVs with current UFV generation techniques presents difficulties associated with chromaticity. Recent advances report added tunability of TC and energy, but not in shortness. Spiral phase plates, astigmatic mode converters, computer generated holograms, diffractive elements displayed in spatial light modulators, and others, present a variety of dispersive effects with broadband radiation including angular, group velocity, delay, and TC dispersion that enlarge or deteriorate UFVs. Many techniques have been implemented to eliminate them, often at the price of introducing expensive, specially designed optical components.

Advances would involve the development of achromatic vortex generators. An intrinsically achromatic vortex generator based on interference of two polychromatic Gaussian beams in a Sagnac interferometer and employing only mirrors and beam-splitters has been proposed [10]. It does not present limitations with regard to energy and wavelength, but still is limited to single-charged vortices.

Alternate routes will be improvement of nonlinear compression techniques, and, in strong-field, broadband light-matter interactions, improved control of OAM content of high-harmonics in attosecond pulse generation experiments. The new frequencies in supercontinuum spectra generated from self-phase modulation of femtosecond vortices in multiple thin plates have been shown to preserve the vortex structure, opening a road for the generation of near single-cycle UFVs of different TCs in the near infrared [11]. In turn, in strong-field light-matter interactions with few-cycle, near-infrared UFVs, recent developments in high-harmonic and attosecond pulse generation allow the control of the OAM content of the harmonics using multiple, non-collinear driving UFVs with equal or unequal TCs, and via spin-orbit coupling. It appears to be possible the generate, in principle, a spectrum of harmonics with a single-valued OAM, which points to a new way to achieve extremely short vortices with a single-valued TC.

**Concluding Remarks.** Other STCs considered in this roadmap may be unwanted, but are in principle removable. Others are intentionally introduced for certain purposes, as in the space-time wave packets in Chapters 7 and 10. The STCs considered here are fundamental, mere consequences of the ondulatory nature of light. Broadening, distortion, separation of colors in UFVs of very short duration and high OAM propagating in vacuum, are multifaceted manifestations of approximating the limit of existence of these wave objects, and can ultimately be understood as a ``reaction" of waves to too high structuring at too short time scales. For the time being, only small manifestations have been observed, but as technology in UFV generation advances, it will be necessary to face up to them in order to improve the myriad of applications of UFVs.

As a last remark, pulsed LG beams underlie the discussion in this Chapter because of their broad use and since they can be free of STCs at long durations. Similar limits and associated intrinsic STCs do exist for other wave packets carrying OAM such as diffraction-free, space-time wave packets (in addition to the STCs needed for diffraction-free propagation), which will be observable when synthesized in the few-cycle regime.

**Acknowledgements**

I acknowledge support from Projects No. PGC2018-093854-B-I00 and No. FIS2017-87360-P of the Spanish Ministerio de Ciencia, Innovación y Universidades.

## 5. Spatiotemporal vortices of light

Andy Chong[1], Chenhao Wan[2,3], and Qiwen Zhan[2,4]

[1]Pusan National University

[2]University of Shanghai for Science and Technology

[3]Huazhong University of Science and Technology

[4]Zhangjiang Laboratory

### Status

Optical vortex beams possess zero intensity phase singularities with spiral phases. Recently, new optical vortices with phase singularities in space-time have been demonstrated [1]. Such optical vortices are referred to as spatiotemporal optical vortices. In contrast to the longitudinal orbital angular momentum (OAM) of vortex beams, spatiotemporal vortices possess the transverse OAM.

The generation of the spatiotemporal optical vortices is based on the Fourier synthesis by a pulse shaper with a two-dimensional (2D) spatial light modulator (SLM) (figure 1) [1]. In the setup, the optical frequencies are spread on the SLM and recombined which can be understood as a Fourier transform and an inverse Fourier transform, respectively. By applying a spiral phase profile on the SLM, which is the spatial frequency – frequency domain, a spatiotemporal phase singularity in space-time occurs by the recombination inverse Fourier transform process.

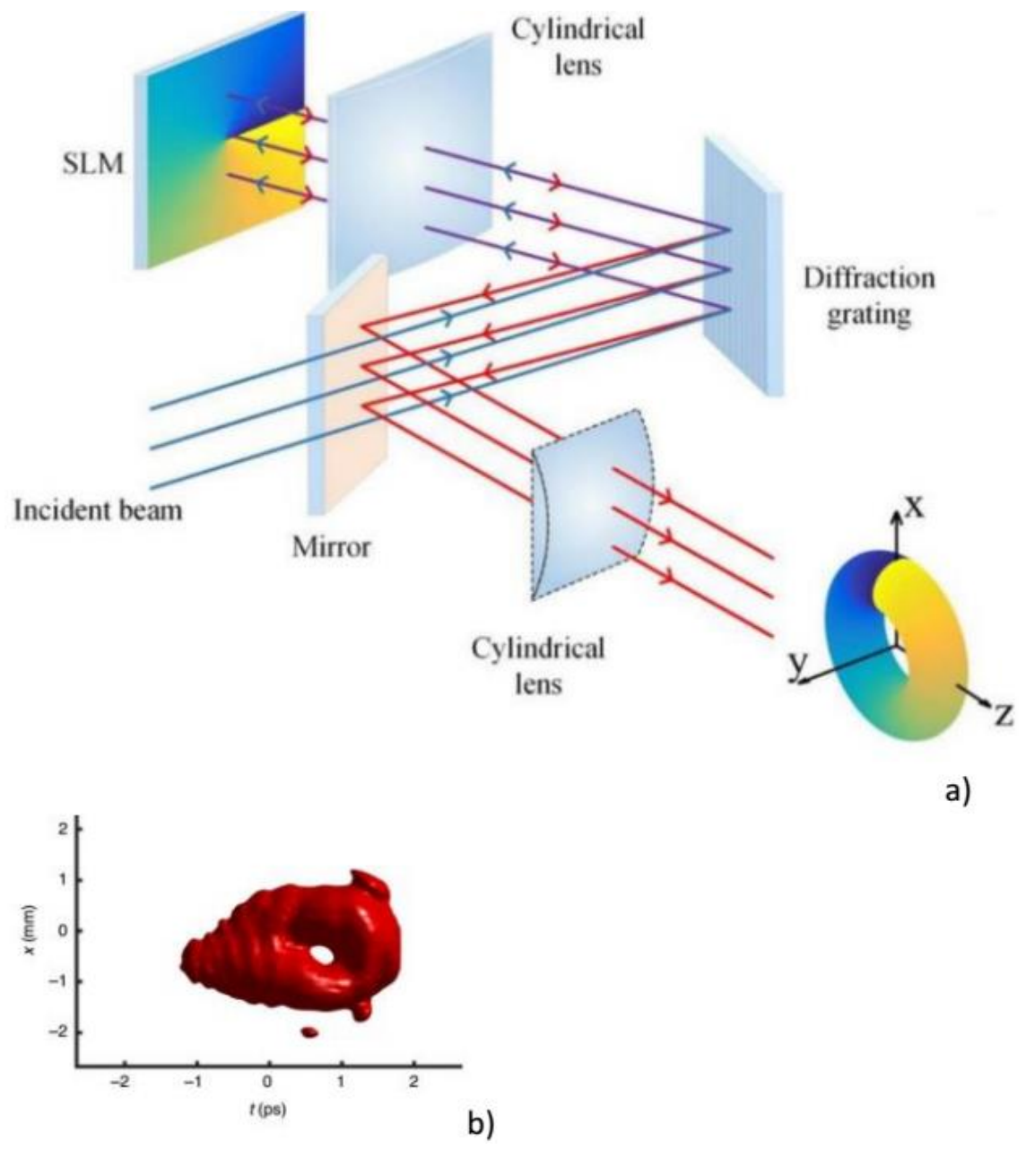

**Figure 1.** The experimental schematic to generate the spatiotemporal optical vortex. The bottom figure shows the measured iso-intensity profile of the spatiotemporal vortex. From [1].

Even though the spatiotemporal vortex has been demonstrated very recently, it has attracted significant attention already. Lately, the research progress on the spatiotemporal vortex has advanced noticeably. For example, the nonlinear processes, such as the second harmonic generation (SHG), on the spatiotemporal vortex have been studied [2]. In the SHG process, it was clearly demonstrated that the topological charge of the spatiotemporal phase is doubled as predicted. Forming a spatiotemporal vortex at a tight focus also has been theoretically studied. It is believed that such tightly focused transverse OAM will induce unique light-matter interactions.

More sophisticated optical wave packets with spatiotemporal vortices have been studied in a variety of aspects. Multiple spatiotemporal vortices with designed topological charges varying in time have been demonstrated [3]. It is believed that such wave packets can be useful for telecommunications. The vortex and therefore the direction of the OAM can be also adjusted [4]. It has been demonstrated that the spatiotemporal vortex may have a tilted OAM direction by a simple lens system [5]. A spatiotemporal vortex can have an exotic Bessel profile in space-time. Such a Bessel vortex is a localized wave packet that resists the dispersive and diffractive effects. As the vortex line is bent to form a ring, an optical toroidal vortex is formed (figure 2) [6]. Toroidal vortices, such as smoke rings, are prolific in fluids, but it has been elusive in optics. However, the transverse phase singularity has led the way to the first demonstration of the optical toroidal vortex.

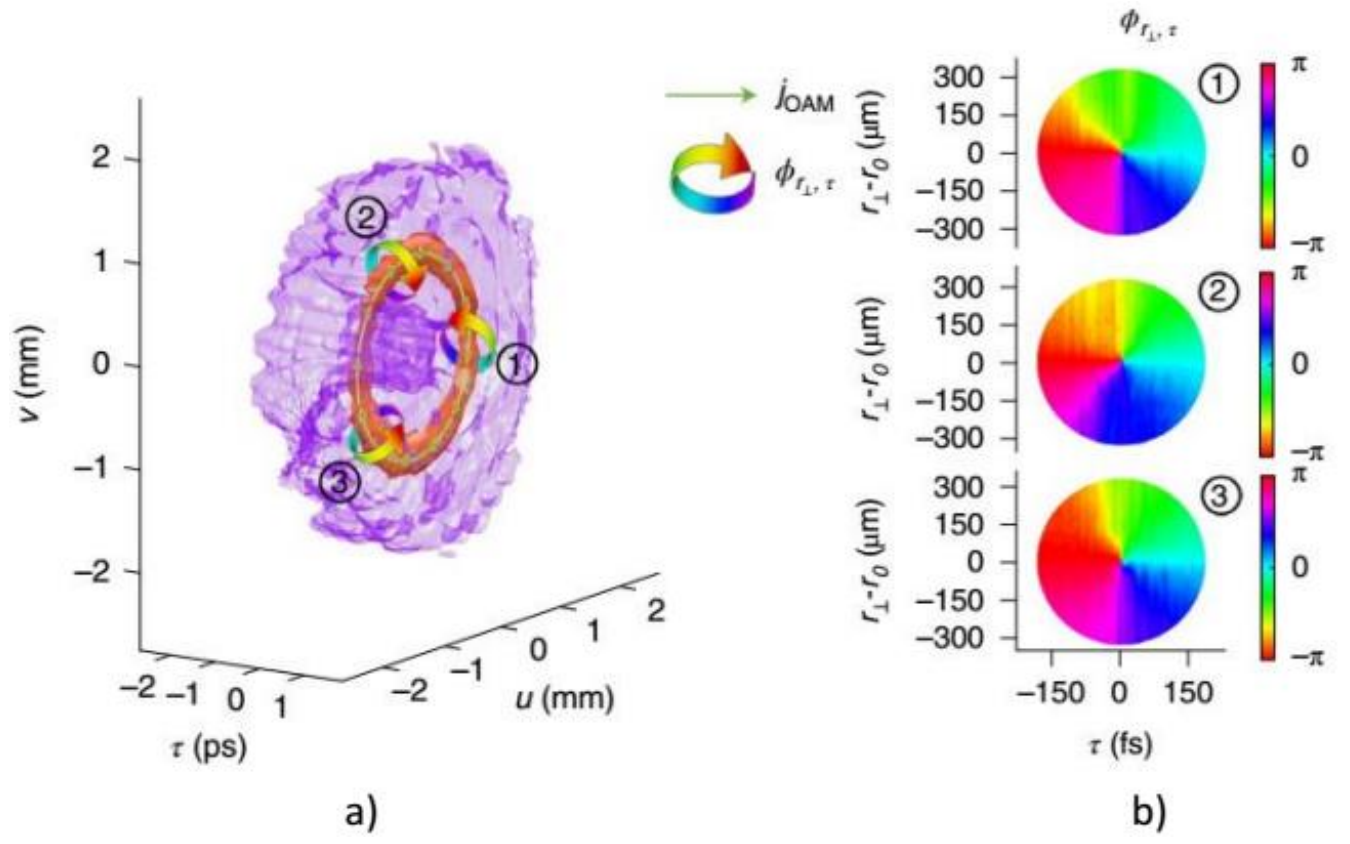

Figure 2. Measured profile of the optical toroidal vortex. The right figure shows measured spiral phases at various locations. From [5].

## Current and Future Challenges

The spatiotemporal vortex has a spatiotemporally coupled wave packet structure and such structures are quite challenging to attain in general. As the spatiotemporal vortex is achieved, it is believed that it has a variety of applications in nanoparticle manipulation, telecommunications, quantum optics, unique light-matter interaction, nonlinear interaction, etc. due to its unique structure with a transverse OAM property.

However, the future research direction is rather clear. The current research desires even more complicated optical wave packets. For example, the localized waves with spherical harmonic symmetries have very complicated structures with optical vortices [7]. Interestingly, such wave packets mimic Hydrogen atom orbitals and therefore, they are referred to as hydrogen orbital-like wave packets. Optical wave packets with spherical harmonic symmetries are three-dimensional (3D) localized waves that maintain the 3D profiles in dispersive and diffractive propagation. Apparently, achieving such novel states of light will be a noticeable achievement. However, there has not been much experimental progress to achieve them even though they have been theoretically proposed quite a while ago (>10 years ago). The reason for lacking experimental progress is that such a complicated optical structure has been a significant experimental challenge. The Hydrogen orbital-like wave is merely one example. In fact, there are numerous theoretically studied 3D wave packets without experimental realizations. Some examples of such wave packets are quasi-nonspreading wave packets which are localized 3D wave packets with very complicated structures [8], abruptly autofocusing 3D wave packets which can focus to enhance the intensity by 4 orders of magnitude by the intricate wave packet profiles, ellipsoidal wave packets which can be the ideal source for the free electron lasers, etc. Again, there are no clear synthesis methods for such wave packets due to complex spatiotemporally coupled structures.

To achieve such wave packets, it is necessary to develop a method to control the optical wave packets in amplitude, phase, and even polarization spatiotemporally in a three-dimensional fashion. In parallel, it is crucial to develop techniques to precisely measure the 3D profiles of optical wave packets. While some 3D intensity measurements are available in various techniques, precise 3D phase and polarization measurements are strongly desired to diagnose such sophisticated wave packets. The 3D profile measurement is desired even on a very small scale such as a tight focus of an optical wave packet.

**Advances in Science and Technology to Meet Challenges**

Such complicated spatiotemporally coupled optical wave packets demand the control of many parameters three-dimensionally. The Fourier synthesis is the viable candidate to achieve such controls (figure1). By adjusting the amplitude, phase, and polarization of light in the pulse shaper setup [9], the Fourier synthesized wave packet can be controlled in 2D spatiotemporal amplitude, phase, and polarization. Even though the Fourier synthesis is a promising candidate to generate a variety of spatiotemporally coupled wave packets, it can control only up to two-dimensional space-time. To accomplish the full 3D control, an additional dimension control is required. This extra dimension control can be provided by an extra pulse shaper setup or a beam shaping which will be a bit simpler than a series of pulse shapers.

An important beam shaping skill to add an additional dimension control is the optical conformal mapping. The optical conformal mapping is the technique to convert the beam shape in some coordinates into a more useful coordinate system. For example, forming a symmetric beam in the cylindrical coordinates is sensible since the diffraction effect is in the radial direction of the cylindrical coordinates. In fact, such conformal mapping has been successfully implemented already in various optics research. For example, for the toroidal optical vortex, an elongated spatiotemporal vortex line in the Cartesian coordinates was mapped into the cylindrical coordinates to form a ring of the phase singularity [6]. So far, most of the optical conformal mappings are from the Cartesian to the cylindrical or vice versa. However, other types of conformal mappings can be explored for some exotic optical wave packets.

Another candidate to achieve the spatiotemporally coupled 3D wave packet is the metasurfaces/metamaterials. Since the metasurface can control the amplitude, phase, and polarization of light by the design, it is strongly believed that the metasurfaces are suitable to generate some interesting wave packets. It has been already proposed that the spatiotemporal vortex can be generated by a photonic crystal slab [10].

Developing the 3D measurement capability is essential to diagnosing the wave packet profile. In fact, there are a variety of 3D profile measurement techniques available (see Chapter 12). However, the 3D profile measurement for a very small scale in the order of the wavelength is not available up to date. It is strongly believed that such small-scale measurement capability will enhance the understanding and applications of 3D wave packets.

**Concluding Remarks**

Due to its exotic spatiotemporal structures with a transverse OAM, the interest in spatiotemporal optical vortices has risen lately. While future applications of spatiotemporal vortices are anticipated, much more complicated wave packets are desired for future research. To attain such complicated optical wave packets, it is necessary to develop reliable methods to control amplitude, phase, and polarization in 3D. In this roadmap article, some feasible 3D wave packet control strategies such as Fourier synthesis, conformal mappings, and metamaterials have been briefly discussed.

The 3D optical wave packet measurement is indispensable in studying spatiotemporal optical wave packets. Besides available 3D measurement techniques, the 3D measurement in the wavelength scale will be a significant accomplishment to widen the understanding and the applications of the spatiotemporal optical wave packets.

**Acknowledgements**

This work is supported by NSFC (#92050202, #61875245), States Administration of Foreign Experts Affairs (G2022013001), Shanghai Science and Technology Committee (#19060502500), Shanghai Administration of Foreign Experts Affairs (21WZ2503100, 22WZ2502700), Wuhan Science and Technology Bureau (2020010601012169).

## 6. Orbital angular momentum of spatiotemporal vortices

*Konstantin Y. Bliokh*, RIKEN

### Status

Monochromatic vortex beams carrying an intrinsic orbital angular momentum (OAM) along their propagation direction play significant role in modern optics and other areas of wave physics, both classical and quantum [1–4]. Such beams have found numerous applications in optical manipulations, imaging, microscopy, classical and quantum communications, etc. Vortex beams have a circularly-symmetric intensity distribution in the transverse $(x, y)$-plane and a screw dislocation of the phase front (vortex) on the beam $z$-axis. The intrinsic OAM of a vortex beam is $L_z = \ell$ per photon in the $\hbar = 1$ units, where $\ell$ is the integer topological charge of the vortex.

Recently, there was a great theoretical [5–7] and experimental [8–12] progress in studies of spatiotemporal vortex pulses (STVPs) carrying an intrinsic OAM orthogonal (or, generally, tilted) to their propagation direction. One can anticipate that such pulses will provide new geometries and functionalities to vortex states, extending their applications to the space-time domain and ultrafast processes. STVPs have an elliptical intensity distribution in the propagation $(x, z)$-plane and a *spatiotemporal* vortex formed by the edge (or, generally, mixed edge-screw) phase dislocation in the pulse centre, Fig. 1. Such vortex propagates with the pulse and is actually located in the centre of the $(x, z - vt)$-plane, where $v$ is the velocity of the pulse propagation.

Although spatiotemporal vortices are also characterised by the integer topological charge $\ell$, calculation of the OAM carried by STVPs is rather nontrivial. On the one hand, one can expect the intrinsic OAM $L_y = \ell$ per photon in a STVP with a circular intensity profile [5,6,9]; generalising this expression to an elliptical profile yields $L_y = \ell \frac{\gamma+\gamma^{-1}}{2}$ [5,6], where $\gamma$ is the ratio of the ellipse semiaxes along the $z$ and $x$ axes. On the other hand, recent calculations of the OAM of an optical STVP [7] resulted in $L_y = \ell \frac{\gamma}{2}$, which yields *half-integer* OAM in circular pulses. This result is in sharp contrast to the previous three-decade-long experience with monochromatic vortex beams [1–4].

To understand the origin of this controversy, we consider paraxial linearly-polarised optical pulses, where one can neglect spin and spin-orbit interaction effects. The normalized (per photon) energy, momentum, and OAM of such pulses can be written as

$$W = \frac{1}{2N}\int (\boldsymbol{E}^2 + \boldsymbol{H}^2) d^3\boldsymbol{r} = \frac{\int |\tilde{\mathbf{E}}|^2 d^3\boldsymbol{k}}{\int \omega^{-1}|\tilde{\mathbf{E}}|^2 d^3\boldsymbol{k}} \simeq \omega_0 \,, \tag{1}$$

$$\boldsymbol{P} = \frac{1}{N}\int (\boldsymbol{E}\times\boldsymbol{H}) d^3\boldsymbol{r} = \frac{\int \omega^{-1}\boldsymbol{k}|\tilde{\mathbf{E}}|^2 d^3\boldsymbol{k}}{\int \omega^{-1}|\tilde{\mathbf{E}}|^2 d^3\boldsymbol{k}} \simeq \boldsymbol{k}_0 \,, \tag{2}$$

$$\boldsymbol{L} \simeq \frac{1}{N}\int \boldsymbol{r}\times(\boldsymbol{E}\times\boldsymbol{H}) d^3\boldsymbol{r} \simeq \frac{\int \omega^{-1}\tilde{\mathbf{E}}^*(-i\boldsymbol{k}\times\nabla_{\boldsymbol{k}})\tilde{\mathbf{E}}\, d^3\boldsymbol{k}}{\int \omega^{-1}|\tilde{\mathbf{E}}|^2 d^3\boldsymbol{k}} \,. \tag{3}$$

Here $\boldsymbol{E}(\boldsymbol{r}, t)$ and $\boldsymbol{H}(\boldsymbol{r}, t)$ are the real-valued electric and magnetic fields, $\tilde{\mathbf{E}}(\boldsymbol{k})e^{-i\omega(\boldsymbol{k})t}$ and $\tilde{\mathbf{H}}(\boldsymbol{k})e^{-i\omega(\boldsymbol{k})t}$ are their complex plane-wave Fourier amplitudes related as $\tilde{\mathbf{H}} = \omega^{-1}\boldsymbol{k}\times\tilde{\mathbf{E}}$, $\omega(\boldsymbol{k}) = k$, we use the $c = 1$ units, $N \propto \int \omega^{-1}|\tilde{\mathbf{E}}|^2 d^3\boldsymbol{k}$ is the number of photons, and $\omega_0 = \omega(\boldsymbol{k}_0)$ and $\boldsymbol{k}_0$ are the central frequency and the wavevector of the pulse. Equations (1)–(3) have the form of the normalised expectation values of the canonical energy, $\omega$, momentum, $\hat{\mathbf{p}} = \boldsymbol{k}$, and OAM $\hat{\mathbf{L}} = \hat{\mathbf{r}} \times \hat{\mathbf{p}} = -i(\boldsymbol{k}\times\nabla_{\boldsymbol{k}})$, operators with the 'photon wavefunction' $\tilde{\mathbf{E}}(\boldsymbol{k})/\sqrt{\omega(\boldsymbol{k})}$ in the momentum representation. The fact that the photon wavefunction is well-defined only in $\boldsymbol{k}$-space is well known [13,14] (the photon probability density is ill-defined due to the nonlocality of the $\omega^{-1}$ operator in real space), and the corresponding $\omega^{-1}$ factors in $\boldsymbol{k}$-space integrals in (1)–(3) are crucial for our consideration.

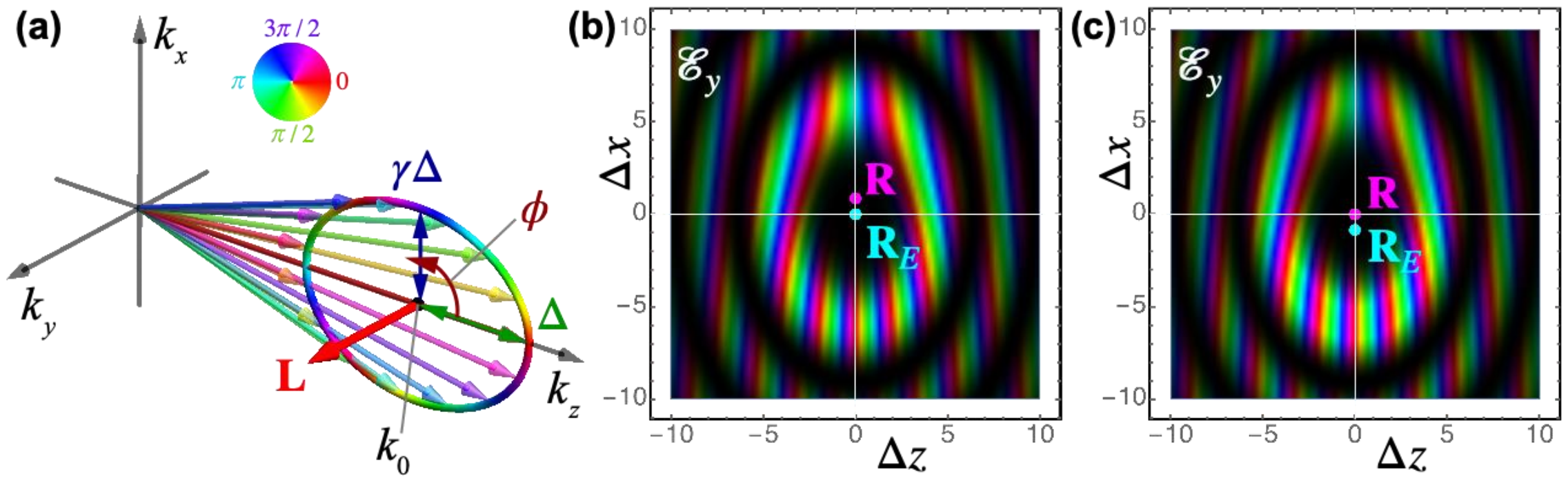

**Figure 1.** **(a)** The plane-wave spectrum (4) of a linearly $y$-polarized Bessel-like STVP with $\tilde{E}_y \propto exp(i\ell\phi)$. The phases of the plane waves are color-coded. **(b)** The instantaneous ($t = 0$) real-space distribution of the intensity (brightness) and phase (colour) of the complex electric field $\mathcal{E}_y$ (the real electric field is $E_y = Re\mathcal{E}_y$). The positions of the photon probability centroid $\boldsymbol{R}$, Eqs. (7) and (8), and the energy centroid $\boldsymbol{R}_E$ are shown. **(c)** Same as **(b)** but for the STVP with $\tilde{E}_y \propto \sqrt{\omega}\, exp(i\ell\phi)$ providing for equal densities of photons in each of the plane waves in the spectrum. The parameters are: $\ell = 3$, $\Delta/k_0 = 0.4$, and $\gamma = 0.7$.

For simplicity, we consider Bessel-type STVPs, with the plane-wave spectrum lying on an ellipse in $\boldsymbol{k}$-space, Fig. 1(a), and with the linear $y$-polarization of the electric-field bearing a vortex of the topological charge $\ell$[6]:

$$k_z = k_0 + \Delta\, cos\phi\,, \quad k_x = \gamma\Delta\, sin\phi\,, \quad \tilde{E}_y \propto exp(i\ell\phi)\,. \tag{4}$$

Here $\Delta \ll k_0$ and $\gamma\Delta \ll k_0$ are the ellipse semiaxes along the $z$ and $x$ directions, respectively, and $\phi$ is the azimuthal angle with respect to the ellipse centre. Substituting the field (4) into Eq. (3) and using relations

$$\frac{\partial}{\partial k_x} = \frac{cos\phi}{\gamma\Delta}\frac{\partial}{\partial\phi}\,, \quad \frac{\partial}{\partial k_z} = -\frac{sin\phi}{\Delta}\frac{\partial}{\partial\phi}\,, \quad \omega^{-1} \simeq k_0^{-1}\left(1 - \frac{\Delta\, cos\phi}{k_0}\right),$$

we calculate the $y$-component of the OAM of the STVP:

$$L_y \simeq \ell\frac{\gamma}{2}\,. \tag{5}$$

This unusual value agrees with the result of [7].

However, expression (5) is the *total* OAM calculated with respect to the chosen coordinate origin, and it does not mean the *intrinsic* OAM carrying by the STVP. In the case of monochromatic vortex beams, the well-defined intrinsic OAM $L_z = \ell$ is invariant with respect to parallel translations of the $z$-axis [15,3]. In the case of STVPs, a parallel translation of the $y$-axis along the $x$-direction by a distance $a$ transforms the OAM (5) as $L_y \to L_y + aP_z \simeq L_y + k_0 a$.

Following the prescription of nonrelativistic classical mechanics [16], the intrinsic and extrinsic parts of the OAM of a complex object can be separated using the position of the mass *centroid* of this object, $\boldsymbol{R}$, and its total momentum $\boldsymbol{P}$:

$$\boldsymbol{L}^{ext} = \boldsymbol{R} \times \boldsymbol{P}\,, \quad \boldsymbol{L}^{int} = \boldsymbol{L} - \boldsymbol{L}^{ext}\,. \tag{6}$$

Using this definition, the intrinsic OAM is invariant under translations of the origin, $\boldsymbol{r} \to \boldsymbol{r} + \boldsymbol{a}$: $\boldsymbol{L}^{int} \to \boldsymbol{L}^{int}$, while the extrinsic OAM is transformed according to the parallel-axis theorem: $\boldsymbol{L}^{ext} \to \boldsymbol{L}^{ext} + \boldsymbol{a} \times \boldsymbol{P}$. The separation (6) works well for monochromatic optical beams [3,17], because the beam centroid and momentum are well defined quantities.

However, the centroid of a non-monochromatic pulse is a subtle notion, which can be defined in different ways. In particular, despite the photon probability density is ill-defined in real space, the photon *probability centroid* is well defined as the expectation value of the position operator $\hat{\mathbf{r}} = i\nabla_{\boldsymbol{k}}$, similarly to Eqs. (1)–(3) [18,19]:

$$\boldsymbol{R} = \frac{\int \omega^{-1} \tilde{\mathbf{E}}^* e^{i\omega t} (i\nabla_{\boldsymbol{k}}) \tilde{\mathbf{E}} e^{-i\omega t} d^3\boldsymbol{k}}{\int \omega^{-1} |\tilde{\mathbf{E}}|^2 d^3\boldsymbol{k}}\,. \tag{7}$$

(Here the $e^{-i\omega t}$ factors in the Fourier amplitude are responsible for the propagation of the centroid with the group velocity along the $z$-axis.) Substituting here the STVP (4), we find that the photon centroid experiences a *vortex-induced shift* along the $x$-axis, Fig. 1(b):

$$X \simeq \frac{\ell}{2\gamma k_0}\,. \tag{8}$$

From Eqs. (5), (6), and (8), we obtain the intrinsic OAM of the STVP:

$$L_y^{int} \simeq L_y + k_0 X \simeq \ell \frac{\gamma + \gamma^{-1}}{2}\,, \tag{9}$$

which agrees with the OAM obtained in [5,6].

Alternatively, one can define the intrinsic OAM with respect to the *energy centroid* $\boldsymbol{R}_E$, which is defined similarly to Eq. (7) but without $\omega^{-1}$ factors in the integrands. The energy centroid is not shifted for the STVP under consideration, $X_E = 0$, and the corresponding intrinsic OAM coincides with (5): $L_y^{int'} \simeq L_y + k_0 X_E = L_y$.

The shift of the probability centroid (8) originates from the fact that the pulse spectrum (4) consists of plane waves with equal amplitudes but with *different densities of photons*, $\rho \propto \omega^{-1}|\tilde{\mathbf{E}}|^2$ (well-defined for plane waves). Considering a modified STVP with $\tilde{E}_y \propto \sqrt{\omega}\, exp(i\ell\phi)$, having the same density of photons in each plane wave in the spectrum, Fig. 1(c), we obtain: $X = 0$, $L_y = L_y^{int}$ given by (9), but $X_E \simeq -\frac{\ell}{2\gamma k_0}$ and $L_y^{int'} \simeq L_y + k_0 X_E$ given by (5).

Thus, the intrinsic OAM is highly sensitive to the definition of the pulse centre, whereas position of this centre is highly sensitive to the choice of plane-wave amplitudes in the pulse spectrum. This stems from the fact that the centroid of a relativistic extended body is not a uniquely defined notion [20]. It depends on the reference frame and on whether we calculate the centroid of relativistic masses/energy ($\boldsymbol{R}_E$) or that of the rest masses ($\boldsymbol{R}$) [18,19]. In the nonrelativistic case, e.g. for STVPs in the Schrödinger equation, both definitions coincide yielding the well-defined intrinsic OAM (9). In the general relativistic case, including optical STVPs, it is even *impossible* to construct a circularly-symmetric object with uniquely defined intrinsic OAM. Indeed, in the above examples of STVPs, the probability and energy centroids are always mutually shifted along the $x$-axis, and, hence, at least one of them is shifted with respect to the phase singularity (vortex centre). (Moreover, the singularity in the field $\boldsymbol{E}(\boldsymbol{r}, t)$ does not generally coincide with, e.g., singularities in the vector-potential field $\boldsymbol{A}(\boldsymbol{r}, t)$, i.e. the singularity is not a uniquely defined single point anymore.) This situation is in sharp contrast to symmetric monochromatic vortex beams, where all these centres coincide. A great advantage of defining the intrinsic OAM with respect to the probability centroid $\boldsymbol{R}$, Eqs. (6) and (7), as suggested in [5,18,19], is that it yields Eq. (9), *universal* for both spatial and spatiotemporal vortices (i.e., longitudinal and transverse OAM), in both the relativistic and nonrelativistic cases.

## Current and Future Challenges

The above consideration shows that defining the intrinsic OAM of optical STVPs is a subtle and tricky problem. Perhaps it does not have a unique fundamental solution and is a matter of theoretical convention. This requires further investigation. Notably, acoustic STVPs for sound in fluids/gases have the same peculiarities. This problem can also be considered within relativistic wave equations for massive quantum particles.

## Advances in Science and Technology to Meet Challenges

Of course, it would be extremely desirable to find suitable experimental methods for measurements of the OAM of optical or other STVPs. Interaction with atoms and STVP-induced atomic transitions could be one of such possibilities. Even if the experimental realization is challenging, theoretical calculations of light-matter interactions with STVPs could provide valuable data.

## Concluding Remarks

We have analysed the OAM properties of optical STVPs. Our results elucidate the origin of the recent controversy [5–7] and suggests a convenient definition of the intrinsic OAM. This results in a universal expression (9) [5,6] consistent with the longitudinal OAM of monochromatic vortex beams [1–4] and valid for any types of waves, both relativistic and nonrelativistic.

## Acknowledgements

I acknowledge fruitful debates with Howard M. Milchberg, which stimulated this work, and helpful discussions with Miguel A. Alonso.

# 7. Ultra-short vortex pulses and their nonlinear frequency conversion

Chen-Ting Liao[1*], Carlos Hernández-García[2], Margaret M. Murnane[1]

[1]JILA and Department of Physics, University of Colorado and NIST, Boulder, Colorado 80309, USA;
[2]Grupo de Investigación en Aplicaciones del Láser y Fotónica, Departamento de Física Aplicada, Universidad de Salamanca, Salamanca E-37008, Spain
*chenting.liao@colorado.edu

## Status

In recent years, research on ultrashort pulses with complex spatiotemporal shapes has focused on generating such light beams with topological phase structures. Building upon the conventional orbital angular momentum (OAM) beams discovered three decades ago, as introduced in prior chapter. nontrivial phases carrying singularities that are time-varying or are defined in space-time can now be mapped onto ultrafast pulses. Moreover, the light science underlying the extreme nonlinear process of high harmonic generation (HHG) makes it possible to up-convert structured light from the infrared (IR) into the extreme-ultraviolet (EUV) and soft X-ray (SXR) spectral regions. During the last decade, the relevant up-conversion rules for angular momenta have been established, opening up an exciting array of possibilities for imparting spin angular momentum (SAM) and OAM into short wavelength light, with femtosecond to attosecond pulse durations.

In the simplest case, a linearly polarized IR vortex beam with OAM $l_1$ is coherently up-converted into high-frequency (short wavelength) harmonic beams with charge $l_q=ql_1$, where $q$ is the harmonic order [1]. By harnessing this simple law and SAM conservation, new and powerful schemes are now available for tailoring EUV/SXR light. For example, by driving HHG with a necklace-shaped beam derived from the interference of two pulses with different OAM, HHG combs with tunable spectral spacing can be produced [2]. It is also possible to create attosecond pulse trains with controllable polarization states from pulse to pulse [3], harmonic vector-vortex beams [4], or circularly polarized harmonic vortices where the different polarization states have different divergences [5]. Very recently, a new property of light was uncovered via HHG—manifested as a time-varying OAM, that is changing during the pulse (Fig. 4.1) [6].

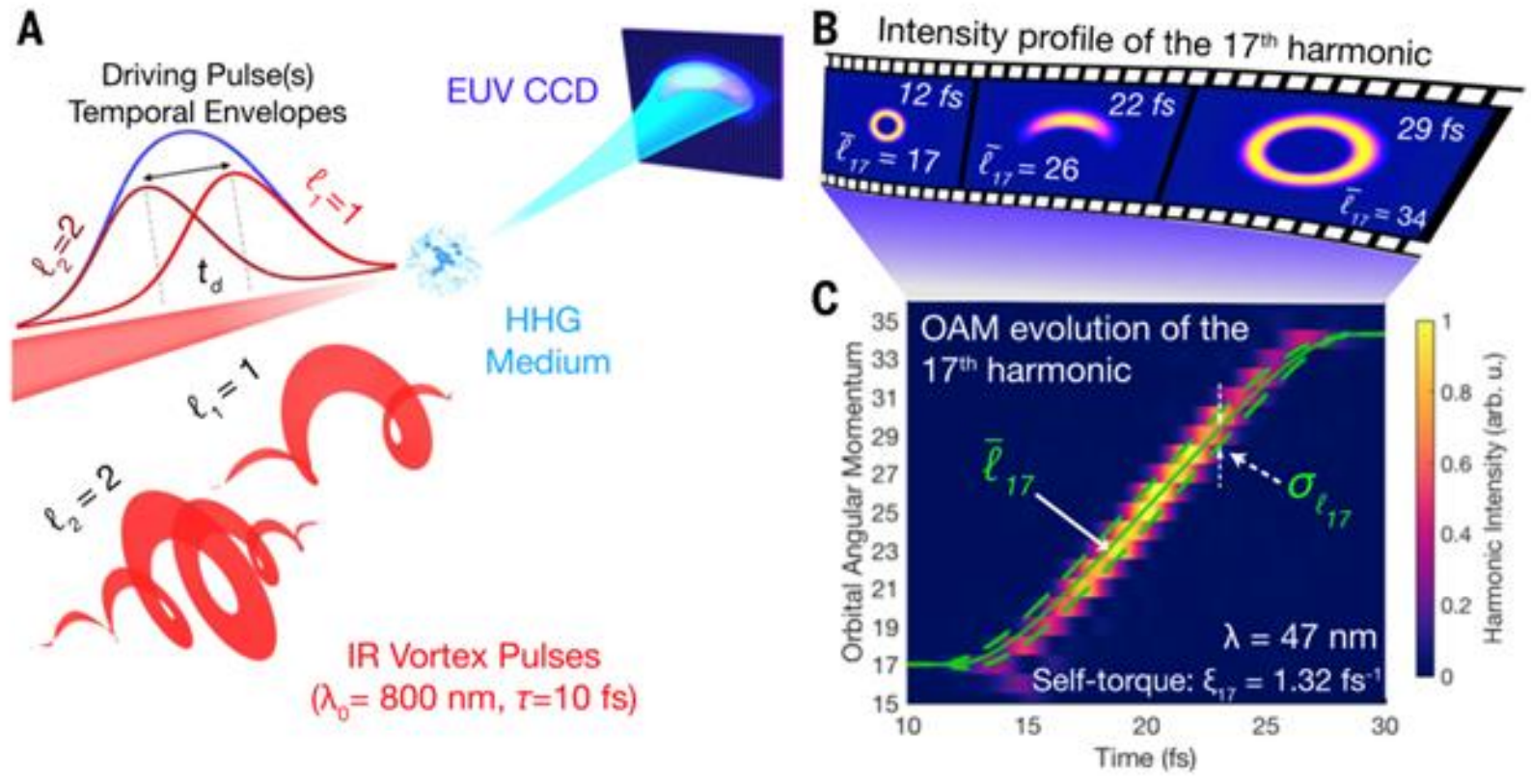

**Figure 4.1.** Generation of a time-varying OAM beam, called a self-torque. (**A**) Two time-delayed, collinear IR pulses with the same wavelength (800 nm), but different OAM values (spatial topological charges), are focused into an argon gas target (HHG medium) to produce harmonic beams with self-torque. The spatial profile of the complete, time-integrated, HHG beam from full quantum simulations is shown on the EUV CCD. (**B**) Predicted evolution of the intensity profile of the 17th harmonic at three instants in time during the emission process. (**C**) Temporal evolution of the OAM of the 17th harmonic, for two driving pulses with the same duration (10 fs), at a relative time delay of $t_d$. The average OAM of the 17th harmonic (solid green), and the width of the OAM distribution (distance between the solid and dashed-green lines), are obtained from theories. This figure is adapted from [6].

In addition, ultrafast pulses with controllable spatiotemporal OAM (ST-OAM) have been generated recently, as introduced in prior two chapters. Such transverse, spatiotemporal optical vortices (also known as STOV) present topological structures residing in *both* space and time. Recent experiments demonstrated that such ST-OAM follows (transverse) OAM conservation rules through nonlinear up-conversions (Fig. 4.2) [7]. For instance, through a second-harmonic generation process, the space-time topological charge $\ell_q^{st}$ of the fundamental field is doubled along with the optical frequency. Thus, $\ell_q^{st}=q\ell_q^{st}$, where $q=2$ for second-harmonic, reflecting the nature of ST-OAM conservation and following a general ST-OAM nonlinear scaling rule that resembles that of the (spatial) topological charge. Indeed, spatiotemporal phase singularities in ST-OAM can be interpreted as space-time topological charges carrying transverse OAM—the term coined in analogy to conventional (longitudinal) OAM.

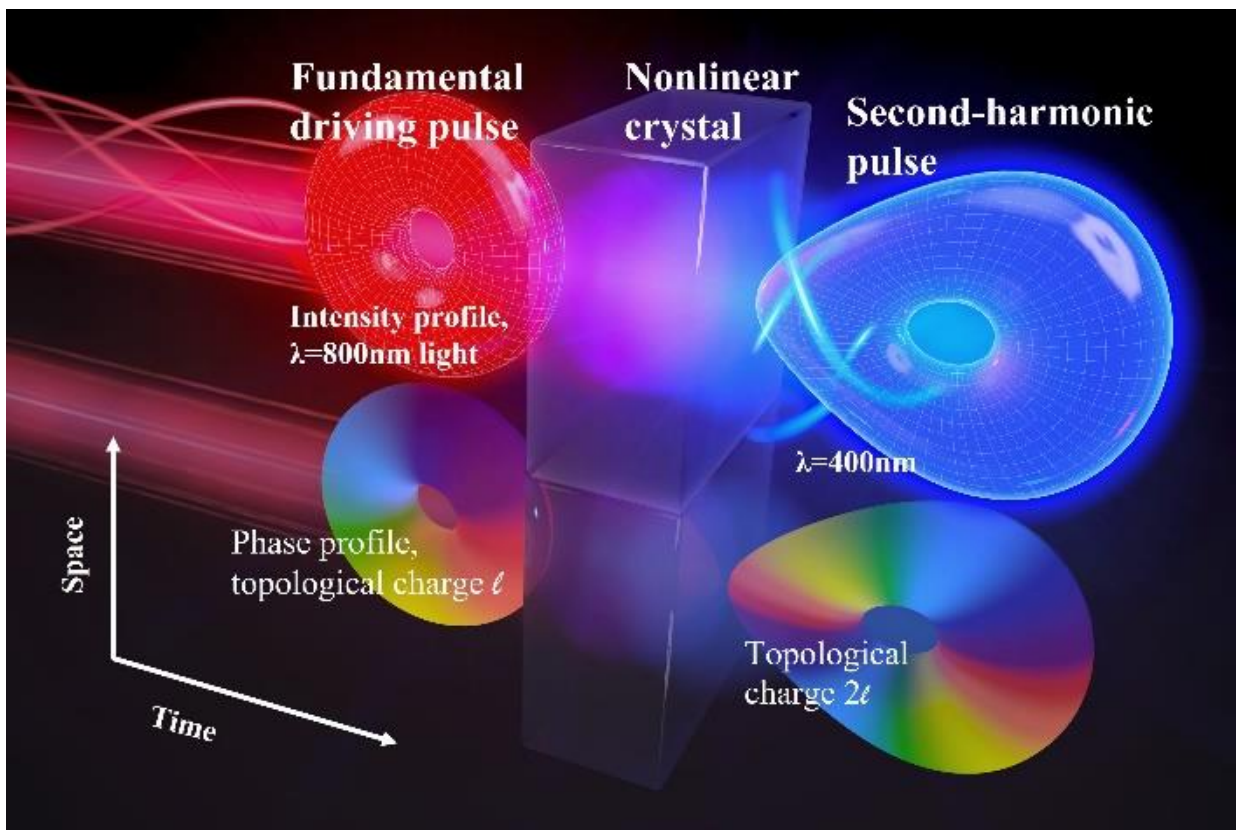

**Figure 4.2**. Schematic harmonic generation and topological charge conservation through a nonlinear process of spatiotemporal vortex pulses. An IR fundamental femtosecond spatiotemporal vortex pulses (left, red torus, 800nm wavelength) is converted into a blue, second-harmonic femtosecond pulse carrying ST-OAM (right, blue torus, 400nm wavelength) through a nonlinear crystal. In addition to optical frequency doubling, the space-time topological charge (rainbow-colored phase structures represented in the reflection) is also doubled and conserved in the process. Note that the topology of a second-harmonic ST-OAM pulse, namely, the number of holes or singularities in the pulse, may not be conserved. Therefore, a second-harmonic process can generate additional phase singularities separated in space-time, depending on spatiotemporal astigmatism due to group velocity mismatch or the phase-mismatch condition in a dispersive medium.

## Current and Future Challenges

The generation and application of short wavelength ultrashort vortex pulses with controllable phase, polarization, and spectral properties, or with time-dependent OAM or space-time OAM, is still in its infancy. For example, a full understanding of the nonlinear conversion and conservation rules of ST-OAM beams still remains to be achieved. While time-dependent OAM or self-torque has already been generated in the EUV using OAM-driven HHG, up-conversion of ST-OAM beams to high-order harmonics has only been proposed theoretically, as introduced in prior chapter, and thus remains to be experimentally validated. In addition, the exploitation of crystal symmetries through HHG in solids may open up new avenues for the generation of ultrashort pulse and short wavelength spatiotemporal OAM beams.

One of the most prominent experimental challenges in ultrashort pulse structured light generation is their extension to the SXR region. This capability would have several benefits: SXR photon energies span the core absorption edges of many magnetic and chiral materials; SXR light penetrates relatively thick samples; and the broad SXR HHG that are naturally generated can support pulse durations from femtosecond to attoseconds. To achieve this, further developments in the technology of structured mid-infrared driving lasers is required, which have been shown to efficiently phase match broadband SXR high harmonic generation.

To enable applications, the generation of topologically structured ultrashort pulses at shorter wavelengths must be accompanied by accurate and fast characterization techniques. While complete characterization techniques are being developed in the visible-to-infrared spectral regions, see chapters 13 & 14, their extension into the EUV/SXR remains very challenging. To date, wavefront sensor measurements [4], ptychography [8], and spectral interferometry [9] techniques present some of the most promising routes. Finally, we note that we have only discussed scalar beams (linearly and circularly polarized beams). Structured vector beams, which can possess space-varying and/or time-varying electric field directions [4], provide additional control knobs for tuning light-matter interactions. For instance, similar to the ST-OAM beams, whose phase structure resides in space and time, light beams can present electric field structures (or polarization states) residing in space and time, forming beams with non-transverse fields ($\nabla \cdot \vec{E} \neq 0$). The nonlinear up-conversion, topological charge conservation, and spin-orbit coupling of such non-paraxial beams are largely elusive.

**Advances in Science and Technology to Meet Challenges**

A grand challenge ahead includes the development of a complete understanding of light-matter interactions involving both time-dependent and spatiotemporal OAM. Our understanding of fundamental excitations when a matter is perturbed by such spatiotemporally structured light is still in its early stage. For instance, recent experimental findings show that chiral and/or topological magnetic materials will respond to spatial OAM due to the so-called OAM dichroism (also termed helicoidal dichroism) in the EUV and x-ray regimes [10]. A better understanding of these interactions is critical for potential applications for sensing and controlling magnetic, topological, and quantum excitations of materials and for manipulating molecules and nanostructures on their natural time and length scales. Gaining more insights into such interactions will also help to develop new applications in metrology, sensing, and imaging that could benefit from tailoring ultrashort visible and infrared pulses and their up-converted light into EUV, x-rays, and other pulsed particle beams such as electrons (see chapter 9) in both space and time.

**Concluding Remarks**

In the visible region of the spectrum, OAM beams have enabled applications in super-resolution imaging, optical communications, quantum optics, and microparticle manipulation. Many of these applications show promise for translation to short wavelength OAM beams. In the case of visible spatiotemporal OAM beams, the demonstration of an optical spatiotemporal differentiator is expected, see chapter 15. Spatiotemporal structured short wavelength OAM beams have promise as extraordinary tools for laser-matter manipulation on attosecond time and nanometer spatial scales.

**Acknowledgments**

The JILA authors acknowledge funding from the Department of Energy BES Award No. DE-FG02-99ER14982 for the development of the new characterization technique, and a MURI grant from the Air Force Office of Scientific Research under Award No. FA9550-16-1-0121 for the experimental setup. C.H.-G. acknowledges support from Ministerio de Ciencia e Innovación, y Universidades, Spain for a Ramón y Cajal contract (RYC-2017-22745), and the European Research Council (ERC) under the European Union's Horizon 2020 Research and Innovation Program (Grant Agreement No. 851201).

## 8. Vector space-time wave packets localized in all dimensions

Murat Yessenov[1] and Ayman F. Abouraddy[1]

[1]CREOL, The College of Optics & Photonics, University of Central Florida, Orlando, FL, United States of America

**Status**

Structuring pulsed optical fields jointly in space and time has been a long-standing goal in optics [1]. Spatio-temporal structuring can yield a host of useful characteristics, such as propagation invariance, whereby the spatio-temporally structured pulsed field propagates without diffraction or dispersion in linear media. This was first recognized by Brittingham in 1983 [2] who introduced the focus-wave mode (FWM), which propagates in free space invariantly at a group velocity $c$. To date the FWM has not been convincingly demonstrated. Subsequently, X-waves were introduced in ultrasonics in 1992 and then realized in optics in 1997 by P. Saari [3]. The X-wave is another spatio-temporally structured propagation-invariant wave packet whose group velocity however is superluminal. Nevertheless, both the FWMs X-waves require extremely large bandwidths and numerical apertures to deviate in any observable way from a conventional wave packet that is separable with respect to its spatial and temporal degrees of freedom [4].

This state of affairs changed about 6 years ago when our group at CREOL investigated a new class of propagation-invariant spatio-temporally structured pulsed beams that we have called 'space-time wave packets' (STWPs) [5,6]. In contrast to previous examples, STWPs can be produced with narrow spectra and small numerical apertures; they have demonstrated diffraction-free behaviour over extended distances; their group velocity can be tuned over an unprecedented span (subluminal, superluminal, and negatived-valued); and display a host of intriguing and useful characteristics [4]. This has led to rapid growth in the study of spatio-temporally structured optical fields as evinced in this Roadmap article.

These recent demonstrations of STWPs have a simplified field structure that is localized along one transverse dimension only; i.e., the STWP takes the form of a light-sheet. Interestingly, there are no *monochromatic* diffraction-free beams in one transverse dimension (except for the trivial cases of plane or cosine waves, which are not localized [4]). Indeed, a Bessel light sheet diffracts in contrast to its two-dimensional counterpart. The one exception of course is the monochromatic Airy beam in one dimension that is diffraction-free but travels along a parabolic trajectory. In contrast, because STWPs are pulsed rather than monochromatic fields, they do not suffer from this restriction [4], and thus represent the only truly propagation-invariant fields in one transverse dimension (diffraction-free *and* propagate along a straight line). Nevertheless, many applications in nonlinear optics and free-space propagation require field localization along both transverse dimensions – an optical needle rather than a light sheet.

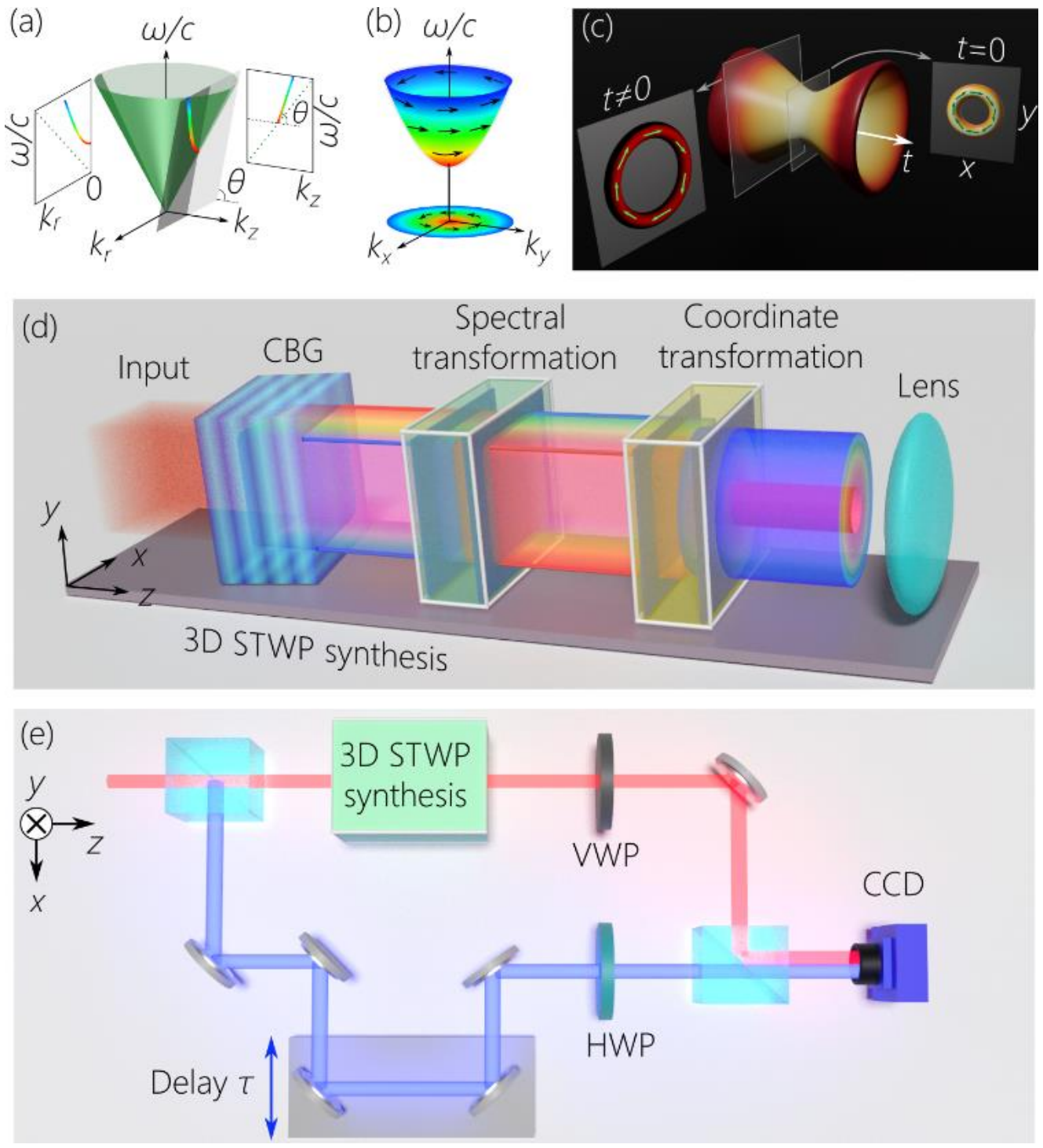

**Figure 1.** (a) The spectral support for a superluminal vector 3D STWP on the surface of the free-space light-cone $k_x^2+k_y^2+k_z^2=\left(\frac{\omega}{c}\right)^2$ and its spectral projections in $(k_r,k_z,\omega/c)$-space, and (b) in $(k_x,k_y,\omega/c)$-space. (c) Structure of a vector STWP with azimuthal polarization symmetry at a fixed axial plane, where we plot the spatio-temporal intensity profile and superimpose the distribution of the polarization vector at $t=0$ and $t\neq 0$. (d) Schematic of the optical setup for synthesizing 3D ST wave packets, and (e) for interferometrically characterizing 3D ST wave packets. STWP: Space-time wave packet; VWP: vortex wave plate; HWP: half-wave plate; CBG: volume chirped Bragg grating.

## Current and Future Challenges

The challenge in synthesizing STWPs that are localized in all dimensions is the difficulty or producing non-differentiable angular dispersion (AD) in two dimensions. AD refers to optical fields in which each frequency $\omega$ travels at a different angle $\varphi(\omega)$. Diffractive and dispersive devices, such as gratings and prisms, introduce AD into collimated pulses. In general, $\varphi(\omega)$ is assumed to be differentiable; i.e., $\frac{d\varphi}{d\omega}$ is defined everywhere. This is a natural assumption that is never stated explicitly, and the conventional theory of AD is based on a Taylor expansion of $\varphi(\omega)$ that takes this for granted.

This is surprising because AD also undergirds STWPs; each $\omega$ travels at a different angle $\varphi(\omega)$. Nevertheless, the attributes of STWPs are in direct contradistinction to those predicted for conventional AD. Our recent work has revealed that the AD underlying STWPs is non-differentiable: $\frac{d\varphi}{d\omega}$ is not defined at some frequency $\omega_o$ [7,8]. Such an AD-profile can be produced by the 'universal AD synthesizer' developed in our previous work [4], but only along one transverse spatial dimension. To produce an STWP that is localized along all dimensions, the radial wave number $k_r$ must be related to the frequency $\omega$ in the paraxial regime via the relationship: $\frac{\Omega}{\omega_o}=\frac{k_r^2}{2k_o^2(1-\cot\theta)}$, where $\Omega=\omega-\omega_o$, $\omega_o$ is the non-differentiable frequency, $k_o=\omega_o/c$, and $\theta$ is referred to as the spectral tilt angle. A useful visualization tool to understand the structure of STWPs is their spectral support on the light-cone surface $k_x^2+k_y^2+k_z^2=\left(\frac{\omega}{c}\right)^2$. Restricted to azimuthally symmetric fields structures, the AD profile for an STWP is associated with the spectral support in the form of a conic section at the intersection of the light-cone $k_r^2 + k_z^2 = \left(\frac{\omega}{c}\right)^2$ with a plane $\Omega = (k_z - k_o)\tilde{v}$, which represents a propagation-invariant pulsed field transported rigidly in free space at a group velocity $\tilde{v} = c\tan\theta$; see Figure 1(a,b). The resulting field structure is X-shaped in any plane passing through the meridional propagation axis as shown in Figure 1(c).

How can such a field structure be synthesized? We have recently demonstrated that the experimental strategy outlined in Figure 1(d) can achieve this task [9]: spectral analysis via a volume Bragg grating, 'reshuffling' the spatially resolved wavelengths in a desired order, followed by a log-polar transformation to arrange the wavelengths in an annulus, and finally a Fourier-transforming lens. Using this setup, we have produced the first propagation-invariant STWP that is localized in all dimensions [9].

**Advances in Science and Technology to Meet Challenges**

Although interest in spatio-temporally structured wave packets was initially driven by the desire for producing propagation-invariant wave packets, a host of other useful properties can also be realized. Examples include acceleration of the wave packet along the propagation axis, dispersion-free propagation in dispersive media (whether normal or anomalous), long-distance propagation for free-space communications, anomalous refraction (where the group velocity can increase or remain unchanged after traversing a planar interface to a higher-index medium), isochronous propagation (taking the same group delay to traverse a slab of different thickness by changing the angle of incidence), blind synchronization of remote optical clocks (without knowing the distance of multiple clocks from the same source), among other possibilities [4]. The first tests of these applications have been recently performed with STWPs in the form of light sheets. It is now possible to extend these applications to the typically more desirable form of a beam localized in both transverse dimensions.

Localization of STWPs along both transverse dimensions allows for exploring more complex spatio-temporally structures fields. For example, orbital angular momentum (OAM) can no be included in the STWP, which was not possible in the case of light sheets. The setup shown in Figure 1(d) is indeed capable of introducing OAM into the STWP, and time-resolved measurements performed via the interferometric configuration depicted in Figure 1(e) are plotted in Figure 2(a-d). Off-axis holography helps reconstruct the amplitude and phase of the field at multiple axial planes, and linear interferometry enables resolving the wave packet in time [9]. Figure 2(a,b) corresponds to an STWP endowed with no OAM, whereas Figure 2(c,d) corresponds to an STWP endowed with OAM of order 1. To the best of our knowledge, this is the first propagation-invariant pulsed OAM wave packet. Furthermore, vector beam structures [10] can be realized as shown in Figure 2(e,f) by placing a vortex wave plate in the path of the spatio-temporal spectrum as shown in Figure 1(e). Here, the polarization structure is distributed over the 3D structure of the STWP, and off-axis holography helps reconstructs this polarization distribution in space and time.

A critical avenue to be explored is the development of compact and stable arrangements for synthesizing STWPs. Two possible routes are adapting metasurfaces or volumetric Bragg gratings for this purpose; much more work needs to be done along these lines.

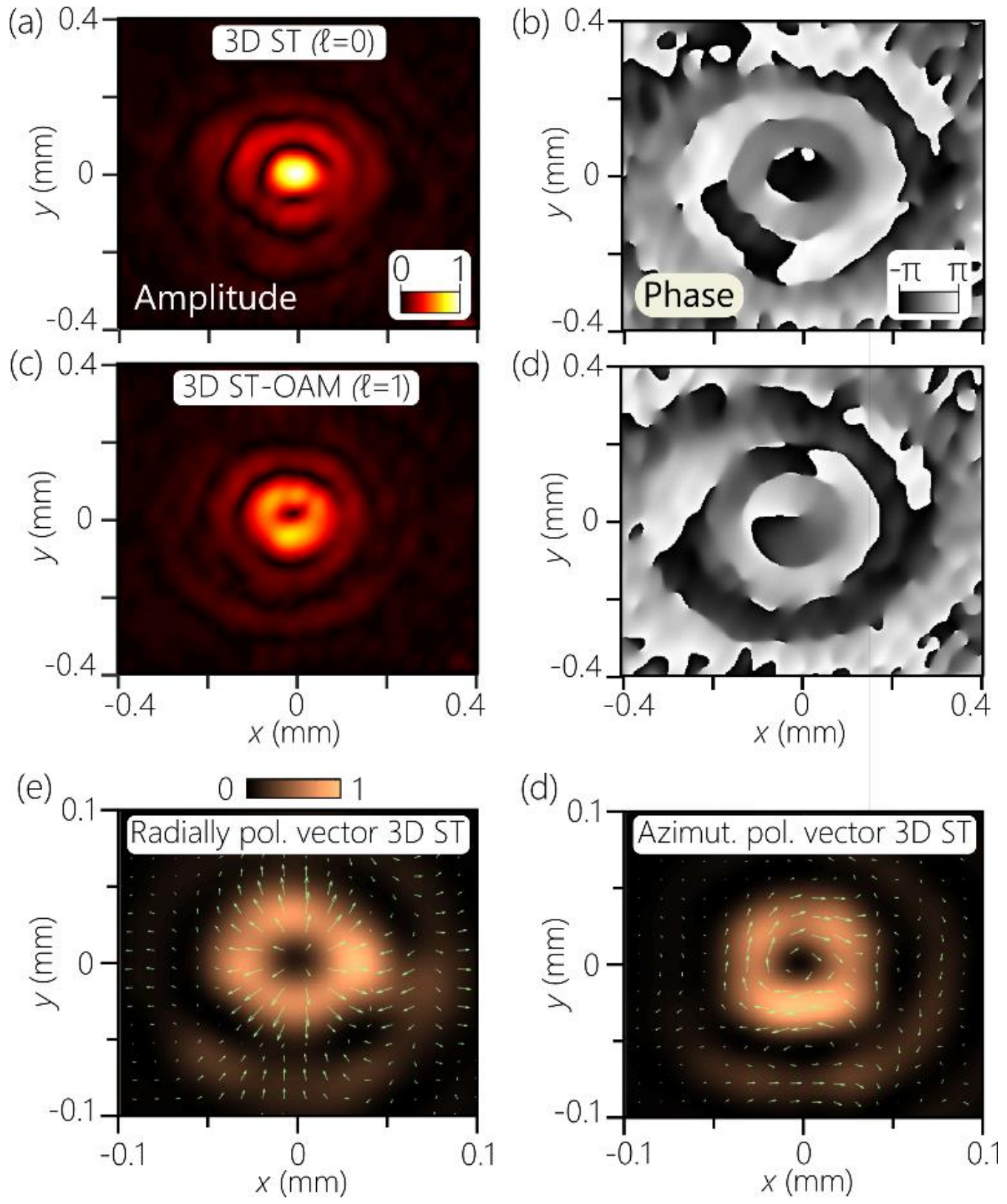

Figure 2. (a) Measured complex-field amplitude $|\boldsymbol{E}(x,y)|$ and (b) phase profile $\phi(x,y)$ at the fixed axial position $z$ and pulse center $\tau = 0$ for scalar 3D STWPs. (c-d) Same as (a-b) for 3D STWPs with OAM number $\ell = 1$. (e) Reconstructed polarization structure $\boldsymbol{E}(x,y)$ at the pulse center $\tau = 0$ and axial position $z$ for radially polarized and (f) azimuthally polarized vector STWPs. To improve the visualization, we plot as a background the total intensity $I = |E|^2$and the polarization is represented by green arrows whose direction and length correspond to those of the vector field $\boldsymbol{E}$.

## Concluding Remarks

The recent developments in synthesizing STWPs have opened new vistas for the study of spatio-temporally structured optical fields [4]. This has culminated most recently with the synthesis of STWPs that are localized in all dimensions and are endowed with orbital angular momentum and/or vector beam polarization structure. In many ways, this field of study is still in its very early stages of development, and it is anticipated that this research enterprise will expand dramatically in the near future. In addition to ongoing investigations of their basic physics, the vast array of unique attributes of STWPs provides ample opportunities for applications in free-space optical communications, metrology, nonlinear optics, and biomedical imaging. Advancement in the photonic technologies that reduce the footprint of spatio-temporal synthesis systems in particular will be crucial for the success of these applications.

## Acknowledgements

The work was funded by the U.S. Office of Naval Research (ONR) under Contract N00014-17-1-2458 and ONR MURI Contract N00014-20-1-2789.

## 9. Shaping Light by Shaping Free Electrons

Liang Jie Wong*,[1], Michael Go[1], Suraj Kumar[1]
[1] School of Electrical and Electronic Engineering, Nanyang Technological University, 50 Nanyang Avenue, Singapore 639798, Singapore
* liangjie.wong@ntu.edu.sg

### Status

Free electron-driven light-matter interactions play a central role in modern science and technology. Free electron radiation has attracted much interest by revealing unprecedented physics in processes like Cherenkov radiation, Smith-Purcell radiation, superradiance, inverse Compton scattering, etc. [1,2]. At the same time, free electrons have proven an invaluable tool for device characterization and diagnostics through mechanisms like cathodoluminescence and photon-induced near-field electron microscopy (PINEM) [3,4]. Free electron radiation plays an indispensable role at extreme wavelengths such as the X-ray and gamma ray regimes of the electromagnetic spectrum [5]. It is noteworthy that practically all X-ray sources today – from commercial X-ray tubes to stadium-sized synchrotrons and kilometers-long free electron lasers – use free electrons as their basic operating principle. Therefore, the wealth of applications that depend on X-rays – including medical diagnostics, manufacturing inspection, security scanning, forensics, biological imaging and crystallography – is testament to the usefulness and importance of free electron science in the today’s world.

There are fundamental reasons why free electrons have advantages over bound electrons (e.g., in atoms and materials) and other particles/quasi-particles in certain scenarios. Firstly, free electrons are readily accelerated to relatively high energies with just a direct current (DC) power supply. This is used in X-ray tubes to produce electron kinetic energies of several 10s of keV or more, needed for collisional inner-shell ionization and the consequent generation of characteristic X-ray peaks. Secondly, the free electron has a continuum of energy states, unlike bound electrons which are restricted to discrete energy states. This allows free electrons to emit radiation at any frequency, not being confined to the differences between discrete energy levels. Thirdly, the free electron effectively functions as a highly nonlinear optical medium: a moving electron readily up-converts or down-converts photon frequencies via the Doppler effect. This is exploited in X-ray synchrotrons and free electron lasers, where relativistic free electrons up-convert the periodic magnetostatic field of undulators into X-ray photons via the relativistic Doppler effect.

The importance of shaped light can be seen through numerous advances in imaging, sensing, manufacturing, communications, and computation. For free electron-based radiation processes, the shaping of free electrons provides extra degrees of freedom to shape the emitted light or control light-matter interactions. In this article, we discuss the promises of free electron shaping for creating exotic electromagnetic modes and photon statistics (Fig. 1) that are typically inaccessible via traditional methods of light shaping, across the spectrum from microwaves to X-rays and gamma rays.

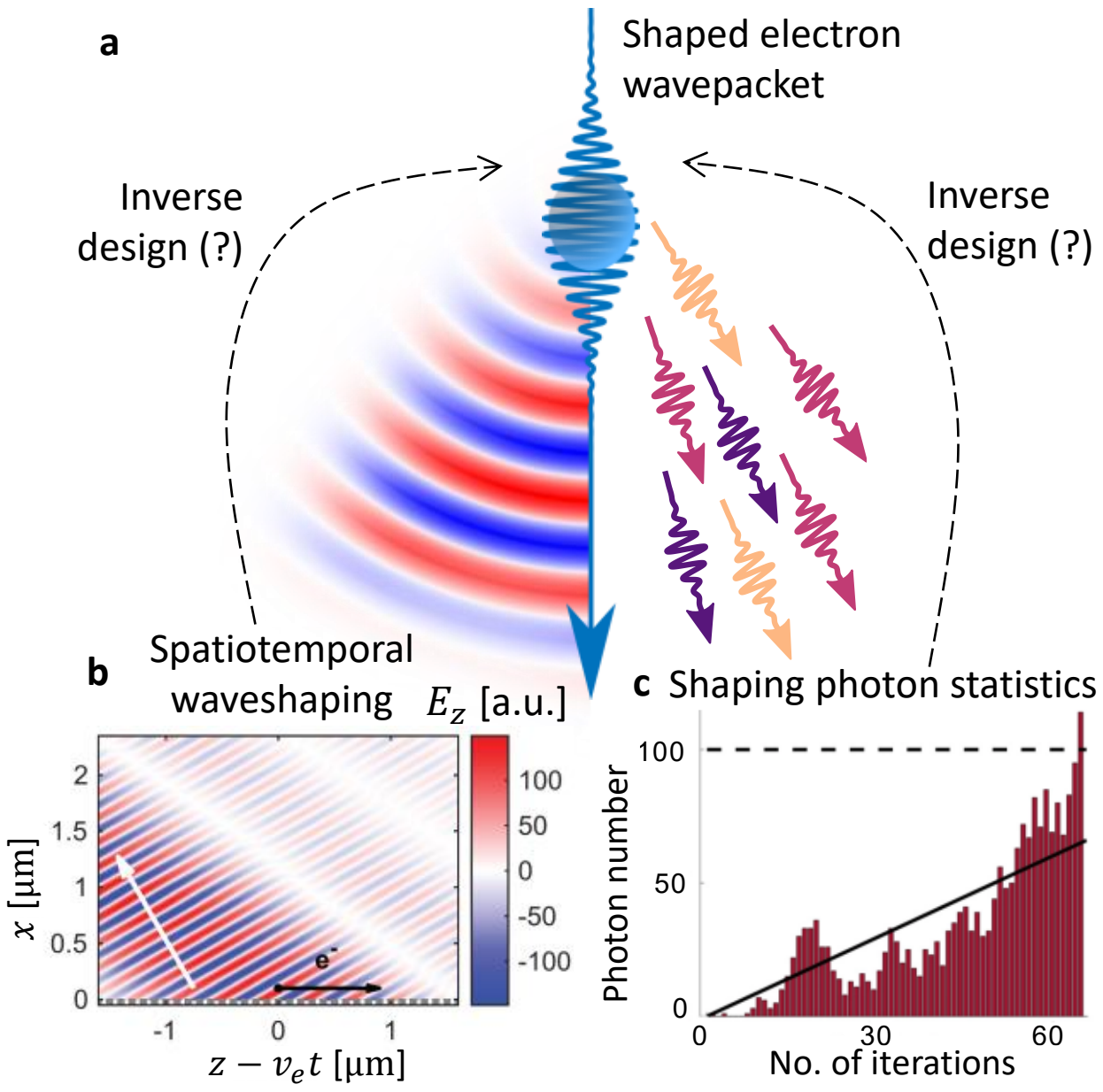

**Figure 1.** (a) Free electron waveshaping provides a means to tailor the spatiotemporal and spectral profile of the emitted electromagnetic field (left half) as well as the quantum statistics of emitted photons (right half). An open problem is that of inverse design: how to determine the required electron wavepacket shape for a desired spatiotemporal profile and/or statistical distribution. (b) shows the electric field profile of a space-time wavepacket with central wavelength 193 nm generated through Smith-Purcell radiation from a 200 keV electron ($e^-$) bunch of speed $v_e$. The white arrow indicates direction of propagation of the emitted radiation's wavefront. The grating is drawn in grey at the bottom. (c) Proposed stochastic process of obtaining a target Fock state with 100 photons (dashed line) from multiple free electron-light interactions [6]. The photon number of the Fock state is plotted against the number of interactions, showing the gradual evolution to a large-number Fock state. (b) used with permission from Y. J. Tan *et al.*, *Advanced Science*, vol. 8, no. 22, p. 2100925, 2021 [2]. Copyright 2021 Authors, licensed under a Creative Commons Attribution (CC-BY) license. (c) used with permission from A. Ben Hayun *et al.*, *Science Advances*, vol. 7, no. 11, 2021 [6]. Copyright 2021 Authors, licensed under a Creative Commons Attribution (CC-BY) license/ Cropped from original.

## Current and Future Challenges

Spatiotemporal shaping of electromagnetic fields is readily achieved in the microwave to optical regimes through an arsenal of light shaping elements, including programmable spatial light modulators (SLMs), metasurfaces, waveguides, etc. Complex photonic circuits have also been realised for quantum shaping of optical light, which involves shaping of the photon's wavefunction (equivalently, its probability distribution) and of photon statistics. Although these light-shaping techniques have enjoyed rapid advances, many desired electromagnetic profiles and photon modes remain out of reach.

One example is entangled high-number Fock states, which have been predicted to improve quantum metrology, sensing and imaging. The creation of such states, however, remains elusive as there is no efficient, on-demand means to generate Fock states, especially for high photon numbers. As shown recently, free electrons can provide a versatile platform for customizing photon statistics, for instance, to create deterministic, high-number Fock states [6].

Another example is the challenge of tailoring and focusing extreme wavelength radiation such as X-rays. Due to the near-unity refractive indices of materials in the X-ray regime, large, defect-free surfaces and multilayer structures fabricated with ultra-high precision are often needed to steer and mold an X-ray beam. Furthermore, traditional X-ray optics requires X-rays that are substantially coherent over large areas, but such X-rays are only available from large facilities like synchrotrons. Due to these constraints, the question arises as to how to achieve the versatility and robustness of visible-light optics in the X-ray regime.

As we discuss in the next section, shaping free electrons could be key to resolving this challenge. Since free electrons are already an essential component in practically all X-ray sources, it stands to reason that manipulating the electrons themselves could be an effective means of controlling the emitted X-ray radiation [5]. This is in fact the principle behind free electron lasers, where an electron bunch interacts with its own emitted X-ray radiation, ultimately evolving into electron nano-bunches in a process known as self-amplified spontaneous emission [1]. These nano-bunches function like a nanoscale antenna array by emitting X-rays that constructively interfere, producing highly directional, ultra-bright X-rays. The foregoing, however, involves only the control of electrons as a collection of classical point charges. The fundamental question arises as to whether we can shape light by manipulating the wavefunction of a single electron emitter. In the next section, we discuss how this is indeed the case.

**Advances in Science and Technology to Meet Challenges**

As nicely put by Feynman, an electron behaves ultimately as a point-like particle, regardless of its wavefunction. It is therefore valid to question whether shaping the wavefunction of a single free electron could possibly affect the emitted radiation from said free electron. One might expect that this is not the case, since the electron would be emitting as a point-like particle regardless of its wavefunction. Quantum theory, however, reveals surprisingly that shaping the wavefunction has a substantial impact on the emitted photons for both stimulated and spontaneous emission [7]. In particular, it has been theoretically shown using quantum electrodynamics that the free electron wavefunction can be shaped to make bremsstrahlung more directional and undulator radiation more monochromatic [7]. These theoretical findings remain to be experimentally demonstrated. The wealth of possible free electron waveshapes also opens new areas of exploration where the polarization, orbital angular momentum and other properties of electron waves can be harnessed to realize unique and/or bespoke output radiation profiles and photon statistics, especially at extreme wavelengths like X-rays and gamma rays [5].

It is also intriguing to consider the shaping of multi-electron pulses. Disruptive technologies remain to be discovered just by considering the classical picture of electrons as point charges, exemplified by a recent study where shaped electron bunches were used to suppress microbunching instability endemic in free electron lasers [8]. Even more fascinating is the question of whether and how we can combine the notion of classically shaping a distribution of point charges with that of shaping the quantum wavefunction of each individual electron. Some inroads have been made into this in a recent theoretical study where a pair of entangled input electrons were considered [9]. The study found that different free-electron Bell-states leads to substantially different spectra and spatial patterns of the emitted light, in a manner that cannot be accounted for by a classical mixed state. However, the practical question of how to shape each electron individually in a multi-electron pulse for experimental demonstration remains. A recent study created two-electron pulses from a nanotip to study their non-Poissonian nature using coincidence measurements [10], paving the way to further studies of multi-electron effects in electron pulses.

Finally, the inverse design problem remains to be tackled – to develop frameworks and algorithms for determining the wavefunction of input electron/s in a computationally efficient way based on the spatial, spectral and statistical properties of the desired output.

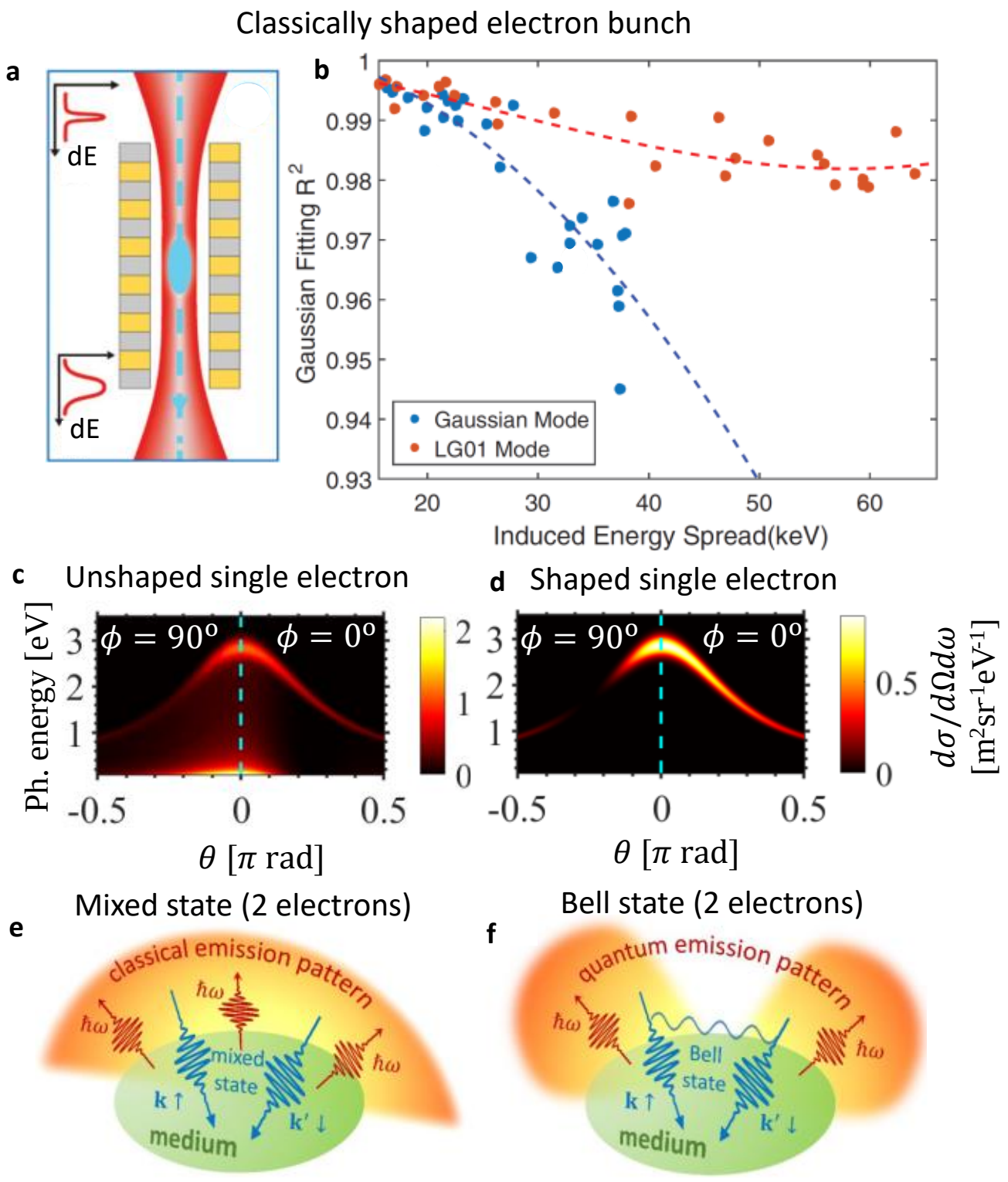

**Figure 2.** (a) Shaping the momentum space of a classical free electron bunch by laser-electron interaction in a laser heater to suppress microbunching instability. (b) shows that the electron bunch deviates much less from the desired Gaussian distribution with increasing induced energy spread under the influence of a Laguerre-Gaussian (LG01) laser mode (red), as opposed to a conventional Gaussian laser mode (blue). (c) An unshaped (single-state) 200 keV free electron incident on a short nanoundulator generates radiation with a narrowband peak as well as a broadband background at lower photon energy (ph. energy). In (d), using a shaped (two-state) free electron of the same kinetic energy makes the output radiation more monochromatic by destructively cancelling the broadband background. (e) The cathodoluminescence emission pattern from a two-electron state is generally non-directional. However, as illustrated in (f), entangling the electrons into a Bell state results in a different angular emission profile due to superradiance and subradiance phenomena. (a,b) Used with permission from J. Tang *et al.*, *Physical Review Letters*, vol. 124, no. 13, 2020 [8]. Copyright 2020 Authors, licensed under a Creative Commons Attribution (CC-BY) license/ Cropped from original image. (c,d) Used with permission from L. J. Wong *et al.*, *Nature Communications*, vol. 12, no. 1, 2021 [7]. Copyright 2021 Authors, licensed under a Creative Commons Attribution (CC-BY) license/ Cropped from original image. (e,f) Figure reprinted with permission from Karneli *et al.*, *Physical Review Letters*, vol. 127, no. 6, 2021 [9]. Copyright 2021 American Physical Society/ Cropped from original image.

## Concluding Remarks

Free electrons play an indispensable role in today's photonics landscape because 1) free electrons can access extreme wavelengths like X-rays (a multi-billion-dollar industry) and gamma rays, which lie beyond the means of purely bound electron-based light sources, and 2) free electrons can realize radiation with spatial, spectral and statistical properties that are not possible or challenging to achieve by other means. The advent of new methods to shape free electrons provides new degrees of freedom with which to shape the emitted light. The shaping of free electrons can take place in two ways: control over the classical distribution of point charges, and the shaping of the electron's quantum wavefunction on a single-electron level. The nascent field of free electron waveshaping has already presented some remarkable ideas in light shaping, including shaping of bremsstrahlung and undulator radiation, and unprecedented control over the quantum statistics of output photons. Future research includes the study of exotic electron waveshapes, the physics of entangled free electrons and multi-electron pulses in light-matter interactions, and the inverse design problem. An improved understanding of the relationship between free electrons and the photons they emit should pave the way to greater robustness and versatility in next-generation light sources.

**Acknowledgements**
This work was supported by the National Research Foundation (Grant No. NRF2020-NRF-ISF004-3525). LJW acknowledges the support of the Nanyang Assistant Professorship start-up grant.

## 10. Space-time correlation by nonlocal nanophotonics

*Cheng Guo and Shanhui Fan*
Stanford University

### Status

Light usually diffracts or disperses under propagation. However, there are unusual solutions to linear Maxwell's equations, called space-time wave packets or light bullets [1], that are propagation-invariant. These wave packets do not suffer from dispersion or diffraction. They can also propagate with an arbitrary group velocity. Such wave packets have been extensively studied for potential applications including communications, bioimaging, lithography, and quantum key distribution.

The fundamental property of all space-time wave packets is the characteristic "space-time correlation" between their temporal and spatial frequencies: $\omega = v_g k_z + b$, where $\omega$ is the angular frequency, $k_z$ the wavevector component in the propagation direction, $v_g$ the group velocity, and $b$ a constant. Generating space-time wave packets requires the accurate synthesis of the space-time correlation. There is substantial progress to create the desired space-time coupling for wide ranges of $v_g$ and $b$ [2] with the use of spatial light modulation and free-space optics. However, these setups are quite bulky. For further developments in this field, it is highly desirable to develop techniques to generate space-time wave packets with a more compact set up.

In recent years, there are significant advances in nonlocal nanophotonics, which enable the control of light in the wavevector space. Such a control can be achieved by using either *static* periodic structure or *dynamic* spatiotemporal modulation. The former includes nonlocal metasurfaces [3] that exhibit wavevector-dependent transfer functions, which are in sharp contrast with conventional local metasurfaces characterized by space-dependent transfer functions. They enable important functionalities including optical differentiation [3], [4], squeezing free space [5], and generating optical pulses with tilted orbital angular momentum (OAM) [6]. The latter includes traveling wave modulators that can induce wavevector modulation in guided waves. They enable significant opportunities in integrated nonreciprocal photonic devices [7], [8].

Here we briefly review recent works which reveals a deep connection between the fields of space-time wavepacket and nonlocal nanophotonics. The basic idea is that the desired space-time correlation can be induced by nonlocal nanophotonics directly in the frequency-wavevector domain. This enables the capability of generating space-time wave packets using compact nanophotonic devices.

### Current and Future Challenges

Generating space-time wave packets is a nontrivial task; it requires simultaneous and precise control of frequency and wavevector for a light pulse. Here, we highlight two important challenges:

1. The generation of space-time wave packets in free space using a compact optical device. The conventional approach to generating space-time wave packets uses a technique combining spatial-beam modulation and ultrafast pulse shaping. Such a technique requires bulky components and complicated operations. For practical applications, it would be preferable to generate space-time wave packets in free space with a simple and compact device.

2. The in-situ generation of space-time wave packets in a waveguide. Most existing works focus on space-time wave packets propagating in free space or bulk media. Creating space-time wave packets in a waveguide may lead to new opportunities in integrated photonics. For example, these wave packets can propagate at arbitrarily low group velocity but still possess broad spectral bandwidth. Previous proposals to generate such wave packets either first synthesized a wavepacket in free space and then coupled it into the waveguide or relied upon strong nonlinearity. For practical applications, it would be preferable to generate space-time wave packets in a waveguide in-situ without the need for nonlinearity.

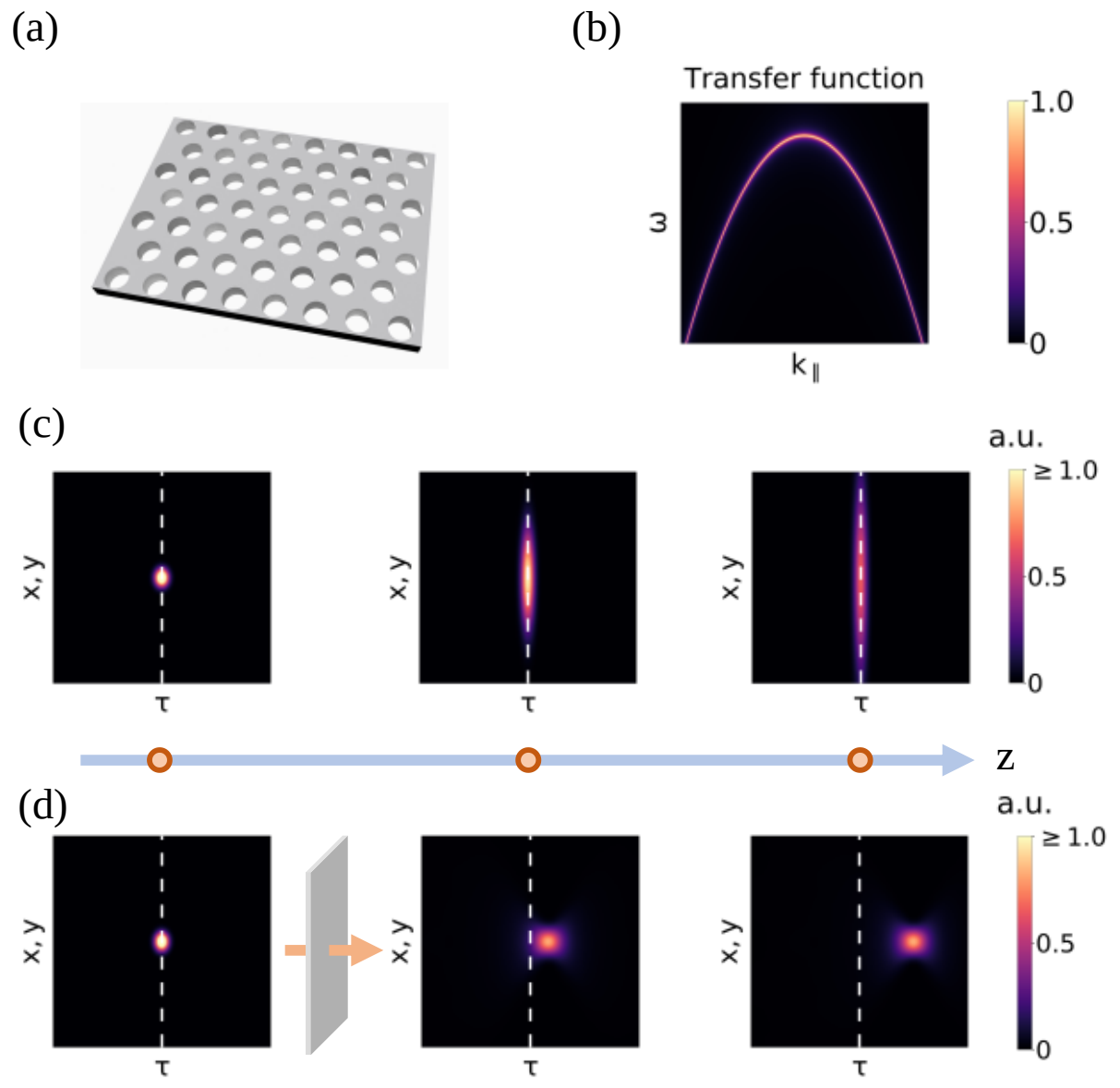

**Figure 1.** (a) A properly designed photonic crystal slab device. (b) The device's transfer function enforces the space-time correlation required by light bullets. (c) A Gaussian wave packet propagates with a group velocity c and diffracts. (d) The photonic crystal slab device transforms the Gaussian wave packet into a light bullet, which propagates rigidly with a group velocity $v_g$ that may differ from c. The plots use a temporal frame τ=t-z/c that moves at c. The vertical dashed lines indicate τ=0. Reprinted from Ref. [9], copyright 2021 Springer Nature.

## Advances in Science and Technology to Meet Challenges

Here we highlight two recent advances in nonlocal nanophotonics that provide new approaches to addressing the challenges mentioned above.

1. Generation of space-time wave packets in free space using nonlocal metasurfaces. Ref. [9] shows that an ultrathin nonlocal nanophotonic layer provides a compact and versatile platform to generate controllable 3D light bullets in free space. This approach consists of sending a Gaussian wave packet [Fig. 1(c)] into a single-layer periodic nanophotonic structure supporting guided resonance [Fig. 1(a)]. By designing the band dispersion of the guided resonance [Fig. 1(b)], one can generate the required space-time correlation to achieve a 3D light bullet as the output [Fig. 1(d)]. This method offers straightforward control of the external degrees of freedom of the light bullet including the group velocity and the propagation distance, and the internal structures such as the spin angular momentum and orbital angular momentum. Such controllable light bullets may inspire broader applications.

2. Generation of space-time wave packets in a waveguide using traveling wave modulation. Ref. [10] shows that in a multimode waveguide, space-time wave packets can be generated using indirect photonic transitions. In this approach [Fig. 1(a)], one first excites an eigenmode of the waveguide [Fig. 1(b, left)], and then transmit the light through a traveling wave modulator. With a suitable design, the modulator can convert such an excitation into a space-time wave packet [Figs. 1(b, middle) and 1(b, right)]. The group velocity of the wave packet equals the phase velocity of the traveling wave modulation and thus can be arbitrarily tuned. Such an approach can generate broadband propagation-invariant space-time wave packets with tailored field profiles. This work reveals a connection between the study of time-varying media and space-time optics and may enable integrated photonic applications of guided space-time wave packets.

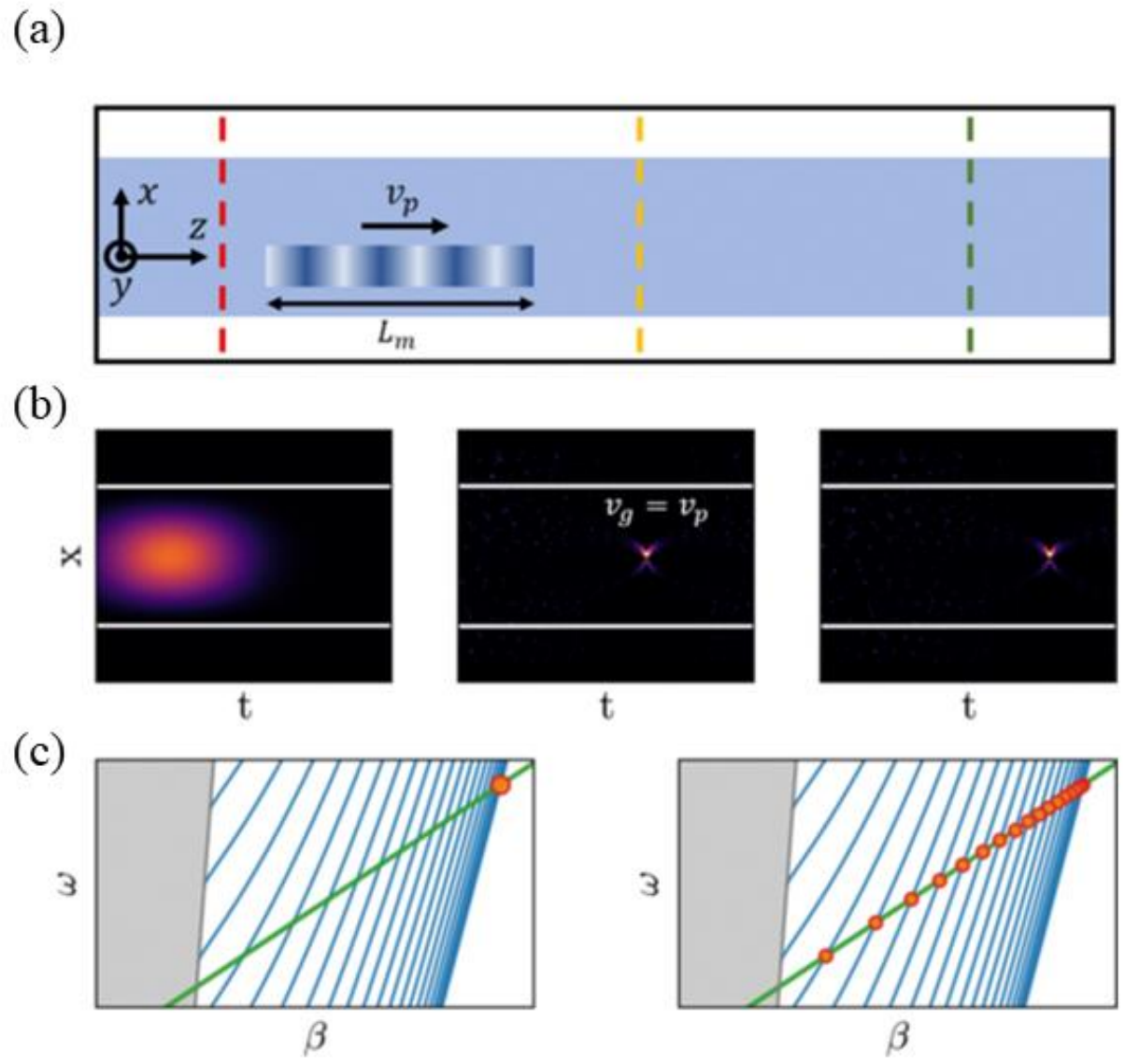

**Figure 2.** (a) A multimode waveguide with a region of length $L_m$ modulated by a traveling wave with phase velocity $v_p$. (b) Left: the incident light [at the red line in (a)] is approximately an eigenstate of the static waveguide. Middle, right: the transmitted light [at the yellow and green lines in (a)] is a propagation-invariant space-time wave packet with a group velocity $v_g = v_p$. (c) The spatiotemporal spectrum of the incident (left) and transmitted light (right) on the dispersion relation of the waveguide (blue curves). ω and β are the frequencies and the wavevector, respectively. The selected eigenmodes (orange dots) lie on a straight line specified by the traveling wave modulation. Reprinted from Ref. [10], copyright 2021 American Physical Society.

## Concluding Remarks

In summary, nonlocal nanophotonics can be utilized to create nontrivial space-time correlations. This provides simple methods to sculpt light pulses in the frequency-wavevector domain with high precision. Such a capability opens significant new opportunities in the emerging field of space-time nanophotonics. We anticipate further development of compact nonlocal nanophotonic devices that can significantly advance the practical applications of space-time wave packets.

## Acknowledgements

This work is supported by a Vannevar Bush Faculty Fellowship from the U. S. Department of Defense (Grant No. N00014-17-1-3030), by the U. S. Office of Naval Research (Grant No. N00014-20-1-2450), and by a MURI project from the U.S. Air Force Office of Scientific Research (FA9550-18-1-0379).

## 11. Toroidal and supertoroidal light pulses

Nikitas Papasimakis, University of Southampton (UK)
Yijie Shen, University of Southampton (UK)
Nikolay I. Zheludev, University of Southampton (UK) & Nanyang Technological University (Singapore)

### Status

Toroidal light pulses (TLPs), termed Flying or Focused Doughnuts, are few-cycle pulses of toroidal topology with space-time non-separable structure and vector polarization (i.e. azimuthally or radially polarized) that were discovered by Hellwarth and Nouchi in 1996 (Fig.a1) [1]. Research interest in toroidal light pulses has recently surged owing to the emergence of toroidal electrodynamics [5], but also due to advances in ultrafast lasers and the wide availability of wavefront and pulse shaping devices (DMDs, SLMs, etc.). In fact, toroidal light pulses have only recently been generated experimentally for the first time [2]. Following the demonstration of their generation, a broader family of pulses has been introduced, termed Supertoroidal Pulses (STPs), of which the Flying Doughnut is the fundamental member.

Toroidal and supertoroidal light pulses exhibit a unique combination of physical characteristics radically different from conventional light waves:

1) Non-transverse electromagnetic field configuration: the magnetic (electric) field traces the body of a torus, while the electric (magnetic) field traces the torus surface, resulting in field components oriented along the direction of pulse propagation (Fig.a2);
2) Few-cycle pulse duration: for instance, fundamental TLPs exhibit single and 1 ½ cycle duration, switching from one to the other upon propagation due to a Gouy phase shift (Fig.a1);
3) Toroidal topology: Toroidal pulses are ideal probes for engaging toroidal and non-radiating anapole excitations in matter [5,6];
4) Skyrmion-like field topology: STPs exhibit topological structure similar to that observed in solid-state (e.g. magnetic skyrmions) (Fig.a3) [3];
5) Self-similar patterns: STPs exhibit a complex, fractal-like singularity structure consisting of nested singular shells (Figs.b1,b2) [3];
6) Space-time non-separability: The spatial dependence of STPs cannot be separated from the temporal one, which has significant effects on their propagation dynamics. For instance, in the fundamental TLPs, space-time non-separablity leads to isodiffracting propagation [7-9], whereas in STPs non-separability results in non-diffracting propagation over arbitrarily long distances (Fig.c1), reminiscent of vortex street flows in fluids (Figs.c2,c3) [4].
7) Space-time superoscillations: STP fields can oscillate faster than the highest spatial and temporal frequency components of the pulses (Fig.b3) [10].

To date, only the fundamental toroidal light pulses have been experimentally generated in the optical and THz part of the spectrum [2], while the other members of the family have only been studied theoretically. Moreover, the propagation dynamics and light-matter interactions of toroidal light pulses remain largely unexplored, in particular with respect to their unique space-time non-separable and skyrmionic field structures.

## Current and Future Challenges

The main challenges facing the field involve the experimental generation and characterization of said pulses. In particular, the generation part requires precise spatial control of amplitude, phase and polarization of the electric field over a broad frequency band. On the other hand, the characterization of such pulses also introduces a number of significant challenges. Similarly to pulse generation, characterization will also require accurate retrieval of the electric field of the generated pulses as a function of space and time. Recent developments on techniques for the full 3D retrieval of the electromagnetic fields of few-cycle vector polarized pulses (e.g. ref. [11] and related techniques shown in Chapter 13 and Chapter 14 of this roadmap) can be expanded and applied to the more complex supertoroidal and space-time superoscillatory pulses.

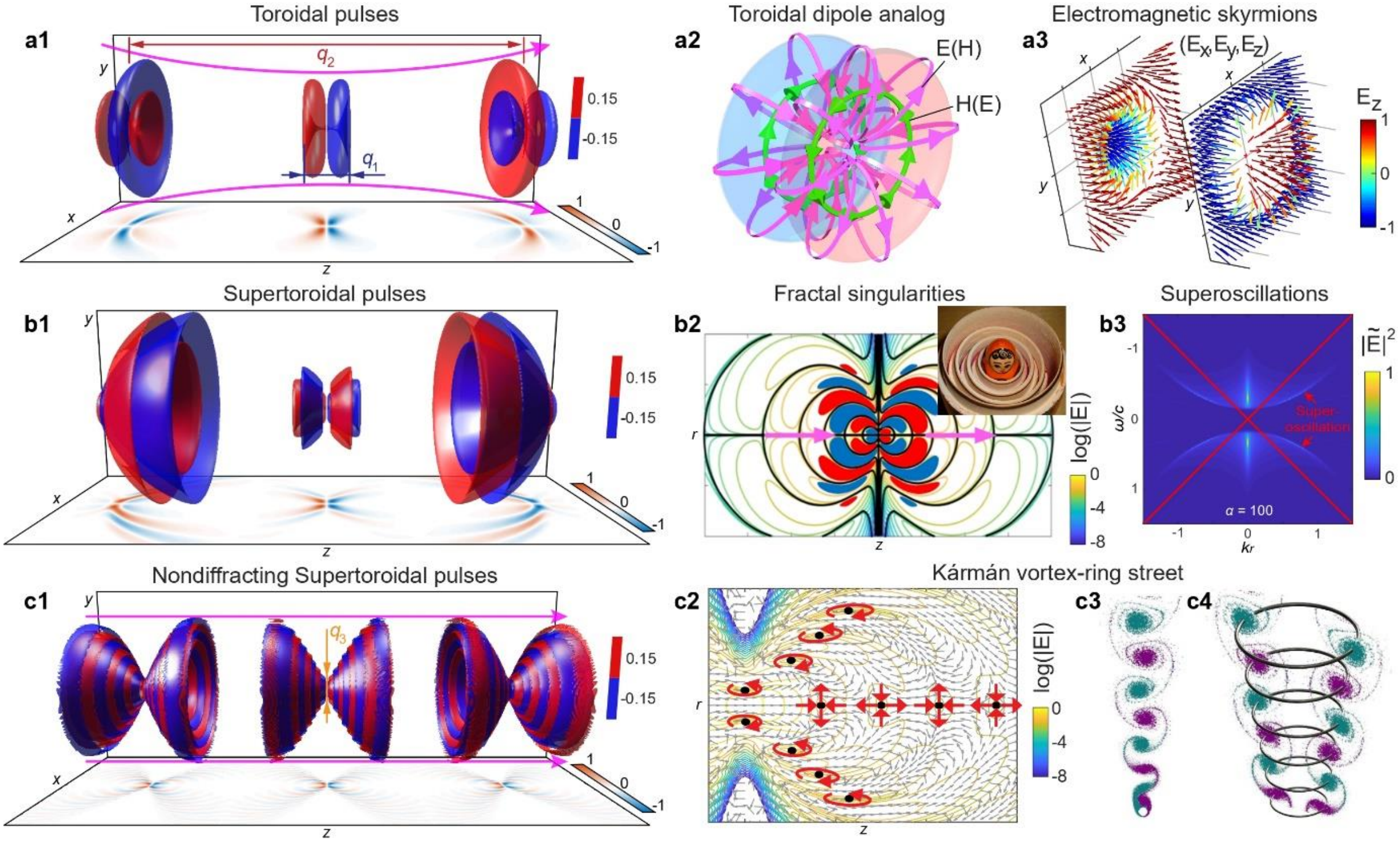

**Figure 1.** **a1**, The spatiotemporal evolution of the electric field structure of an azimuthally polarized toroidal light pulse. **a2**, The electric (magenta) and magnetic (green) field structure of a radially polarized toroidal pulse exhibiting an analogy with toroidal dipole excitations in matter, and, **a3**, skyrmionic field configurations at two transverse cross-sections of the TLP electric field. **b1**, The spatiotemporal evolution of electric field structure of an azimuthally polarized STP. **b2**, The electromagnetic field structure of an azimuthally polarized STP exhibiting a matryoshka-like singularity (black bold lines) structure, the purple arrows show the propagation direction and the inset shows a picture of matryoshka toy, **b3**, The plane wave spectrum of a STP reveals space-time superoscillations, as indicated by components outside the light cone (red lines). **c1**, The spatiotemporal evolution of electric field structure of an azimuthally polarized ND-STP. **c2-c4**, The electromagnetic field singularity (black bold dots with red arrows marking the type of vortex or saddle) structure of a ND-STP **(c2,c4)** exhibiting an analogy of Kármán vortex streets in fluid dynamics (**c3**) but in a 3D cylindrically symmetric vortex-ring street configuration.

## Advances in Science and Technology to Meet Challenges

Key advances required to meet challenges involve the availability of ultrabroadband sources that can generate few- to single-cycle pulse. To date, such sources are routinely available only at lower frequencies, e.g. THz and microwave papers. Moreover, the space-time nonseparability of TLPs and STPs raises intriguing questions regarding the presence of space-time nonseparable excitations in matter and the relation of the latter to free-space propagating STPs. Toroidal pulses can also be employed as information and energy carriers, where information can be encoded in the complex, robust upon propagation, topological structure, while their non-diffracting nature allows energy transfer across arbitrary distances. Further, to date toroidal pulses have only been studied in the classical regime, while the quantum optics of TLPs remain completely unexplored. Finally, a systematic classification and exploration of the family of toroidal pulses is still under way and is expected to lead to new discoveries of spatiotemporal doughnuts pulses.

**Concluding Remarks**

To summarize, supertoroidal pulses exhibit vector singularities of electromagnetic fields with intriguing topological properties, such as skyrmion-like magnetic field configurations and fractal-like distribution of singularities. We note that the topological structure of the STPs is not limited to the focal region and persists upon pulse propagation. Such topological features will be of interest for fundamental studies in topological optics, as well as for applications in encoding and transferring information by topologically structured fields.

**Acknowledgements**

This work was supported by the UK Engineering and Physical Sciences Research Council (grant no. EP/M009122/1), MOE Singapore (grant no. MOE2016-T3- 1-006), the European Research Council (Advanced grant FLEET-786851, funder ID https://doi.org/10.13039/501100000781) and the Defence Advanced Research Projects Agency (DARPA) under the Nascent Light Matter Interactions programme.

## 12. Arbitrary Space-time Wave Packet Synthesis

*Lu Chen[1,2], Wenqi Zhu[1,2], Amit Agrawal[1]*
[1]National Institute of Standards and Technology, Gaithersburg, MD 20899 USA
[2]University of Maryland, College Park, MD, 20742 USA

### Status

Developments in sculpting the spatial phase and amplitude distribution of light has led to numerous advances in areas ranging from imaging and lithography to classical and quantum communication. These advances are largely enabled by development of wavefront shaping tools such as liquid-crystal-based spatial light modulators (LC-SLMs), digital micromirror devices (DMDs) and microfabricated phase-plates [1]. Recently, metasurfaces composed of deep-subwavelength nanostructures have demonstrated exceptional versatility in moulding the spatial wavefront of light [2, 3]. The unique advantages offered by metasurfaces including multifunctional control of light at the nanoscale, ultrathin planar form-factor, ready design flexibility, and complementary metal-oxide semiconductor (CMOS) fabrication compatibility, make them an attractive platform for applications requiring portability, robustness, and scalability.

Similarly, controlling the temporal waveform of light on an ultrafast time scale, commonly referred to as pulse shaping, has had tremendous impact in the field of ultrafast optics. For example, manipulating the spectral phase and amplitude of the pulse has enabled applications such as chirped pulse amplification and optical arbitrary waveform generation, whereas controlling the polarization degree of freedom has provided access to ultrafast chiral light-matter interactions. Temporal shaping has traditionally been achieved either through manipulation of the pulse directly in the time domain, using, for example, chirped mirrors or acousto-optic modulators (AOMs), or alternatively, in the frequency domain by employing Fourier synthesis combined with LC-SLMs, DMDs or AOMs [4]. Recently, finely tailored pulse shaping operations were realized using a dielectric-metasurface based pulse shaper, where simultaneous and independent control of the phase and amplitude of the constituent frequency components of a near-infrared femtosecond pulse over an ultrawide bandwidth was achieved [5].

Space-time wave packets where the phase, amplitude, and polarization of light in both space and time can be simultaneously and independently engineered have the potential to unveil new physics and enable new capabilities in ultrafast light-matter interactions. The past few years have witnessed numerous advances in spatiotemporally structuring optical fields in free-space or in waveguides by treating its spatial and temporal degrees of freedom either as separable or correlated variables. Complex space-time wave packets exhibiting diffraction-free propagation, tunable group velocity, transverse orbital angular momentum (OAM), self-torque, toroidal topology, flying focus, and predesigned three-dimensional trajectory, among others, have been demonstrated [6]. Having the ability to arbitrarily control all the fundamental degrees of freedom of an optical wave in space and in time would enable synthesis of novel classes of optical wave packets opening intriguing possibilities in both basic research and potential real-world applications.

### Current and Future Challenges

Despite rapid progress in synthesizing space-time wave packets, arbitrary control over all the spatiotemporal properties of the light field over a large bandwidth has proven challenging. This requirement generally translates to simultaneous and independent manipulation of a large group of coherent frequencies with ultrahigh spectral and spatial resolution. Generally, a Fourier transform pulse shaper, consisting of dispersive (e.g., lens) and diffractive (e.g., grating) optics, is used to spatially separate the spectral components of an ultrafast optical pulse train at its Fourier plane, and can be engineered to provide a high spectral resolution. However, typically the comb line spacing, determined by the repetition rate of the laser, is on the order of tens of MHz to a few GHz, which is well below the resolvance of gratings. Moreover, high temporal resolution necessitates a large spectral bandwidth, demanding enormous number of units (composed of one or an array of pixels) on the modulation device, placed at the Fourier plane, to achieve parallel and independent manipulation of each spatially dispersed frequency component. At the same time, each pixel is ideally required to provide multifunctional control by being able to simultaneously manipulate the phase, amplitude, and polarization degrees of freedom of the corresponding frequency line. To ensure high modulation efficiency, these operations are expected to be implemented by pixels of subwavelength size and spacing. All of these are nontrivial requirements. Considering practical applications, miniaturization of space-time wave packet synthesizer is also highly desirable. Conventional gratings, lenses, and modulation devices such as LC-SLMs and DMDs are bulky and post a limit on the footprint of the synthesizer. Also, such wave packet synthesizers, depending on the choice of the modulation device, are also often limited to operate in the visible to near- and mid- infrared parts of the electromagnetic spectrum, leaving a gap for technologically important ultraviolet and terahertz wavelengths. Finally, high brightness light sources are cropping up worldwide, with ever-increasing interest in space-time wave packet synthesizers that can operate at high input peak powers, a capability not supported by current modulation technologies due to their low damage threshold.

**Advances in Science and Technology to Meet Challenges**

Significant progress in design, synthesis and characterization of space-time wave packets exhibiting unique characteristics has been made over the last decade. However, each approach only addresses a subset of these challenges primarily because of limitations in the size, number and functionality at the pixel level of traditional approaches based on LC-SLMs or DMDs. By leveraging the multifunctional response at the nanoscale offered by a dielectric metasurfaces, a versatile approach for controlling the four-dimensional properties (temporal phase, amplitude, polarization, and spatial wavefront) of near-infrared femtosecond pulses at will over an ultrawide bandwidth has recently been demonstrated [7]. Here, a single-layer metasurface device, consists of hundreds of modulation units (termed superpixels), each containing a two-dimensional (2D) array of deep-subwavelength birefringent Si nanopillars as modulation pixels. The device is designed to provide simultaneous and independent control over the temporal state of light (i.e., phase, amplitude, and polarization) via engineering the metasurface at the superpixel level, and the spatial wavefront evolution via engineering individual nanopillars within each superpixel (figure 1). Complex space-time wave packets including a single femtosecond pulse carrying both temporal spin and orbital angular momentum, a light-coil exhibiting a helical intensity distribution evolution, and a pulse with coherently multiplexed time-varying OAM orders, are experimentally synthesized. These space-time wave packets have up until either been theoretically proposed or require high-harmonic nonlinear processes. The metasurface-based approach can be extended towards synthesizing other forms of complex space-time wave packets via metasurface design engineering, or other wavelength regimes by simply choosing a compatible dielectric material with an appropriate bandgap. Furthermore, the dispersive and diffractive optics in the synthesizer built using the metasurface approach can enable realization of a compact space-time wave packet synthesizer, and the use of dielectric materials lends the platform a high damage threshold.

To expand the number of frequencies that can be simultaneously addressed, the spectral components of the pulse can be dispersed into a 2D array using advanced cross-dispersion arrangement based on the combination of a virtually imaged phased array (VIPA) and a grating [8]. This combined with the flexibility in metasurface design offers the potential for ultimate realization of line-by-line spatiotemporal shaping. Continued developments in the field of metasurfaces, such as by incorporating multiple nanopillars within one pixel or the combined use of propagation phase and geometric phase [3], further provide more advanced multifunctional control over the fundamental properties of light. Moreover, the rapid progress in metasurface design strategies, for example, leveraging inverse design and machine learning, is also expected to provide novel capabilities not possible with traditional nanopillar library-based approaches [9]. Although a static metasurface response is sufficient for several applications, reconfigurability such as those offered by LC-SLMs is still favourable. While much of the development in the area of active metasurfaces using various transduction mechanisms such as electro-optical, nano-electro-mechanical, nonlinear or via phase change materials, are at an early-stage, we expect them to in time reach a level of maturity required to provide a full programmable control over the light properties [10]. Benefiting from the continued interest from both academia and industry, efforts in developing low-cost, large-scale nanofabrication of metasurfaces also show promise for scaling and commercial viability of metasurface-based space-time wave packet synthesizers.

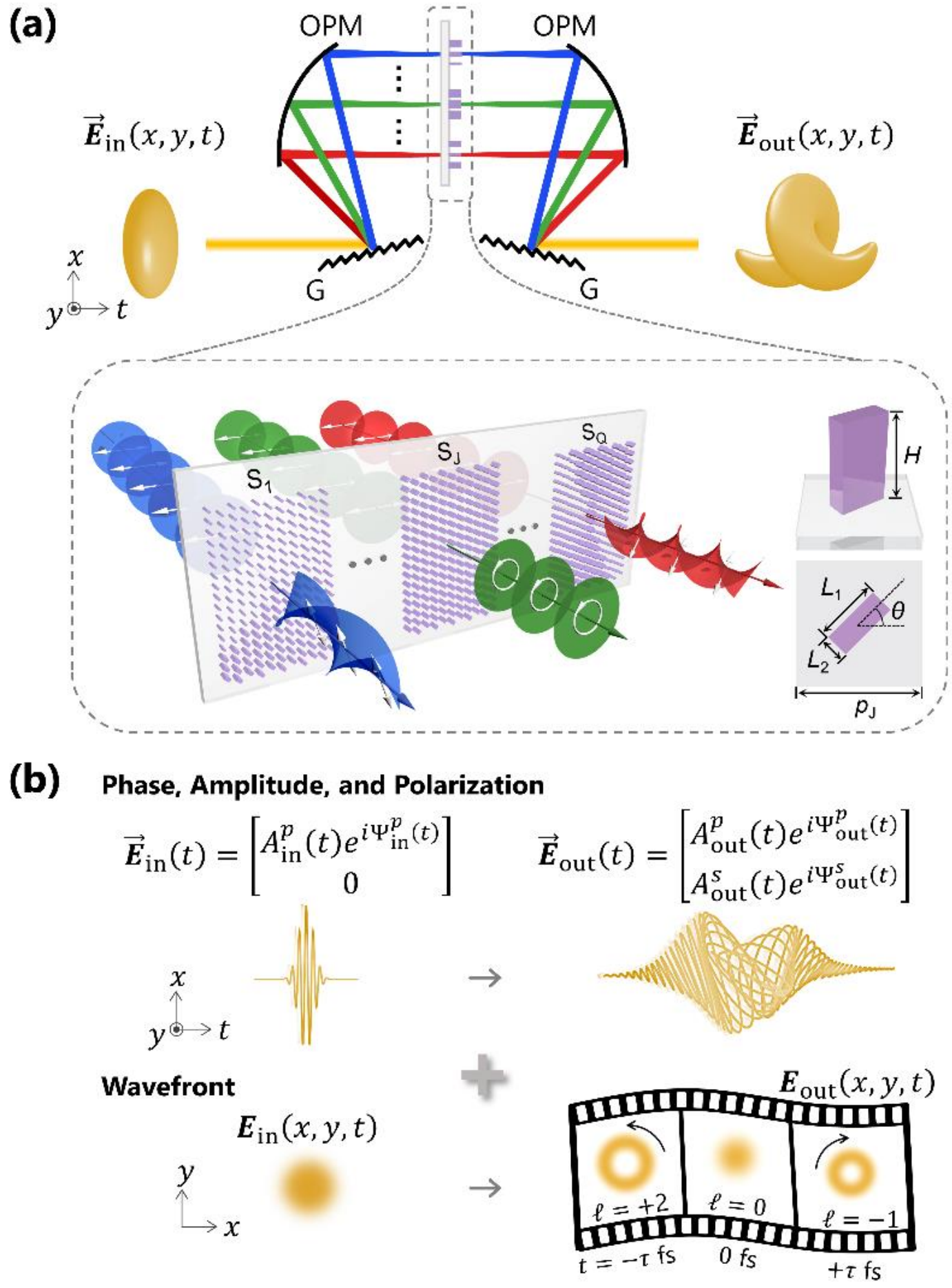

Figure 1. Arbitrary space-time wave packet synthesizer enabled by a dielectric metasurface. (a) Schematic of the Fourier-transform pulse synthesizer. (b) Simultaneous and independent control of the phase, amplitude, polarization, and wavefront as a function of time enabled by the synthesizer.

**Concluding Remarks**

Arbitrary space-time wave packet synthesis is an emerging research field offering the promise of enabling new applications via customized control over all the fundamental properties of light field. We have outlined some outstanding challenges and offered perspective on the requirements for a universal approach towards synthesis of arbitrary space-time wave packets. These capabilities are expected to be further accelerated by continued developments in new theoretical models, metasurface design concepts, nanofabrication technologies, and light-field characterization. In the short term, we expect the versatile optical control offered by metasurfaces to be applied to other spatiotemporal light field concepts and applications discussed in this roadmap. In the long term, development of a large area, lithography friendly, fully reconfigurable modulator offering multidimensional control at the nanoscale is expected to have lasting impacts beyond the field of ultrafast optics.

**Acknowledgements**

L.C. and W.Z. acknowledge support under the Cooperative Research Agreement between the University of Maryland and NIST-PML, Award no. 70NANB10H193.

## 13. Measuring ultrashort spatiotemporal structures

Spencer W. Jolly, Université Libre de Bruxelles, Belgium
Christophe Dorrer, Laboratory for Laser Energetics, University of Rochester, USA

### Status

A crucial aspect of spatiotemporal shaping of ultrashort pulses and the associated probing of laser-matter interactions, complementary to all of the previous sections, is being able to characterise the non-separable space-time aberrations. This field has been growing steadily in the past years, with a number of techniques demonstrated, each with their own key advantages and disadvantages [1,2].

Spatiotemporal measurements are important for at least two reasons. The first is to address the spatiotemporal properties of large and expensive high-power laser systems to maximise the intensity produced on target after tight focusing. Spatiotemporal measurements allow laser scientists to assess how the pulse's properties, e.g., peak intensity, differ from ideal properties calculated using the energy, pulse duration, and focal-spot size. They also allow one to track down optical components that have chromatic aberrations or imperfections, or that are misaligned. The second purpose of spatiotemporal characterization is to confirm and address the results of spatiotemporal shaping, both to understand the shaping process and to understand how exotic spatiotemporal pulses - like those described in the previous sections - propagate and interact with physical systems.

While temporally characterising the pulse shape of an isolated ultrashort pulse requires a time-non-stationary element, often synthesised via a nonlinear interaction [3], most spatiotemporal diagnostics use linear optics to characterise the variations in temporal/spectral field properties across the transverse dimensions, $x$ and $y$, by comparison to a reference field from another source or generated from the test field itself [1,2]. In these approaches, the spatially dependent spectral phase $\varphi(x,y,\omega)$ is typically reconstructed up to an arbitrary spectral phase that corresponds to the phase of the reference pulse or the phase of the pulse under test at the reference point. For example, in STRIPED FISH (Spatially and Temporally Resolved Intensity and Phase Evaluation Device: Full Information from a Single Hologram),[4] spectrally filtered replicas of the pulse under test and a reference pulse interfere on a camera, allowing for the reconstruction of their spatial phase difference at the filtered frequencies [Fig. 1(a)]. In TERMITES (Total E-field Reconstruction using a Michelson Interferometer's TEmporal Scan) [5], the test pulse interferes with a reference pulse generated from a small portion of the test pulse itself after reflection off a spherical mirror; the spatial interference is recorded as a function of the delay between the two pulses, from which a frequency-dependent interference is generated via Fourier transformation [Fig. 1(b)]. A multispectral camera intrinsically provides spectral resolution thanks to an array of dielectric filters located before the photodetection chip, and it can provide spectrally resolved wavefront measurements when combined with conventional wavefront-sensing techniques [Fig. 1(c)], [6]. Similar data can be obtained using a conventional camera with preliminary filtering and demultiplexing of the test pulse [7]. Conversely, spectral interference of the field at any point in the transverse plane, sampled by an optical fiber, with a reference field also yields a spatially and spectrally resolved map of the electric field [Fig. 1(d)], [8]. Implementing spectral resolution, particularly on a single-shot basis to characterise low-repetition-rate laser systems, and generating an adequate reference pulse remain practical challenges.

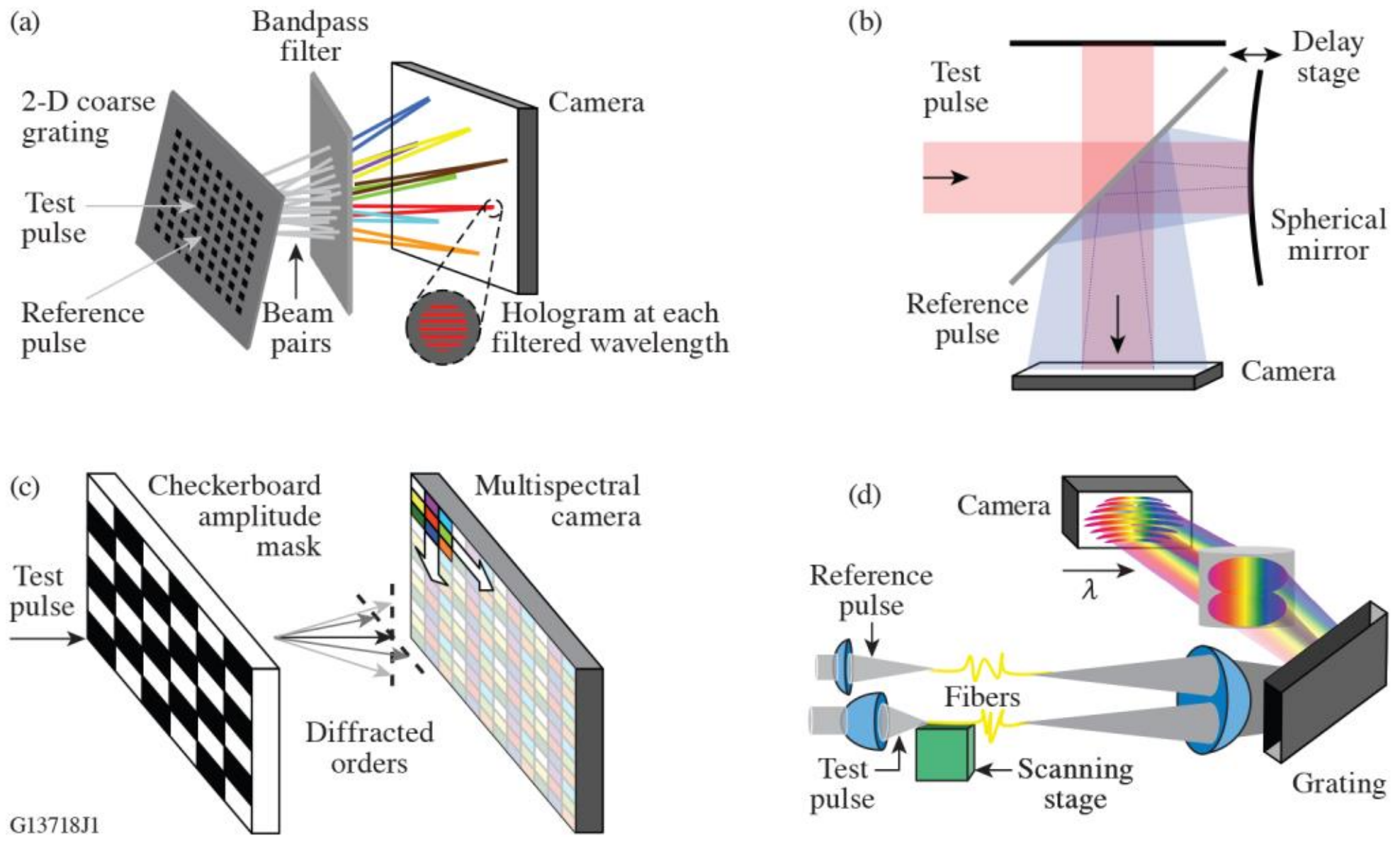

**Figure 1.** Concept of four different techniques for spatio-temporal coupling characterization. (a) STRIPED FISH, based on the spatial interference of the test pulse with a reference pulse after spectral filtering. (b) TERMITES, based on the spatial interference of the test pulse with a reference pulse generated by a spherical mirror, measured using Fourier-transform spectroscopy. (c) Spectrally resolved wavefront measurements obtained by combining a wavefront-sensing technique (in this case, shearing interferometry implemented with a checkerboard amplitude mask) with a multi-spectral camera. (d) Spectral interferometry of the test pulse and a reference pulse sampled by a single-mode fiber. (a) and (c) are adapted with permission from [4] and [8], respectively, © The Optical Society.

## Current and Future Challenges

Spatiotemporal or spatiospectral intensity and phase data are in 3 dimensions, but one is usually limited to a 2-dimensional detector. Full spatio-temporal characterization generally requires taking multiple 2-D measurements while scanning one parameter, which can result in increased noise due to fluctuations in the source, or at the very least results in longer and less convenient measuring times. Multiplexing images or interferograms on a 2-D detector can drastically decrease the required number of acquisitions, but generally results in a similarly drastic reduction in the resolution either in space, frequency/time, or both. This is true for example in [Fig. 1(a)] and [Fig. 1(c)] where both the number of frequencies and spatial resolution is decreased for single shot operation. Operation with a small number of frequencies is however sufficient to detect common spatio-temporal couplings such as radial group delay and pulse front tilt [Fig. 2(a)]. Developing single-shot techniques is especially important for extremely high-power lasers that operate at rep-rates below 1 pulse per second, or for trains of pulses that experience significant fluctuations from shot to shot. In some cases, the generation of an adequate reference pulse is a practical challenge.

Extending spatiotemporal characterisation beyond the somewhat standard near-infrared pulses corresponding to Ti:Sa laser systems is another significant challenge. Measuring pulses in the UV, XUV (e.g., from high-harmonic generation), IR, and THz ranges requires to adapt the existing concepts with adequate optics and detectors. This is true as well for pulses that have an extremely broad spectrum, for example after broadening in a hollow-core fibre or multi-pass cell system, with required operation over a large wavelength range. Additionally, when ultrashort pulses are tightly focused, beyond the paraxial regime, they put additional constraints on spatial resolution and present nuanced challenges related to spatiotemporal propagation, and may introduce vector fields around the focus. Lastly, determining the spatiotemporal properties of ultrashort vector pulses in general, with spatially or temporally-varying polarisation, is an even greater challenge that is treated in the following section of this roadmap.

**Advances in Science and Technology to Meet Challenges**

Beyond the already discussed techniques based on measuring spatial data at different frequencies on a 2-D detector, developing single-shot or few-shot techniques that make no compromises on spectral or spatial resolution is an ongoing technological challenge. It can be addressed by improved hardware and software (algorithms to reconstruct the spatio-temporal field from the measured experimental trace) to fulfil the need for better-characterised laser systems and highly shaped pulses. Technological improvements in the manufacturing of optical components (e.g., better filters) and sensors (e.g., higher pixel count, smaller pixel size) and their commercial availability will allow for improving the current generation of spatiotemporal diagnostics.

Reducing the number of acquisitions in a multi-shot device can also be achieved by compressed sensing, where choosing random sampling positions can produce similarly resolved spectral images with fewer total measurements. This has been applied to hyperspectral imaging that utilised Fourier-transform interferometry [9], where the number of shots was reduced to considerably sub-Nyquist acquisition rates (down to 15% of the Nyquist limit) by combining bandpass and non-uniform compressive sampling.

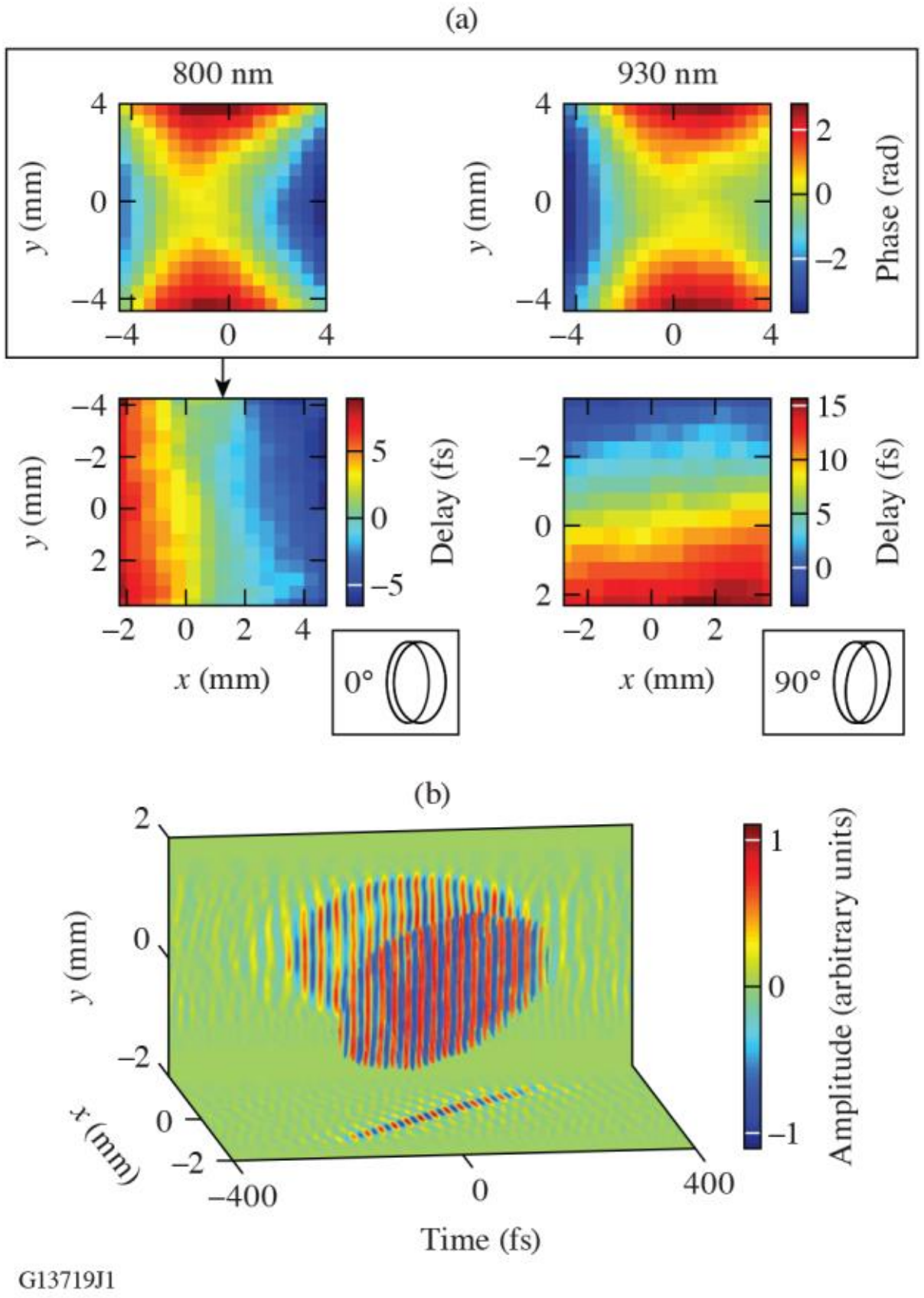

**Figure 2.** Comparison of results from two different single-shot characterization techniques. In (a) the combination of a multispectral camera and a wavefront sensor can reconstruct the wavefront at two wavelengths (inset) from which the group delay can be calculated at each spatial position [6]. When rotating a wedge (left to right) this method correctly shows the rotating pulse-front tilt. In (b) Compressed Optical Field Topography (COFT, [10]) reconstructs an entire spatiotemporal amplitude and phase distribution in a single shot. This allows one to characterize arbitrary couplings in addition to simple ones like pulse-front tilt as shown. (a) is adapted with permission from [6] © The Optical Society. (b) is adapted with permission from [10] © Springer Nature.

Snapshot hyperspectral imaging is a technology in common use for industrial applications (agriculture, cultural heritage, industrial sorting) where frequency-resolved images are produced in a single shot. Inspiration can be drawn from this applied field on how to apply an existing technique to ultrashort pulses, and how to additionally measure the spatiospectral phase. One example of a single-shot method is coded aperture snapshot spectral imaging (CASSI), where a specifically designed (complicated) aperture, a dispersive element, and an advanced algorithm can produce the spatially-resolved spectrum on a 2-D sensor in a single-shot. This has just recently been applied to spatiotemporal pulse measurements [10], producing high fidelity results [Fig. 2(b)].

**Concluding Remarks**

Ultrafast metrology is evolving to answer the experimental need for better-characterised pulses interacting with physical systems. A large range of spatio-temporal diagnostics are now available, with fast progress toward extended operating ranges and more practical implementations. The current and future generation of spatiotemporal diagnostics is paramount to generating high-intensity tightly focused optical pulses but also driving physical phenomena with pulses producing novel interaction conditions via spatio-temporal coupling.

**Acknowledgements**

S.W.J. has received funding from the European Union’s Horizon 2020 research and innovation program under the Marie Skłodowska-Curie grant agreement No 801505. The work of CD is supported by the Department of Energy National Nuclear Security Administration under Award Number DE-NA0003856, the University of Rochester, and the New York State Energy Research and Development Authority.

## 14. Characterization of complex vector pulses

Benjamín Alonso, Ignacio Lopez-Quintas, Miguel López-Ripa, and Íñigo J. Sola.
Universidad de Salamanca, Spain.

### Status

The raising importance of ultrashort light pulse applications increases the need of its characterisation. Since the invention of the laser and the provision of ultrashort pulses in the femtosecond time scale, different challenges in their characterization have been gradually fulfilled, e.g., measuring temporal durations, amplitude and phase of linearly polarized pulses or spatiotemporal couplings. The advances in more and more complex pulse structures, currently presenting simultaneous spatial and temporal polarization shaping, here called ultrafast vector beams (see an example in Fig. 1), have substantially raised the measuring demands. Firstly, when the spatial dependence can be disregarded, in the so-called vector pulses the polarization state is time —and, therefore, frequency— dependent. Their measurement lies in obtaining the amplitude and phase of both polarization components, paying special attention to the accuracy in their relative phase to obtain the full information. Some successful approaches to measure time-dependent polarization pulses are based on dual-channel spectral interferometry [1], where two orthogonal polarization projections are referred to the same reference pulse, and tomographic retrieval [2], where three polarization projections are measured with any temporal diagnostics and two orthogonal polarization projections are connected thanks to the temporal measurement of a third intermediate projection. Other strategies are based on time-resolved ellipsometry or angle-multiplexed spatial-spectral interferometry. More recently, in-line polarization interferometry has been shown to accurately retrieve vector pulses using a birefringent crystal as a bulk common-path interferometer [3]. On the other hand, when the polarization state is constant, i.e., time-independent, different methods can be used to obtain the spatial dependent polarization, e.g., Stokes polarimetry or interferometric measurements based on positive operator value tomographic measurements [4].

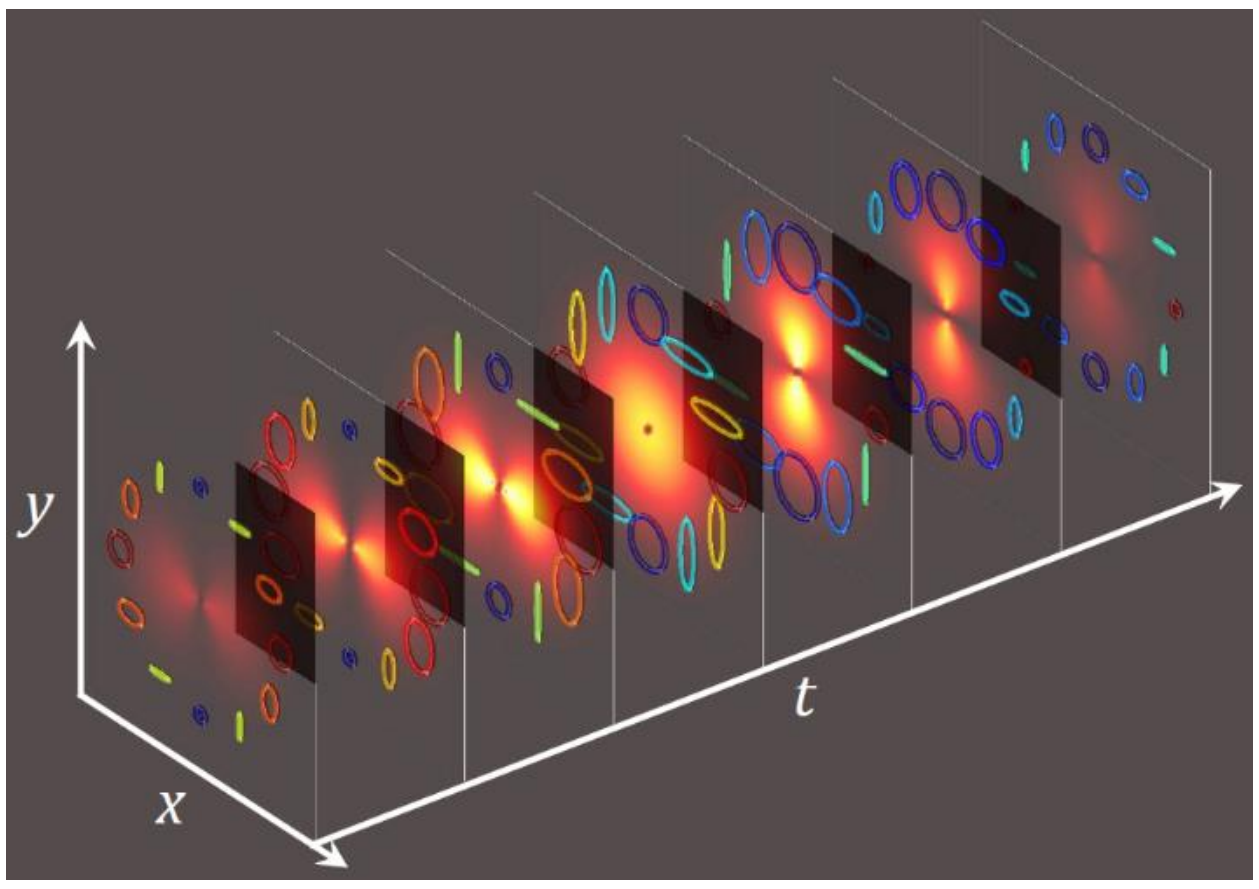

Figure 1. Illustration of a vector pulse with spatial (*x*,*y*) and temporal (*t*) dependent polarization. The example corresponds to spatially varying polarization gates [5]. The colour of the polarization ellipses shows the ellipticity.

As expressed before, some applications make use of ultrafast vector beams with both space- and time-dependent polarization. Their characterization has recently been accomplished by using two-fold spectral interferometry [5], which is based on the combination of spatially resolved spectral interferometry from a fibre coupler interferometer and in-line polarization interferometry (Fig. 2). More recently, a combination of Fourier transform spectroscopy with a Mach-Zehnder interferometer [6] and two-dimensional temporal scanning interferometry [7] have been used for the same purpose.

Since these techniques are very recent, improvements and new techniques are expected to be introduced in the near future to overcome the different challenges in the measurement of ultrafast vector beams.

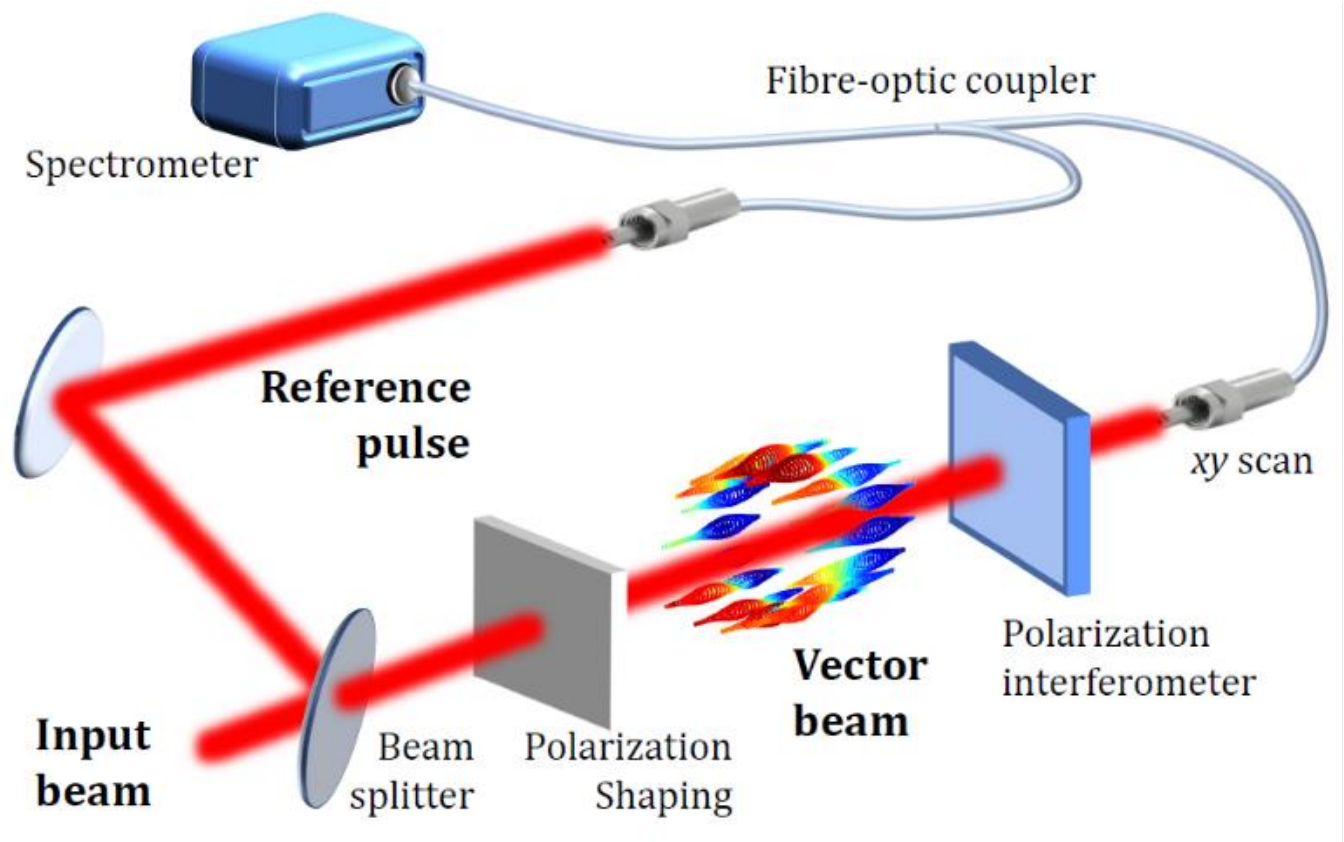

Figure 2. Simplified scheme of the characterization of a vector beam by two-fold interferometry [5]: it is obtained the x-component of the field from spectral interferometry (fibre-coupler) and the y-component from polarization interferometry (birefringent plate and linear polarizer), being both cases spatially resolved.

## Current and Future Challenges

Although the above-mentioned progresses allow to develop the spatiotemporal and polarization characterization of pulsed beams, further challenges and needs should be addressed because of its increasing rate of application.

Apart from the complexity of the space-time polarization structures —which can be reconstructed with the above commented state-of-the-art methods—, a matter to face is the that the amplitude and phase of two polarization components of the electric field must be retrieved as a function of three variables: two spatial variables (in the transverse plane to the propagation) and the time (or, equivalently, the frequency). Related to this fact, there is the need to encode all the information in the detection (e.g., through interferometry or spectroscopy) in the different techniques, which is at the end related to having scanning measurements. A second relevant challenge in the future is to reach single-shot solutions that can speed up the detection. This upgrade will be beneficial, especially for the measurement of low repetition rate sources (e.g., high-power systems), as well as for unstable vector pulse trains [8].

Expected upgrades will surely be related to the spectral and temporal range of application of the current techniques. In this sense, the extension to mid- and far-infrared (MIR/FIR) spectral bands (including THz pulses), as well as to extreme ultraviolet (XUV) or X-ray pulses in the attosecond scale, is highly desirable since nowadays there are many progresses in complex vector beams generated through high harmonics [9]. Their measurement is partially tackled in space and frequency, but it is not temporally resolved simultaneously, thus other detection methods and strategies will be necessary for their full characterization. In the case of single- or few-cycle pulses with carrier-envelope phase (CEP) stabilized, the techniques will have to attain the characterization of this CEP, which is key for electric-field dependent interactions.

Most current techniques only work with completely polarized light sources, so new studies on the measurement of partially polarized pulses must be done, e.g., through spatiotemporal resolved Stokes polarimetry [10].

Also, state-of-the-art methods are usually limited to the transverse electric field, while the longitudinal component is neglected. This challenge must be addressed e.g., for tightly focused beams, being its characterization rather complex, as seen in 1D (temporal) previous diagnostics [11].

Finally, in integrated photonics, miniaturization and simplification of the detection is a challenge for photonic circuits, lab-on-a-chip experiments, or robust industrial detectors, among others.

**Advances in Science and Technology to Meet Challenges**

Naturally, in addition to introducing new strategies, meeting the challenges outlined above will require new developments. Firstly, new schemes and processes to encode the pulse information are necessary (e.g., interferometry, diffraction, nonlinear processes), for example for single-shot detection or partially polarized beams. Secondly, new technology will be needed to accomplish those challenges.

Due to the great interest in measuring the spatial and temporal dependence at once, basic 1D (e.g., a spectrometer) or 2D (e.g., a simple CDD) detectors imply scanning techniques, so sensors resolving simultaneously different dimensions (e.g., hyperspectral detectors, spatially resolved streak cameras, polarization resolving cameras, etc.) will play a key role in the development of new techniques, in particular for single-shot schemes.

In an analogous way, the extension to new spectral regions will demand the use of unconventional materials or processes for the pulse manipulation: adapting the way of creating pulse replicas and their relative delay, required nonlinear processes, polarization management, polarimetry analysis, etc. Thus, transparent materials with broadband operation are needed in general, to include new spectral ranges of interest or to retrieve pulsed vector beam at even shorter time scales. In particular, the detection will also play a significant role to work in the MIR, FIR or the XUV. Some technology is nowadays available and is used for beam characterization purposes, but the characterization of, e.g., polarized XUV pulses is partially done, separately in time and in space coordinates. Therefore, upgrades are necessary whether to merge existing techniques addressing partial characterisation dimensions (e.g., polarization, spatiotemporal, time) or developing new ones including all of them, while keeping the needed resolution.

Regarding the integration of the characterization in optical devices, scientific and, even more, technological developments are required to open the way. Detections using fibres or optical waveguides may be useful for this task. Also, liquid crystals or nonlinear crystals with refractive index tuned by external signals are candidates to ease the integration of the detection. In any case, the fabrication of this miniaturized devices (e.g., through laser processing, lithography...) constitutes a challenge itself due to the resolution required and the high amount of information to be recorded in a measurement. Recent advances in XUV pulses assisted optoelectronics are promising in this field.

Finally, it is worth mentioning not only the amount of data recorded, but also its processing, which is done by devoted computational algorithms. Together with fast data acquisition (ideally, single shot), fast algorithms are desirable to monitor the beam information in real time (preferably, at video rates). Undoubtedly, artificial intelligence will provide foremost breakthroughs on this point.

**Concluding Remarks**

Very recently some techniques have successfully faced the challenge of measuring ultrafast vector beams [5]–[7] —i.e., ultrashort pulses with space and time dependent polarization— using different strategies, in all cases operating in the near infrared region. It is expected that other new techniques will be developed in the next years, improving the performance of the current state of the art. Some of the main challenges that have been identified are related to upgrade the capabilities of present techniques (e.g., the resolution or the spectral range of operation) or to increase the information of the pulses that can be retrieved (for example, cases where the CEP, the longitudinal electric field or pulse train instabilities, among others, are relevant). Also, faster acquisition times (preferably, single-shot detection at high repetition rates) and data processing are targeted in forthcoming developments. Technological challenges have also been highlighted, being truly relevant future advances in multidimensional detection (such as hyperspectral imaging) or micro manufacturing for integrated characterization devices.

**Acknowledgements**

*The authors acknowledge funding from Junta de Castilla y León and European Regional Development Fund (SA136-P20); Ministerio de Economía y Competitividad (EQC2018-004117-P); Ministerio de Ciencia, Innovación y Universidades (PID2020-119818GB-I00); H2020 European Research Council (851201). M. López-Ripa thanks the University of Salamanca for the Ph.D. contract.*

## 15. Strong-field physics with optical vortices

Yiqi Fang[1], Qihuang Gong[1,2] and Yunquan Liu[1,2]
[1]State Key Laboratory for Mesoscopic Physics and Collaborative Innovation Center of Quantum Matter, School of Physics, Peking University, Beijing 100871, China
[2]Beijing Academy of Quantum Information Sciences, Beijing 100193, China

### Status

Light-matter interaction is one of the most important approaches to understand the inherent laws of this world. Exposing of atoms in intense laser fields, high harmonic generation (HHG) [1] and strong-field ionization (SFI) [2] are the typical processes in strong-field community, accompanied by the emission of high-energy photons, photoelectrons and ions. They can be conventionally described by a semi-classical picture: with the effect of external strong laser field, the electrons are tunnelled through the bent Coulomb barrier into the classical region, and could be accelerated back to be recombined with the parent ions with the emission of high-energy photons, or be elastically (inelastically) scattered with the parent ions. On one hand, the products of SFI and HHG provide a robust, high-precision and flexible toolbox for monitoring the ultrafast non-equilibrium behaviours of matters in the process of absorbing photons. On the other hand, it also provides coherent table-top ultrafast extreme ultraviolet (EUV) light sources which have been be further utilized in chemical, biological and medicinal studies.

In strong-field physics, the fundamental Gaussian beam is widely employed both experimentally and theoretically, since it is the routine spatial mode from the lasers. The non-trivial spatial and temporal phase structures have been less studied in SFI and HHG, and thus the plane wave approximation is usually adopted in analysing SFI and HHG with such laser fields. The structured lights, such as the optical vortex beams and vector beams, allow to introduce new degrees of freedom for the interaction between the intense light fields and matters. To date, such spatial degree has provided new opportunities for strong-field physics with auspicious applications in attosecond science.

At present, the generation and control over optical vortices in visible and near infrared domains can be easily realized by using spatial optical components, but it is still challenge to modulate light fields in short wavelengths (i.e., UV, EUV) directly. Fortunately, the advances of HHG driven by the intense optical vortices have opened the door for circumventing this issue. Backed by the intrinsic conservation rules of photon energy, spin angular momentum (SAM) and orbital angular momentum (OAM) in HHG, the spatiotemporal nature of incident photon can be well conserved and be transferred into the up-converted EUV photons. In this way, one can sculpt the EUV beams with spatial phase, polarization and other dynamic structures. It was more recently discovered that the transverse OAM can be imprinted onto EUV radiations in HHG driven by spatiotemporal optical vortices (STOV) with transverse OAM states [3]. It was revealed that the transverse OAM is conserved in HHG and such unique spatiotemporal signature has significant effects on both the microscopic and macroscopic role of HHG. The spatiotemporal structure of optical vortices has also crucial influences on SFI. It has been theoretically revealed that the intense vortex pulses give rise to new selection rules in photoionization and photoexcitation. Because of the non-dipole interaction between atoms and optical singularities, the excitation of electrons is ruled by unique angular momentum transitions. The optical vortices are therefore able to trigger OAM-dependent dichroic photoelectric effect in SFI [4]. Moreover, one can utilize SFI to monitor the dynamics or topology of optical vortices, such as mapping optical spin-orbit interaction of intense light field [5] and realizing non-destructive measurement for the topological charges [6]. Furthermore, the optical vortices have also been shown the potential applications in laser wakefield accelerator [7].

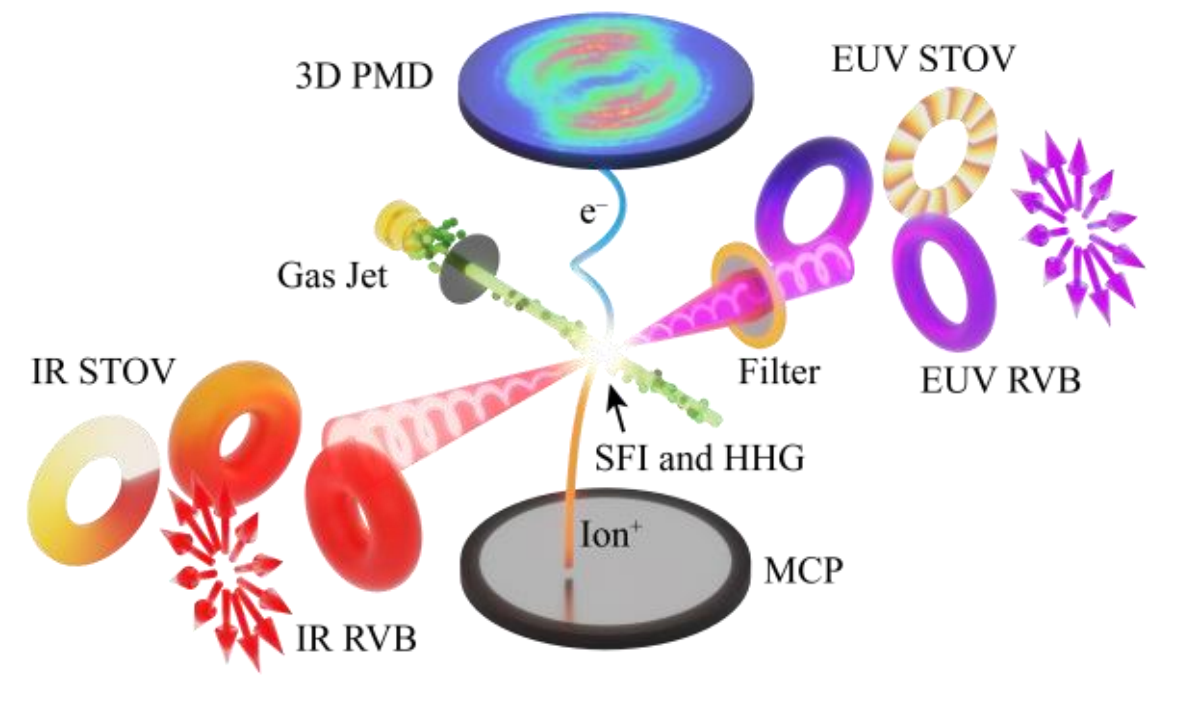

**Figure 1.** Illustration of high harmonic generation (HHG) and strong-field ionization (SFI) experiments with various optical vortices. The products of HHG and SFI are the extreme ultraviolet (EUV) photons and photoelectrons, respectively. IR, infrared; STOV, spatiotemporal optical vortex; RVB, radial vector beam; PMD, photoelectron momentum distribution; MCP, microchannel plate.

**Current and Future Challenges**

An important step toward a closer combination between strong-field physics and optical vortices is to realize the fine manipulation over the sub-light-cycle control of electrons or ions by employing the spatially-varying phase, amplitude or polarization of optical vortices. In principle, there are mainly two field features of optical vortices that is essentially different from plane waves, that is the spiral energy flow and the enhanced longitudinal electric component. Such non-trivial spatial electric field gradient has the promising potential to modify the tunnelling and recollision dynamics of ionized electrons with attosecond temporal resolution and nanometer spatial resolution, accompanied by the generation of EUV radiation or photoelectron momentum distributions (PMDs) that is dependent on the topological charge or topological texture of driving laser fields [8]. It might largely advance the stereo control over the structured electron sources and structured EUV light sources. In the meanwhile, it will open up new avenue for the understanding of complex-light-matter interactions especially in strong-field regime.

The strength of the transverse non-dipole effect or the longitudinal electric field components of optical vortices is determined by the spatial scale of focal light spots. In the non-perturbative regime, the motion of tunnelling electrons scales as tens of nanometers during the interaction with laser pulses, whereas the focal light spot is tens of microns. Due to the significant spatial difference between the electron motion and light field structure, it is hard to directly imprint the spatial structure onto the microscopic behaviour of electrons or ions, and the ejected electrons can only ‘feel’ the local electric field of optical vortices. In addition, compared to the plane waves, the longitudinal electric field component of optical vortices is enhanced very limited in current experimental conditions, so that the longitudinal component is still too trivial to influence ionization or recollision dynamics. Future the non-dipole effects can be utilized to controlling nonlinear dynamics with the structured fields.

**Advances in Science and Technology to Meet Challenges**

In theory, one can resort to various theoretical models to predict the SFI and HHG with optical vortices, such as solving the time-dependent Schrödinger equation and semiclassical models. Currently, the main challenge is how to experimentally reveal the effect of the optical vortices. We here propose three potential experimental solutions:

(i), the most straightforward and promising scenario to circumvent this difficulty is to employ the tightly focused optical vortices [9] in SFI and HHG experiments. By using high numerical aperture (NA $\approx$ 1) optical focusing systems, the spatial scale of focal light spots becomes comparable with the optical wavelength of driving fields. In this case, the transverse non-dipole effect taken by optical singularities might be stronger than the traditional longitudinal electric non-dipole effect. The non-dipole effect can be revealed by measuring the spatial and phase distributions of HHG and PMDs. Another unique feature of the tightly focused optical vortices is that it has very strong longitudinal electric fields, and such electric field component can form the transverse SAM in the focal plane. This will be certain to bring an intriguing aspect of strong-field physics.

(ii), it is also probable to enhance the effect of optical vortices by using an ultra-intense laser field or perturbative laser field in a pump-probe scenario. The electron motion in the ultra-intense laser field can be largely amplified, so that the transverse non-dipole effect can be significantly enhanced. Then one can first use a fundamental Gaussian beam to excite the atoms to a high Rydberg state. As well known, the electron wavepacket lying at high Rydberg states can be easily extended to several microns. Thus, it is possible to probe the excited electron wavepacket with optical vortices [4].

(iii), using STOV, it is an alternative method to imprint optical singularities onto the electron sub-light-cycle motion in SFI and HHG experiments. The helical phase of STOV resides in space-time plane so that the vortex structure has certainly an influence on the dynamics of electrons. At present, there is only a theoretical study about the HHG with STOV [3], and the experimental realization is still lacking. The most important issue of STOV HHG experiments is the generation and characterization of the STOV pulse at focus.

**Concluding Remarks**

In general, we are convinced that the combination between strong-field physics and optical vortices is an innovation-driving field with excellent future prospects. At present, researchers have tried to explore the macroscopic effect of optical vortices in both HHG and SFI. The microscopic effect still remains unexplored. Much efforts are needed from the communities of the ultrafast science and the structured light optics. As such, an exciting future for next-generation applications and fundamental physical mechanisms can be foreseen.

**Acknowledgements**

We thank the support of the National Natural Science Foundation of China (Grant No. 92050201 and 11774013).

## 16. Spatiotemporal differentiators

Junyi Huang[1], Hongliang Zhang[1], and Zhichao Ruan[1,*]

[1]Interdisciplinary Center of Quantum Information, State Key Laboratory of Modern Optical Instrumentation, and Zhejiang Province Key Laboratory of Quantum Technology and Device, Department of Physics, Zhejiang University, Hangzhou 310027, China.

[*] zhichao@zju.edu.cn

### Status

Over the past decades, optical computing has been of great interests due to high speed and low power consumption in comparison with traditional electronic digital counterparts. With the advantage of optical parallel processing, Fourier optics setups with bulky systems of lenses and filters have been widely explored to enable high-throughput real-time image processing. In order to integrate with optoelectronic devices, deliberate efforts have been made to miniaturize such optical computing elements with on-chip photonic circuits and nanophotonic structures. Differentiation is a fundamental mathematical operation in any field of science or engineering. In particular, spatial differentiation enables edge detection, which provides a robust image processing to extract important information about the boundary of objects in an image. In the past several years, there have been substantial efforts on the design of various nanophotonic structures for spatial differential operation [1-5]. For example, based on a simple surface plasmonic structure with a single layer of 50-nm-thick silver, optical spatial differentiation was experimentally demonstrated for image processing of edge detection [2]. In addition, on account of the polarization degree of freedom, the generalized spatial differentiator was proposed with a single planar interface from the spin Hall effect of light, which is an intrinsic optical effect with broad bandwidth [3]. On the other hand, all-optical computation of time-domain differentiations was realized and proposed as all-optical differential equation solvers or pulse reshapers for sensing and control. Here, we propose a differentiator to couple both the spatial and temporal computation in the spatiotemporal domain. We design such a differentiator with single metasurface nanostructures and show that spatiotemporal optical vortices (STOVs) [6-10] can be generated by illuminating a Gaussian-shape optical pulse. Moreover, the spatiotemporal differentiators with the generated STOVs can be used to detect the sharp changes of pulse envelopes, in both spatial and temporal dimensions [9].

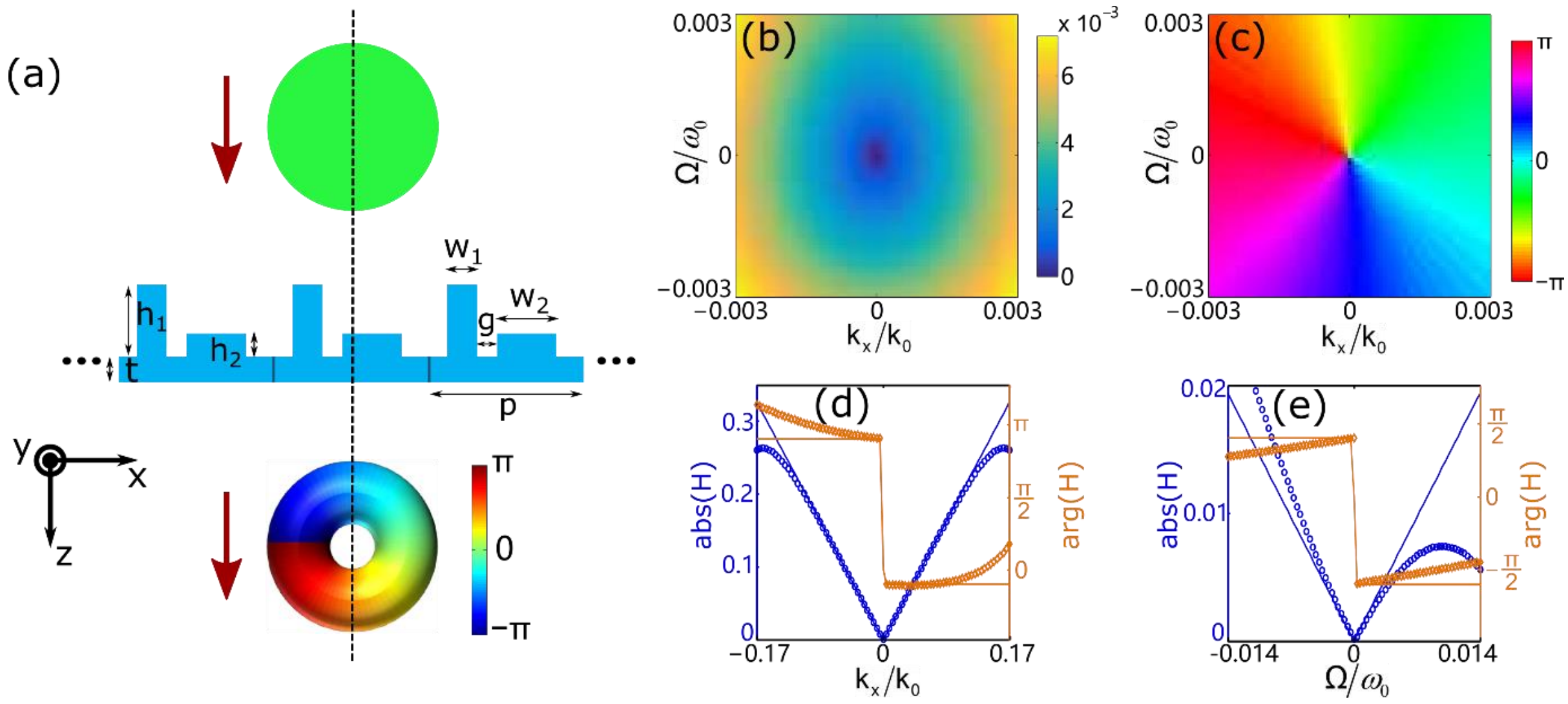

**Figure 1.** Spatial-symmetry analysis of spatiotemporal differentiator to generate STOVs. (a) Breaking mirror symmetry of a spatiotemporal differentiator is necessary for generating the phase singularity, where the transmitted pulse corresponds to a STOV carrying transverse orbital angular momentum. Moreover, the mirror symmetry cannot exist for any normal incident plane. (b) Amplitude and (c) phase distributions of the transmission spectrum function for the designed spatiotemporal differentiator with respect to $k_x$ and $\Omega$. The amplitudes and phases along (d) $\Omega = 0$ and (e) $k_x = 0$. The blue circles and lines correspond to the amplitudes of simulation results and the fitting ones, and the orange rhombi and lines correspond to the phases, respectively. $\omega_0$ and $k_0$ are the frequency and the wavenumber of light in vacuum at the central wavelength. Figure reprinted with permission from [9]. 

## Current and Future Challenges

To show general requirements and design principles for spatiotemporal differentiators, without loss of generality, we consider a one-dimensional periodic grating structure by breaking spatial mirror symmetry [9]. Figure 1(a) schematically shows the spatial-symmetry analysis of spatiotemporal differentiators. Suppose that a pulse with both Gaussian envelopes in the spatial and temporal domains impinges the structure at normal incidence. If the structure has mirror symmetry about the plane x = 0 (indicated by the dashed line) [Fig. 1(a)], due to the mirror symmetry, the phase distribution of the transmitted pulse must also be symmetric about the mirror plane. In this case, the output pulse cannot be a first-order Hermite-Gaussian in $x$ direction, and thus the device cannot perform first-order differential operation in spatial dimension. For example, in order to break the mirror symmetry, the spatiotemporal differentiator is designed by the two silicon rods in each period etched on a silicon substrate with different sizes.

We design the spatiotemporal differentiator by calculating the transmission spectrum function $T$ of the device. Figures 1(b) and 1(c) show the amplitude and phase distributions of $T$ with respect to $k_x$ and $\Omega$, respectively. Here $k_x$ is the wavevector component parallel to the structure interface, and $\Omega$ is the sideband angular frequency from the central one $\omega_0$, i.e. $\Omega = \omega - \omega_0$. Furthermore, the blue circles and orange rhombi in Figs. 1(d) and 1(e) correspond to the amplitudes and phases of the transmission spectrum function $T$ along $\Omega = 0$ and $k_x = 0$, respectively. They show that $T$ exhibits good linear dependence on $k_x$ and $\Omega$ within $-0.07 \le k_x/k_0 \le 0.07$ and $-0.007 \le \Omega/\omega_0 \le 0.007$, and the phase shifts with $\pi$ at the minima occurring at $k_x = 0$ and $\Omega = 0$, respectively, where $\omega_0$ and $k_x=0$ are the frequency and the wavenumber of light in vacuum at the central wavelength. It indicates that the structure enables the first-order differentiation in both the spatial and temporal domains. With such transmission spectrum function, the output amplitude distribution of transmitted pulse will be a typical first-order Hermite-Gaussian profile along each direction, and the phase distribution is a vortex carrying a singularity, corresponding to the central zero amplitude schematically shown in Fig. 1(a). Since the optical vortex exists in $(x,t)$ coordinate, it is a STOV carrying transverse orbital angular momentum. In addition, there are much more degrees of freedom for the parameters to design the spatiotemporal differentiator for the normal incident light, by considering the spatial mirror symmetry breaking.

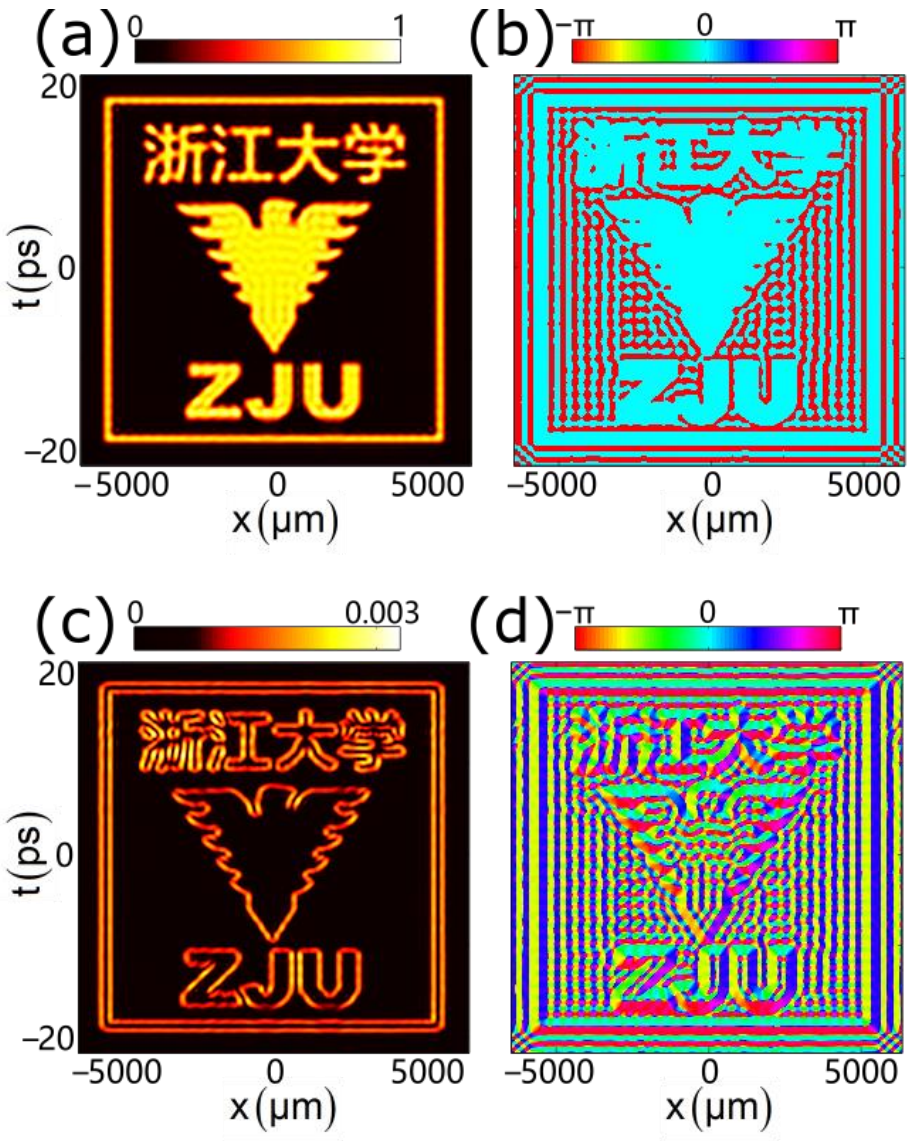

**Figure 2.** (a) Amplitude and (b) phase distributions of an incident pulse envelope as the Zhejiang University logo. The phases of the incident pulse envelope are binary with only 0 or π, without phase singularities. (c) Amplitude and (d) phase distributions of the transmitted pulse envelope. Figure reprinted with permission from [9]. Copyright 2022 John Wiley and Sons/ Cropped from original image.

The application of topological spatiotemporal differentiator and the generation of STOVs is shown in Fig. 2, which is used to detect sharp changes of pulse envelopes. The incident pulse is with the amplitude modulation as the Zhejiang University logo in the spatiotemporal domain [Fig. 2(a)]. Correspondingly, the phases of the incident pulse envelope are binary with only 0 or $\pi$ [Fig. 2(b)]. Figures 2(c) and 2(d) correspond to the amplitude and phase distributions of the transmitted pulse envelope, respectively. Remarkably, Fig. 2(d) shows the generation of large numbers of adjacent STOVs in the spatiotemporal domain. Moreover, as the generated STOVs interfere with each other, the amplitude distribution of Fig. 2(c) shows that the constructive interference occurs at the sharp changes of the incident pulse envelope in both the spatial and temporal domains, while the destructive one strongly takes place where the amplitudes have slight variations. Therefore, the differentiation in the spatiotemporal domain results in the highlighting-sharpness effect of the incident pulse field.

## Advances in Science and Technology to Meet Challenges

To fabricate such spatiotemporal differentiators, we recently show that most of the challenges for fabrication defects can be overcome with the topological protection feature of spatiotemporal differentiator [10]. We elucidate that spatial mirror symmetry breaking introduces a synthetic parameter dimension associated with the metasurface geometry. More importantly, the STOV generation through spatiotemporal differentiator can be topologically protected by the configuration of vortices emerging with the synthetic parameter dimension. The strength of the topological protection is evaluated by the distance between the central vortex of the transmission spectrum function $T$ for the device and its nearest vortex in the $k_x - \omega$ domain. When the distance is small, two vortices with zero points in $T$ are close to each other, which means the spatiotemporal differentiator is much more sensitive, because the vortices can be annihilated together in pairs of opposite charges. In contrast, when the distance is large enough, the vortices are stable due to topological protection from the perturbation: Since the continuous structural perturbation only continuously perturbs the contours $\mathrm{Re}(T) = 0$ and $\mathrm{Im}(T) = 0$, the intersection of the two contours still gives zero points in the $k_x - \omega$ domain. Therefore, we can design the topological spatiotemporal differentiator with topological protection, which has robust immunity to fabrication defects when integrated with other optoelectronic devices for potential applications.

**Concluding Remarks**

The spatiotemporal differentiator is an optical device coupling both spatial and temporal dimension, with applications to detect the sharp changes in the spatiotemporal domain. Practically, the input spatiotemporal modulated pulse could be generated by ultrafast moving objects, and the differentiator has potential applications in detecting the movements of objects. Furthermore, the transmission spectrum function has an isolated zero amplitude in the frequency and wavevector domain, corresponding to the vortex phase distribution carrying a topological charge. Therefore, the device can be used to generate STOVs, which has applications in optical tweezers and optical communications. In addition, we propose that the strength of the topological protection is evaluated by the distance between the two vortices associated with a synthetic dimension. We believe that there are high-order vortices emerging in the synthetic parameter dimension by including more geometry parameters, which could be important in the applications of STOVs and spatiotemporal differentiators.

**Acknowledgements**

The authors acknowledge funding through the National Key Research and Development Program of China (Grant No. 2017YFA0205700), the National Natural Science Foundation of China (NSFC Grants Nos. 12174340 and 91850108), the Open Foundation of the State Key Laboratory of Modern Optical Instrumentation, and the Open Research Program of Key Laboratory of 3D Micro/Nano Fabrication and Characterization of Zhejiang Province.

## 17. Arbitrary vector spatiotemporal beam shaping

Mickael Mounaix, Nicolas K. Fontaine and Joel Carpenter
(The University of Queensland)

### Status

Historically, it has been possible to create reconfigurable optical beams that are functions of space, $F(x,y)$, using components such as spatial light modulators (SLM), or reconfigurable functions of time, $F(t)$ using components such as electro-optic modulators or spectral pulse shapers. By using these respective technologies together in series it is possible to create separable spatiotemporal beams of the form $F(x,y)F(t)$, which have the same transverse spatial profile, $F(x,y)$ for every point in time, $F(t)$. A traditional spectral pulse shaper based on a 2D SLM can create 2D beams of the form $F(x,t)$ [1]–[3], where one transverse spatial dimension, and one longitudinal time/frequency dimension can be controlled independently.

However the general case where the optical beam is fully 3D of the form, $F(x,y,t)$, and could have any amplitude, phase and polarisation at any point in space and time was not experimentally realisable. Hence this class of beams encompassing the majority of spatiotemporal beams, were not experimentally achievable. It would be possible to create such 3D beams directly in the time domain, as is done for lower frequency wave phenomena like microwaves[4]. However attempting to spatially and temporally stitch together such a 3D optical wave in the time domain on femtosecond timescales using tens of thousands of coherently driven electro-optic modulators would be very difficult. Frequency-domain approaches based on spectral pulse shaping are appropriate for femtosecond pulses, however a traditional implementation using a 2D SLM can only create 2D spatiotemporal beams of the form, $F(x,t)$ [1]–[3].

To generate a fully 3D optical beam of the form $F(x,y,t)$ using a spectral pulse shaper, we employed a special multi-plane light conversion (MPLC) system to implement a 1D-to-2D spatial transformation between a 1D array of spots and a 2D set of Hermite-Gaussian modes [5], [6] . This 1D-to-2D transformation can be thought of as a kind of lookup table, which allows a traditional multi-port spectral pulse shaper to select from a 1D list which selects 2D output spatial components (Hermite-Gaussian modes). In this way a 3D beam can be mapped onto the 2D surface of a traditional SLM and completely arbitrary vector spatiotemporal beams can be controllably created for the first time [7]. Potentially opening up new applications in deep imaging through complex media[8], as well as nonlinear optics [9] and light-matter interactions [10] as traditionally unrealisable spatiotemporal phenomena could now be observed.

### Current and Future challenges

This arbitrary vector spatiotemporal beamshaper [7] enables a new class of optical beam, the most general class, where the beam is an arbitrary function of space, time and polarisation. We believe the fact that previously unrealisable spatiotemporal beams are now available experimentally could open up applications in both fundamental and applied optics, particularly in areas of nonlinear optics and light-matter interactions. Yet only a single device [7] has been demonstrated at time of writing, and wide adoption would benefit from several advances.

Lower Loss : The first device was based on a simplified, single SLM design which was relatively lossy (~17 dB). The first prototype was not optimised for loss, however given commercial spectral pulse shapers typically have ~4 dB of loss, and this spatiotemporal beamshaper has multiple additional components, we believe an implementation below 8 dB of loss would be difficult in practice.

Higher power handling: The device is composed mostly of free-space optics which could be suitably designed for very high power operation. The maximum peak and average power such a device could handle would ultimately be constrained by the power handling capabilities of the SLM.

More spatial degrees of freedom : The total spatial complexity of the beams, are primarily set by two aspects of the optical system. First, a large number of spatial degrees of freedom must be available as independent pixels on the SLM. Second, this single axis of pixels on the SLM must be convertible to 2D spatial modes using a 1D-to-2D spatial transformation that supports the required number of modes. Extending beyond approximately 100 spatial modes would be challenging both from the perspective of a traditional ~2k resolution SLM, but also for the 1D-to-2D transform.

More spectral/temporal degrees of freedom : It is relatively straightforward to adjust the device design to support either shorter temporal features, *or* longer delays. However the ability to do both simultaneously and control a larger total number of temporal features, is ultimately set by the number of addressable pixels on the SLM along the frequency/time axis.

Different wavelength bands : The first device was the combination of a wavelength selective switch (WSS) and MPLC mode multiplexer [5] both taken from optical fibre communications, and hence designed for operation at 1550 nm. Implementing devices for other wavelength bands such as 800 nm (for use with Ti:sapphire lasers) and 1 µm (for use with ytterbium based high-power lasers) would open up applications in these areas.

Ultrawide bandwidth : Modifying the optical design for operation at different wavelength bands is largely a matter of appropriately scaling dimensions, changing optical coatings, etc. What is more challenging is designing a device for a large fractional bandwidth. For example, for applications such as few cycle pulse manipulation, with octave(s) of bandwidth, the design becomes difficult. In particular, the 1D-to-2D spatial transformation is difficult in practice to scale to very large fractional bandwidths as doing so requires a large number of phase planes.

**Advances in Science and Technology to Meet Challenges**

Higher resolution, physically larger SLMs : SLMs are based on LCoS technology often derived from consumer displays. Although there is a drive for consumers towards larger pixel counts for 4k and 8k video, which benefits spatiotemporal beamshaping, there is also a consumer drive towards physically smaller pixels, which reduces cost by allowing more panels to be fabricated on the same wafer. Smaller pixels means more crosstalk between adjacent pixels particularly when working with phase-only modulation and longer wavelengths. Physically larger pixels would be desirable not only to minimise crosstalk but also as the associated larger form factor of the panel would mean better heat dissipation properties for high optical power operation.

Spatial/spectral slicing using multiple SLMs : However the number of total pixels on a single SLM panel will always have limits, and at some scale of spatiotemporal complexity the addressable degrees of freedom will have to be split amongst multiple SLMs. For distributing the spatial degrees of freedom amongst multiple SLMs, one or more panels along the longitudinal direction of the optical path could redistribute light amongst later SLMs each responsible for a subset of the total spatial/polarisation modes. This could have the additional advantage of interfacing with a 2D spot array to 2D HG mode sorter[6], which is inherently more efficient than the 1D-to-2D spatial transformation approach. Stitching together the frequency/time axis across multiple SLMs would likely be more difficult and tedious.

Low-loss, high plane count MPLC : To scale the 1D-to-2D spatial transformation to higher mode counts and/or larger optical bandwidths, it will be necessary to increase the number of phase planes. Which in practice means more loss and tighter mechanical tolerances. To decrease loss, it is expected that the transformation will have to move towards dielectric rather than metallic reflective coatings and the smoothness of each plane will have to be increased to decrease scattering losses. The transformation would ideally be nanofabricated as a whole, rather than assembled manually from parts to meet the stricter mechanical tolerances required by larger plane counts. It is also possible that some combination of MPLC and waveguide structures may be necessary. Particularly for fanning out 2D spot arrays to 1D spot arrays, as this is relatively easy for waveguides to perform but difficult to perform diffractively or refractively using discrete planes.

Fixed mask : For applications that do not require reconfigurability, and simply require the same spatiotemporal beam every time, the phase mask typically displayed on the SLM need not be implemented on an SLM at all, and could be implemented as a fixed mask. This would have the advantage of supporting far more pixels than an SLM and supporting higher powers.

**Concluding Remarks**

Although research into spatial and temporal beamshaping are separately mature technologies, arbitrary vector spatiotemporal beamshaping is still in its infancy. As this technology advances, we hope this new ability to completely control all light's degrees of freedom simultaneously and independently will open up new practical applications related to the delivery of light spatially and temporally deep into complex media and/or to excite new nonlinear processes previously experimentally out of reach. More exotic spatiotemporal, or indeed spatiospectral beams of theoretical interest but experimentally unrealisable could also be explored for the first time.

**Acknowledgements**

*DP170101400, DE180100009, DE210100934*

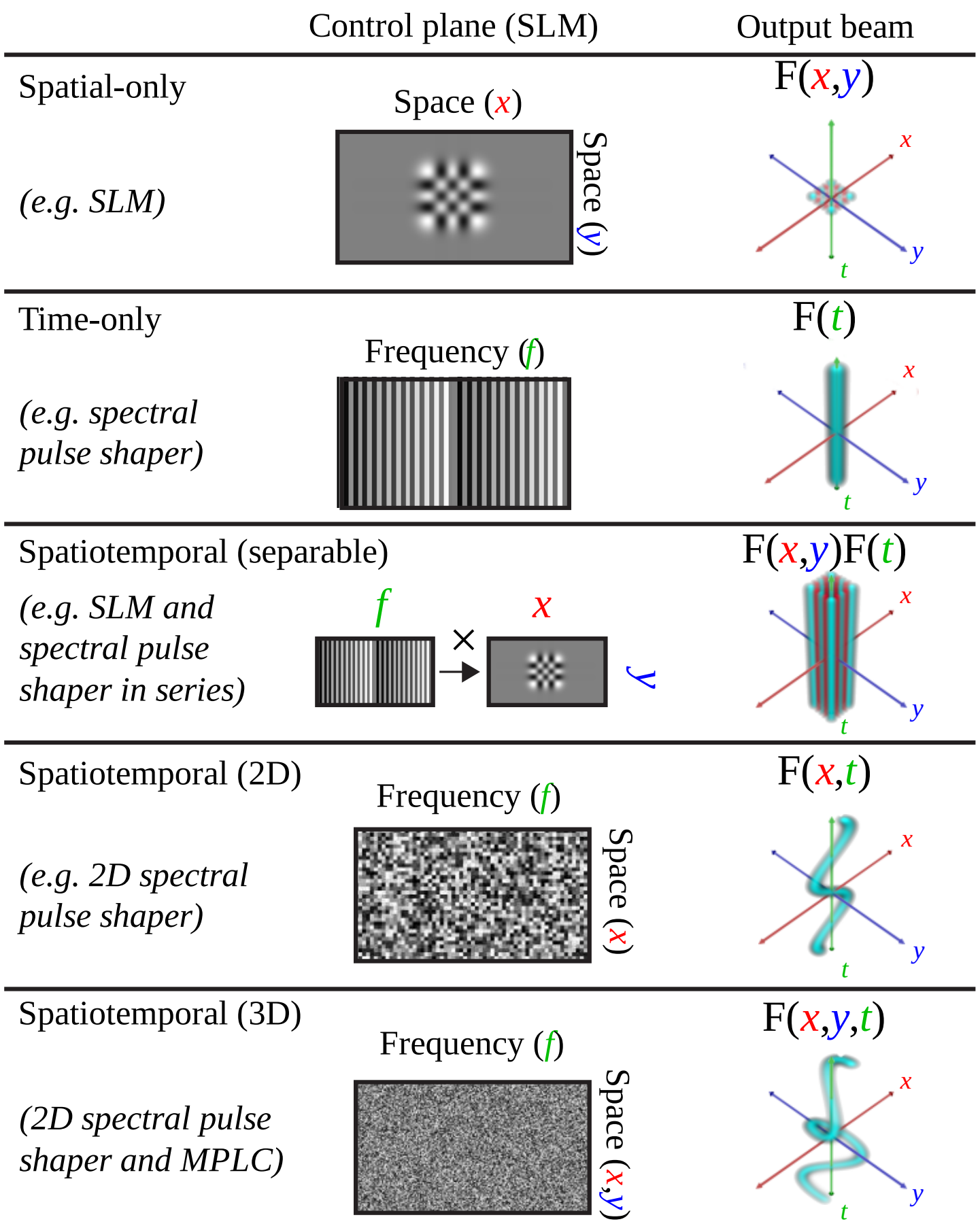

*Figure 1 – Different techniques have been developed to spatially and/or temporally manipulate an optical wave, but only recently have fully 3D arbitrary spatiotemporal beams become possible experimentally. Where any position in space and time can have any amplitude, phase and polarisation.*

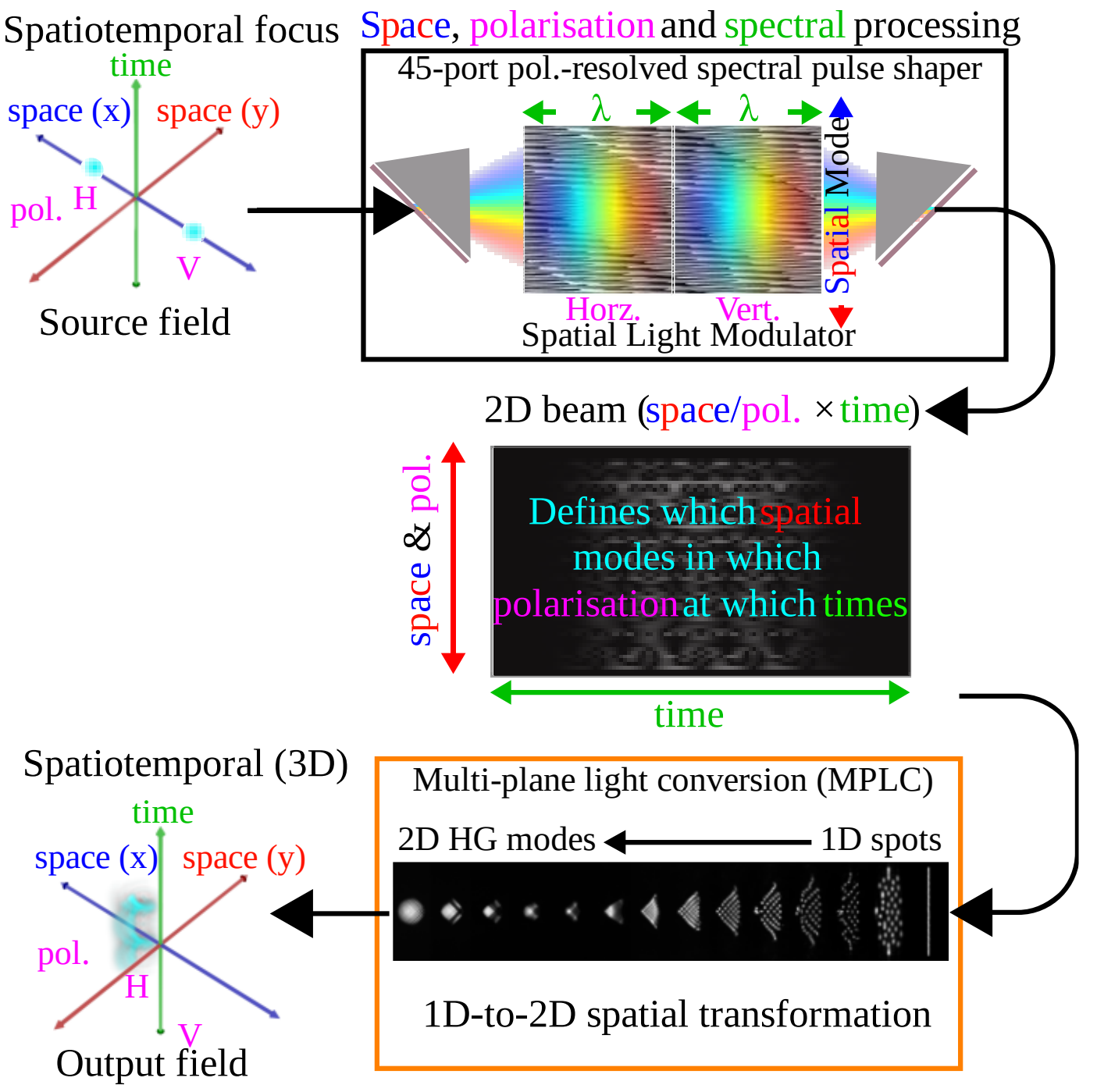

*Figure 2 – Arbitrary vector spatiotemporal beam shaper* [7]. A device which takes an input source (single point in space and time) and remaps the field into an arbitrary distribution in space, time and polarisation.

## 18. Spatiotemporal light control with metasurfaces

Ahmed H. Dorrah and Federico Capasso, Harvard University

### Status

Recent developments in wavefront shaping tools such as spatial light modulators (SLMs) and metasurfaces have enabled versatile control over light's phase, amplitude, and polarization, point-by-point in space, creating complex and exotic classes of structured light [1,2]. While relatively at its infancy, orchestrating structured light in time has also attracted wide attention owing to the massive possibilities it can bring. For instance, structuring light over ultrashort time scales can be widely exploited in probing matter, thus revealing rich physics, as alluded to earlier in this roadmap. The common strategy for synthesizing spatiotemporal modes has traditionally relied on Fourier transform -based pulse synthesizers which mandate pulse spreading and combining as discussed in Sections 6, 7 and 11. At the heart of these setups lies a phase mask that can impart an independent spatially varying phase profile onto each spectral component, thereby sculpting the output spatiotemporal waveform at-will. While SLMs have been deployed as phase masks in most of these setups, metasurfaces have also been widely utilized owing to their subwavelength features, ease of integration, and compact footprint, as schematically illustrated in Fig. 1(a) [3]. Dispersion-engineered metasurfaces are poised to dramatically simplify existing setups even further. For example, instead of spreading the spectral components of the input pulse into different regions of the metasurface (or the phase mask) then combining them with a second grating at the output, a dispersion-engineered metasurface can readily achieve the same task via single transmission. This can be realized by engineering the dispersion behaviour of each meta-atom — the subwavelength-spaced scatterers comprising the metasurface; essentially allowing each wavelength of the incident pulse to "see" a different phase shift and group delay, after a single interaction, as depicted in Fig. 1(b) [4]. While the two former examples were intended for light control in the temporal domain (1D), wavelength selectivity can similarly be achieved in the spatial domain (in 2D). To illustrate this picture, Fig. 1(c) depicts a meta-optic which can impart different phase responses onto three discrete wavelengths, i.e., projecting a wavelength-dependent hologram. Multifunctional response of this kind can be realized by spatially multiplexing the nano scatterers, each optimized for a particular wavelength, as shown in Fig. 1(d) [5]. Spatial interleaving of this kind, however, often leads to low efficiency and undesired diffraction losses that deteriorates the overall efficiency. Dispersion-engineered metasurfaces, on the other hand, can produce this wavelength-selective response on the level of the single meta-atom [6]. One prominent approach relies on guided mode resonances which are highly sensitive to the wavelength of incident light. A meta-atom of this kind is shown in Fig. 1(e). Using this technique, three completely independent spatial profiles can be encoded on red, green, and blue wavelengths, respectively, as depicted in Fig. 1(f) without the need for spatial interleaving [7]. In analogy, by sculpting the spatial profile of each "color" of an incoming pulse, a single meta-optic can in principle produce a spatiotemporal waveform using a very compact and integrated setup. Optics of this kind can simply transform an existing pulsed source to a complex spatiotemporal waveform generator, pushing forward the science and applications of this field.

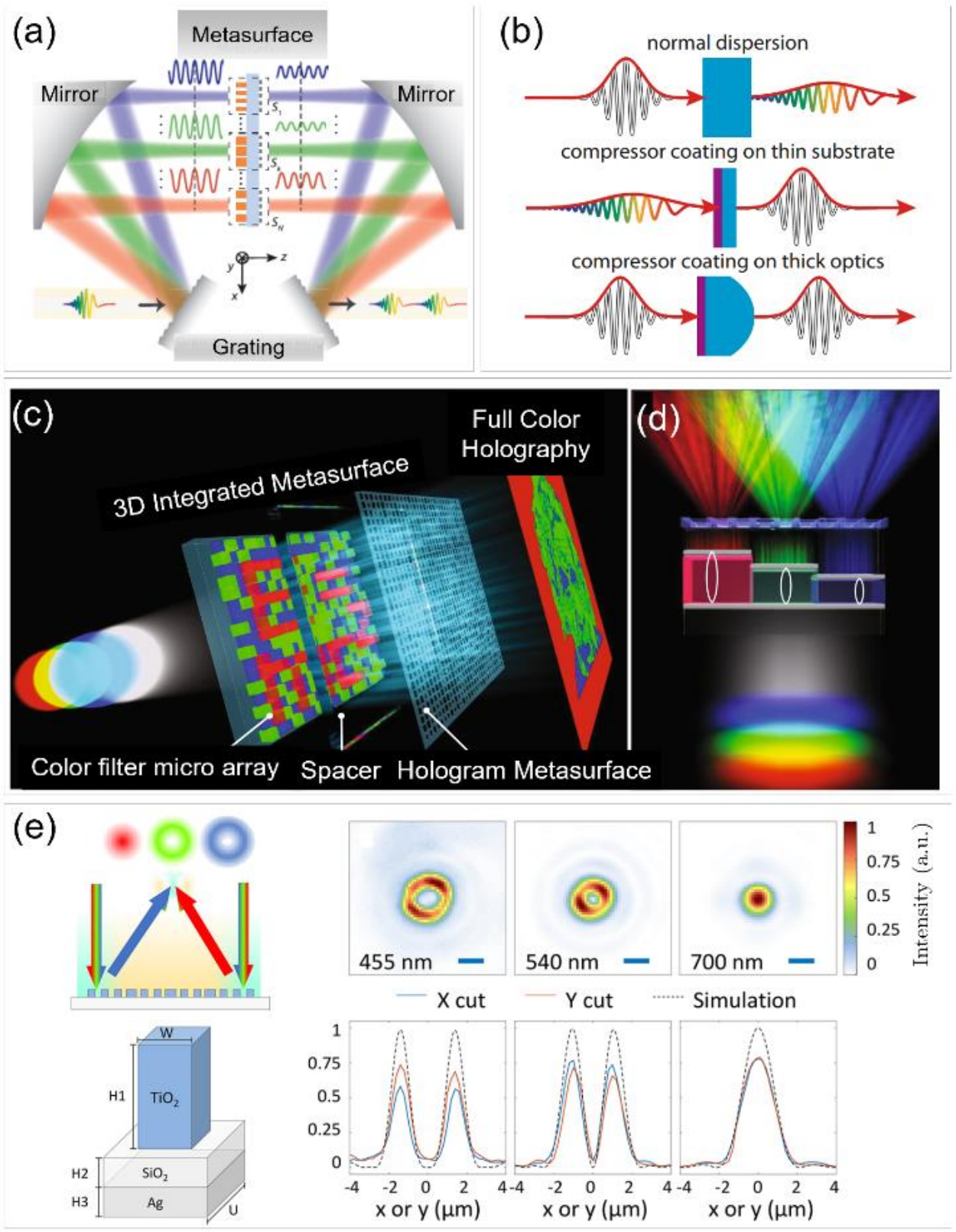

**Figure 1.** Meta-optics with multichromatic response. a) Metasurface-based Fourier transform synthesis setup for ultrafast pulse shaping. It consists of two diffraction gratings, a pair of parabolic mirrors, and a metasurface, partitioned into N super pixels along the x-direction to tailor the temporal characteristics of the output pulse in 1D [3]. b) Metasurface-based pulse compressor can eliminate normal chromatic dispersion experienced by ultrashort pulses, overcoming pulse stretching. The nanocoating (purple) can be stacked onto a thin substrate to shorten the output pulse or onto a thick optic to compensate for its group delay dispersion [4]. c) Schematic of 3D-integrated metasurface for full-color holography by vertically stacking a color-filter microarray underneath and using broadband illumination [5]. d) The three micro-units of the 3D-integrated metasurfaces. The color filters consist of metal/dielectric/metal Fabry–Pérot cavity resonators. Each RGB component will transmit through its adequate filer to excite the correct pixel on the metasurface, generating three independent far-field images [5]. e) Wavelength-selective beam generator focuses incoming R G B light into OAM states $\ell$ of 0, 1, and 2, respectively. The device is composed of phase shifters made of titanium dioxide (TiO2) square nanopillars on a silver (Ag) substrate with a thin layer of silicon dioxide (SiO2) in between. The reflection phase can be controlled by adjusting the width of the nanopillar. Measured output intensity profiles are shown on the right (in 1D and 2D) for each incident wavelength. Scale bars are 2 mm. **H**, height; **U**, unit-cell length; **W**, width of the nanopillar [7].

## Current and Future Challenges

Dispersion-engineered metasurfaces stand as a very promising platform for creating wavelength-dependent response as demonstrated in various applications, from multi-color holography to achromatic and wavelength-selective focusing [6]. In contrast to these applications, which typically deploy a CW laser, spatiotemporal pulse shaping inherently mandates the use of a pulsed source. This imposes a limitation on the smallest spectral resolution that can be achieved; that is the maximum number of phase profiles that can imparted by the meta-optic onto the spectral lines, independently. In multi-color holography, for instance, in which the spectral components (R, G, B) are sufficiently separated —a few tens of nano meters apart— it is feasible to decouple the phase profiles assigned to each wavelength. However, this is not the case in spatiotemporal shaping of an incoming pulse with a continuous spectrum. Broadband focusing with metasurfaces tackles this by engineering not only the target phase profile at the desired wavelength but also its higher order derivative (group delay and group delay dispersion), thereby tailoring the dispersion relation over a spectral continuum. However, in spatiotemporal light shaping —where the target phase can be more complex than a lens function, or where the dispersion relation is not a continuous or analytic function— it becomes more adequate to impart independent phase profiles onto a discrete set of closely separated wavelengths. Obtaining an independent phase response for two closely separated spectral lines (a few nanometers apart), on the level of each meta-atom, becomes very challenging as it typically mandates the use of resonant structures which are highly sensitive to fabrication intolerances, often altering the target response. Another challenge lies in the need for high aspect ratio meta-atoms; that is using taller metasurface unit cells to achieve the phase redundancy required for multi-wavelength response. This often poses a limitation in terms of the fabrication time and robustness of the final device. Multi-wavelength phase coverage can also be achieved using more complex meta-atom topologies based on free-form optimization or multi-layered metasurfaces which often introduce additional challenges in terms of computation, fabrication requirements, and design. Lastly, given their static nature, passive metasurfaces lack the dynamic behaviour afforded by SLMs which can be desired in some applications; for e.g., in sensing or free space communications. While active metasurfaces can mitigate this issue, their spatial resolution and multi-wavelength response is still the subject of ongoing research.

**Advances in Science and Technology to Meet Challenges**

Tremendous efforts are still underway to generate spatiotemporal waveforms using compact and integrated setups that directly operate on the input pulse through a single interaction. Local and nonlocal metasurfaces tackle this task in conceptually different ways, by either operating on incident light in real or k-space, respectively. One example of the former is schematically illustrated in Fig. 2(a); a dispersion engineered meta-optic operates on a pulsed source, transforming it in real space to a time-varying optical vortex —i.e., a beam carrying orbital angular momentum (OAM) while changing its OAM state $\ell$, instantaneously. To achieve this, the meta-optic generates a discrete number of propagating modes —for e.g., higher order Bessel beams— equally spaced in the frequency domain such that their temporal beating modulates the envelope's intensity and OAM state in time, akin to a Fourier series (Fig. 2(b)) [8]. The beating of the copropagating modes can be precisely engineered over an ultrashort time scale by varying the frequency separation and spatial profiles of the copropagating modes. This design principle can be implemented using a frequency comb source as the input pulse whose spectral lines are in turn independently shaped using a dispersion-engineered metasurface, similar to the one depicted in Fig. 1(e), imparting different Bessel profiles onto different wavelengths (Fig. 2(c)). Since the comb lines are generally tightly spaced (~ 1nm or less), then a Fabry-Perot etalon is needed to filter out the desired number of coherent spectral lines, ensuring that their frequency spacing is compatible with the meta-optic. Hence, by integrating a single flat optic, a regular frequency comb source can be transformed to a spatiotemporal pulse generator giving rise to complex and dynamic spatial profiles at ultrashort time scales (Fig. 2(d)). Other variations of this approach that rely on pulse spreading and combining have also been recently demonstrated as discussed in Section 3 [9-10]. Notably, spatiotemporal pulse shaping with metasurfaces has also been exploited in dynamic beam steering without the use of mechanical components, as depicted in Fig. 2(e) [11]. In this configuration, a (virtual) frequency-gradient metasurface is created by combining a frequency comb source ($\lambda = 720$ nm) with a passive metasurface made of silicon, thereby mapping each incoming spectral line into a spatial optical mode with a different wave vector. Since the spectral lines are phase-locked, their corresponding spatial patterns constructively interfere to generate a 4D dynamic pattern that naturally evolves in time, steering $25°$ in $8$ ps. In contrast to these methods which operate on light locally in real space, nonlocal metasurfaces can engineer the dispersion of incoming pulse directly in k-space, akin to photonic crystals, as described more fully in Section 9.

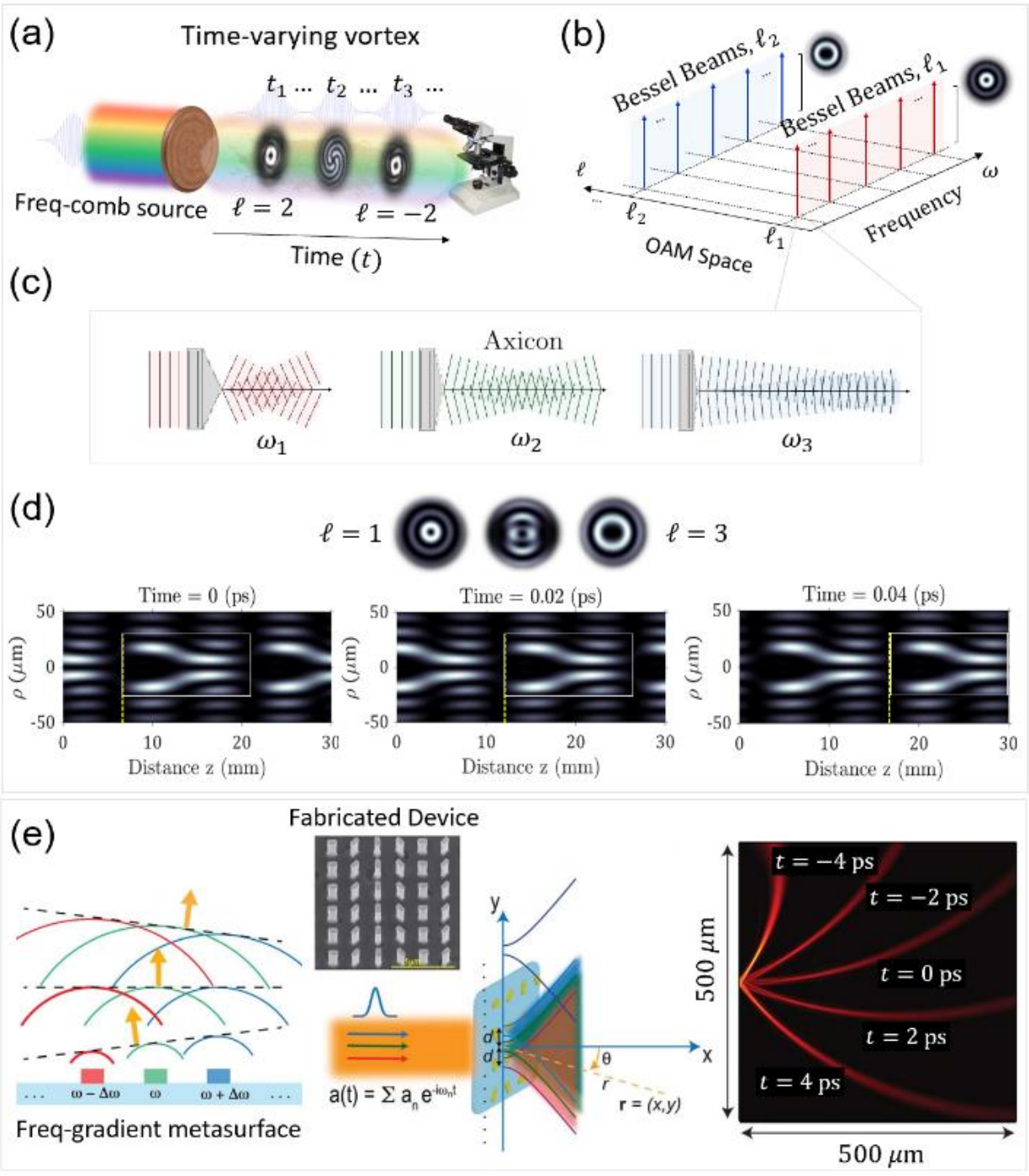

**Figure 2.** Spatiotemporal light control with local and nonlocal metasurfaces. a) A dispersion-engineered meta-optic transforms an input frequency comb source into a time-varying vortex pulse, changing its OAM state $\ell$, dynamically. b) Design principle: the meta-optic implements a superposition of co-propagating higher order Bessel modes with different frequencies and OAM states. The inset depicts the donut-like transverse profile of the modes [8]. c) Each spectral line incident on the metasurface "sees" an axicon profile with a different cone angle. d) The co-propagating Bessel modes interfere with propagation owing to their different propagation constants; their beating produces a vortex mode that changes its OAM state in time (signified by the change in the beam's diameter) as seen in the simulated transverse (top) and longitudinal (bottom) profiles. e) Interaction between an input frequency-comb source and a passive metasurface where each spectral line is mapped to a spatial optical mode, via diffraction and focusing, generating a spatiotemporal optical pattern with time-varying tilt for beam-steering applications. **a**, incoming pulse composed of a discrete superposition of weighted spectral lines; **d**, spatial separation between the output spatial optical modes; **r**, position; **t**, time [11].

## Concluding Remarks

Ultrafast pulse shaping in space and time, simultaneously, is only a first step towards generating multidimensional structured light. Light's polarization and spatial coherence are two additional degrees-of-freedom that, we believe, will be widely studied in this field moving forward as alluded to in Section 17. For instance, by considering the full vectorial nature of light under tight focusing, one can create highly nonparaxial spatiotemporal waveforms with intriguing optical force characteristics. While current experimental setups come in different forms, acting on incident light either locally or nonlocally, we believe that hybrid setups that combine SLMs (for their reconfigurability) and metasurfaces (for their dispersion-engineering capabilities and versatile polarization response) will advance this field of research and accelerate its use in future applications. The abundance of these wavefront shaping tools, ongoing advances in nanotechnology, and emerging trends in inverse design will inevitably enable new modalities for manipulating the fastest magnetic, topological, molecular, and quantum excitations at the nanoscale with high space-time resolution in 4D.

## Acknowledgements

F.C. and A.H.D. acknowledge financial support from NSF (grant no. 1541959), ONR (grant no. N00014-20-1-2450), AFOSR (grant no. FA95550-19-1-0135), and NSERC (grant no. PDF-533013-2019).

## 19. Quantum technologies of spatiotemporally structured light

Andrew Forbes, University of the Witwatersrand (South Africa)

### Status

Light can be tailored, much like cloth, weaving a pattern using all of its degrees of freedom (DoFs), in time and space, for so-called *structured light* [1]. Traditionally this has been done with classical light, for example, amplitude, phase and polarisation control in space [2] and time/frequency control [3] to move to higher-dimensional forms of structured light [4]. The quantum control of structured light is far less developed. A good example is the classical and quantum control of orbital angular momentum (OAM), the former appearing in the early 1990s and the latter only a decade or more later. Quantum states based on spatial modes as a basis can give rise to (in theory) infinite state spaces, described by a bi-photon state given by

$$|\psi\rangle = \sum c_{ij} |M_i\rangle |M_j\rangle \,, \qquad (1)$$

where the states $|M_i\rangle$ and $|M_j\rangle$ describe orthogonal spatial modes (the basis) of two entangled photons, with a probability of detection given by $|c_{ij}|^2$. The dimensionality is determined by the choice of basis (i.e., the modes selected) and the choice of how the modes are created and detected. The benefits of accessing high dimensions (*d*) include a large encoding alphabet that scales as $\log_2 d$ bits per photon, an upper bound to the cloning fidelity that scales as $F = ½ + 1/(1+d)$, with a maximum of 50% for large *d* (it is 83% for $d = 2$ qubits), and thus enhanced security in quantum key distribution (see [5] for a review).

Despite the many advances with spatial control of quantum states, most studies have only considered two-dimensional analogues to polarisation, for qubits. Realising high-dimensional state spaces with structured photons is very much in its infancy [5]. Nevertheless, up to $d = 100$ dimensions has been controlled with one DoF [6] and up to ten photons in two DoFs [7]. Optical cycles are much too fast to allow direct temporal light shaping, and so such light shaping is by frequency. For example, to shape classical light temporally, the frequency components are usually path separated by a dispersive element, mapping frequency to space, shaped in amplitude and phase before a return mapping to reconstruct the desired temporal pulse. In the quantum realm, frequency comb technology has been exploited to produce time-frequency entangled qubits (two dimensional states) across two or more photons [8], extended to ten dimensions as bi-photons [9]. In this case the states $|M_i\rangle$ and $|M_j\rangle$ describe pure single-frequency quantum states of the signal and idler photons, both in a single spatial mode (usually a Gaussian). The phase matching condition of non-linear crystals is a mechanism for producing entangled photons (see Figure 1), and here frequency entanglement is a natural outcome [10]. Time is often used directly as a means of encoding information into quantum states, particularly for quantum key distribution, with so-called time-binning an easily implemented approach to reach high dimensions [11].

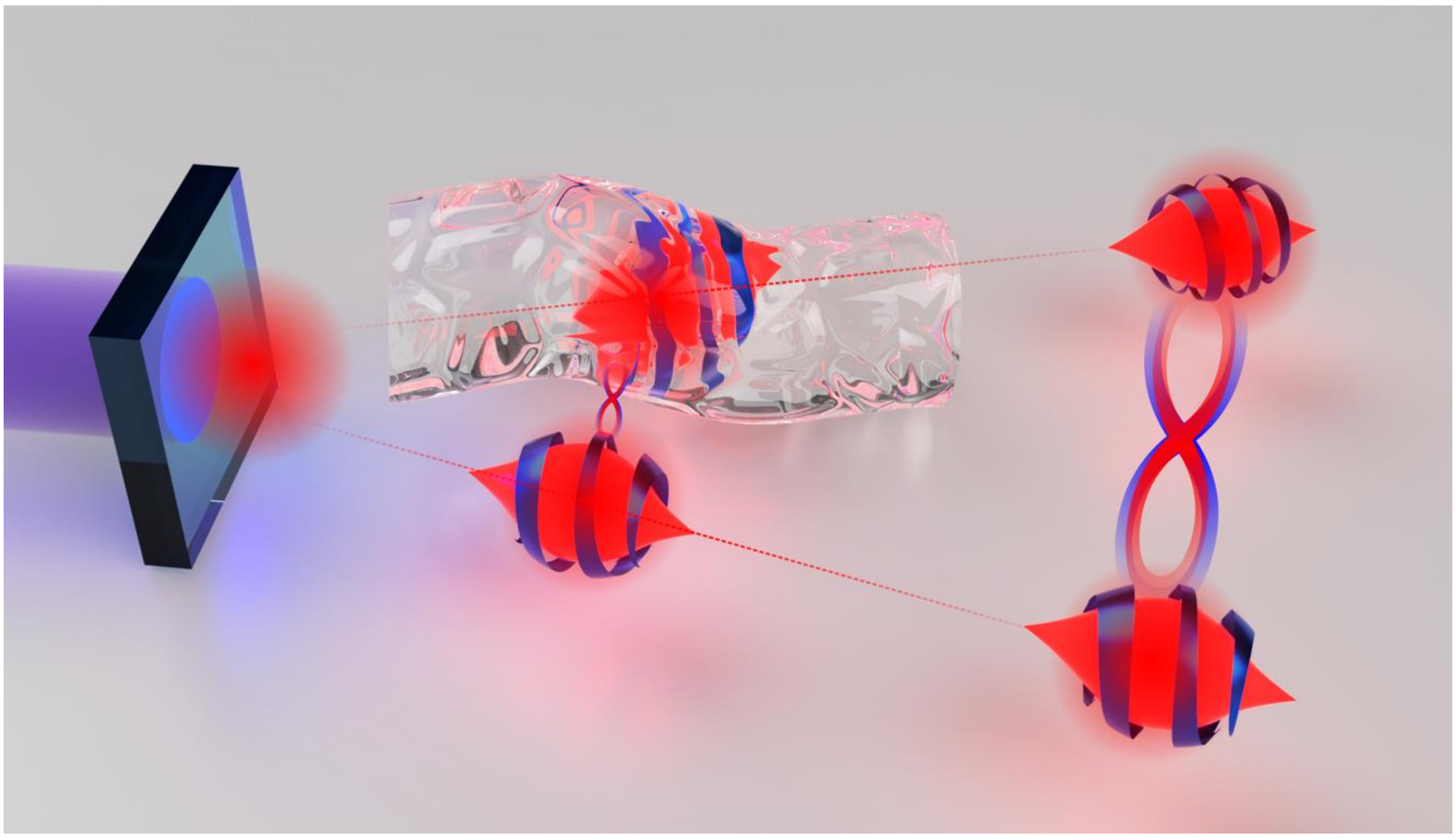

*Figure 1. An illustration of a standard quantum experiment, with a non-linear crystal producing two entangled photons. Can we have the time (indicated by the pulses) and space (indicated by the blue ribbons) DoFs coupled for multi-dimensional spatiotemporal quantum structured light? In general, the challenge is transport of these photons through a noisy channel, denoted by the distorting medium in one arm.*

## Challenges and opportunities

The main challenge in exploiting time binning for quantum information is dispersion, which results in a mixing of states. While dispersion is nominal in free-space, it is rather significant in optical fibre. Further, packing more information into a unit of time means ever more sophisticated detection and control systems. Dispersion compensation fibres do exist, but here the spatial mode associated with the binning is a Gaussian, i.e., a single spatial mode. Multiple spatial modes can be supported in both free-space and optical fibre, but the challenge in both media is modal scattering, which over long distance or after averaging results in mixed states. Compensating this is an ongoing challenge, but with the promise of the quantum equivalent to mode division multiplexing. There exists a tremendous opportunity to merge time with more than one spatial mode for a multi-dimensional state that is high-dimensional in both spatial and temporal degrees of freedom. Temporal control is inherent in all quantum experiments by virtue of measuring in coincidence, but this is a control parameter rather than a controllable DoF. Introducing temporal control in spatial mode entanglement is hindered by the typically slow responses of spatial mode detectors, where patterns are typically detected at rates in the order of display technology (10-100 Hz).

A common limitation to both time and space is that the high-dimensional quantum toolkit for state measurement is still very slow and susceptible to noise. For instance, a quantum state tomography takes on the order of $d^4$ measurements to construct an unknown state, a process than can take weeks for even modest values of $d$. Faster quantitative tools do exist in each DoF, but have yet to be translated across DoFs. Further, dispersion and scattering means that quantum transport of spatiotemporal states is limited in reach, yet the quantum teleportation and repeater tools are not yet developed beyond qubits, very much the focus of cutting edge research the world over.

Despite the challenges, there are many opportunities and exciting prospects to follow to move the state-of-the-art forward. One novel approach is to mix DoFs in such a way that each photon lives in a state space spanned by only one DoF. For instance, it is possible to mix polarisation and spatial modes to realise hybrid entangled states, with photon A polarisation entangled but in a Gaussian mode, and photon B spatial mode entangled but with one polarisation. This has allowed the Gaussian photon (A) to pass through single mode fibre, while exploiting photon B to access an infinite number of spatial mode combinations, for an infinite number of two dimensional states [12]. Such quantum tricks could be extended to time quite easily, allowing an infinite number of *d*-dimensional states to be realised. Another major opportunity is to create and exploit topologically stable quantum states [13]. Topology and quantum have not traditionally been combined, yet here the promise is entanglement robust to perturbations, for long reach and long-lived states in quantum circuits and network. Finally, the low efficiencies inherent in most quantum experiments could be enhanced by a move from natural materials to artificial atoms that make the building blocks of metamaterials and metasurfaces, for ultra-high non-linearity and unprecedented spatial and frequency control for on-demand entanglement in space and time. Indeed, spatiotemporal states have been suggested to enhance sensitivity in metrology when combined with non-linear optical approaches [14].

## Concluding Remarks

Photonics is an exciting prospect in the context of the emerging second quantum revolution, where entanglement is harnessed for quantum information processing, imaging and sensing. The opportunity is to exploit the weakly interacting nature of photons for robust transport of quantum information, but packing information densely into quantum states remains an open challenge, as does getting the information out and integrated into solid state storage systems. Leveraging all DoFs of a photon may hold the key to unlocking this potential, to move us from polarisation qubits to multi- and high-dimensional spatiotemporal states.